# JAN KOCHANOWSKI UNIVERSITY

### DOCTORAL SCHOOL

### FACULTY OF NATURAL SCIENCES

### PHYSICAL SCIENCES

## Enrico Trotti

## SCATTERING OF SCALAR AND TENSOR GLUEBALLS

**Doctoral dissertation written
under the supervision of
prof. dr. hab. Francesco Giacosa**

Kielce 2023

This doctoral thesis was prepared within the research project 'Development Accelerator of Jan Kochanowski University in Kielce,' No. POWR.03.05.00-00-Z212/18, co-financed by the European Union funds under the European Social Fund.

# ABSTRACT

The scalar glueball, the lightest state in the gluonic Yang-Mills (YM) sector of QCD, is stable in that framework. The scattering of two scalar glueballs is therefore a well defined process in YM, which can be studied with the tools of quantum field theory and partial wave analysis. By using a dilaton Lagrangian, which contains a single dimensionful parameter $\Lambda_G$, in the context of proper unitarization procedures, we find that a bound state is expected to form in the $S$-wave if $\Lambda_G$ is below a certain critical value. Additionally, we also evaluate the impact of a cutoff function on the obtained results and we discuss possible future comparison of our model with Lattice QCD and, eventually, with experimental searches.

Usually, partial wave analysis is utilized for describing scattering processes, just as the one mentioned above. Here, we show that the expansion into partial wave can also be useful (together with the covariant helicity formalism) in the study of decays of mesons. This method allows us to describe the ratio between two different waves in the same decays. We use this information to describe decay widths and other relevant quantities, such as the value for the strange-nonstrange mixing angle between isoscalar states. Moreover, we show that the results obtained using the covariant helicity formalism do not change when rotating the reference frame centered in the decaying particle. In this way, an alternative calculation scheme is presented.

Finally, we use the so-called Glueball Resonance Gas (GRG) model to describe the thermal properties of YM below the critical temperature for deconfinement. The quantities obtained from this model (such as the pressure) can be compared with those obtained from lattice works. The use of different lattice spectra in our model leads us to $T_c \approx 320 \pm 20$ MeV. The contribution of heavier glueballs and the interaction between scalar and tensor glueballs turns out to be rather small. Within this context, the scattering of two tensor glueballs, required to estimate its contribution in the GRG model, is investigated in detail.

# Zderzenia skalarnych i tensorowych kul gluowych

## STRESZCZENIE

Skalarna kula gluonowa, najlżejszy stan w gluonowym sektorze Yang-Millsa (YM) QCD, jest w nim stabilna. Rozpraszanie dwóch skalarnych kul gluonowych jest zatem dobrze zdefiniowanym procesem w YM, który można badać za pomocą narzędzi kwantowej teorii pola i analizy fal parcjanych. Używając lagragianu dylatonu, zawierającego pojedynczy wymiarowy parametr $\Lambda_G$, w kontekście odpowiednich procedur unitaryzacji, stwierdzamy, że stan związany powinien powstać w fali $S$, jeśli $\Lambda_G$ jest poniżej pewnej wartości krytycznej. Dodatkowo oceniamy również wpływ funkcji odcięcia na uzyskane wyniki i omawiamy możliwe przyszłe porównanie naszego modelu z Lattice QCD i, ewentualnie z badaniami eksperymentalnymi.

Zwykle do opisu procesów rozpraszania wykorzystywana jest, wspomniana powyżej, analiza fal parcjalnych. Tutaj pokazujemy, że rozwinięcie w falę parcjalne może być również użyteczne (wraz z kowariantnym formalizmem helikalności) w badaniu rozpadów mezonów. Metoda ta pozwala opisać stosunek dwóch różnych fal w tych samych rozpadach. Używamy tych informacji do opisu szerokości rozpadów i innych istotnych wielkości, takich jak wartość kąta mieszania dziwny-niedziwny między stanami izoskalarnymi. Co więcej, pokazujemy, że wyniki uzyskane przy użyciu kowariantnego formalizmu helikalności nie zmieniają się przy obracaniu układu odniesienia wyśrodkowanego na rozpadającej się cząstce. W ten sposób przedstawiono alternatywny schemat obliczeń.

Wreszcie, używamy tak zwanego modelu rezonansowego gazu kul gluonowych (ang, Glueball Resonance Gas - GRG) do opisania właściwości termicznych YM poniżej temperatury krytycznej uwolnienia. Wielkości uzyskane z tego modelu (takie jak ciśnienie) można porównać z tymi uzyskanymi z prac lattice. Zastosowanie różnych widm z lattice w naszym modelu prowadzi nas do $T_c \approx 320 \pm 20$ MeV. Wkład cięższych kul gluonowych i oddziaływanie między skalarnymi i tensorowymi kulami gluonowymi okazują się raczej niewielkie. W tym kontekście rozpraszanie dwóch tensorowych kul gluonowych, wymagane do oszacowania ich wkładu w modelu GRG, jest szczegółowo zbadane.

# TABLE OF CONTENTS

# LIST OF FIGURES

# LIST OF TABLES

# CHAPTER 1

# Introduction

## 1.1 The Standard Model

In this dissertation, we analyse some aspects of Quantum ChromoDynamics (QCD) (theory that describes quarks and gluons), focusing on both conventional ($\bar{q}q$ states) and non-conventional (other) mesons, in particular glueballs (bound states of solely gluons). QCD is a part of the Standard Model (SM) of particle physics, a theory whose formalism has its roots in Quantum Field Theory (QFT).

We start by introducing the main features of the SM. The SM describes three fundamental interactions -electromagnetic, weak and strong-, but not gravity. All the three gauge interactions are mediated by spin-1 bosons, while the matter particles are fermions with spin 1/2. The gauge symmetry group of the SM is $U(1) \times SU(2) \times SU(3)$.

- The electromagnetic interaction is described by the group $U(1)$, which is the only abelian one. Its massless gauge boson is called photon.

- The weak force, whose Lagrangian is invariant under the non-abelian group $SU(2)$, is mediated by three gauge bosons, $W^{\pm}$ and $Z^0$, whose masses are: $M_W = 80.377 \pm 0.012$ GeV and $M_Z = 91.1876 \pm 0.0021$ GeV, respectively [1]. At the electroweak scale ($\approx 246$ GeV), the two interactions above are unified into the electroweak interaction, described by the group $SU(2) \times U(1)$. Although massive vector states may give rise to problems (i.e. non-renormalizability of the theory), the solution was found by introducing the Higgs mechanism [2–6]. Through a Mexican-hat potential, this mechanism makes the vacuum state of the theory being invariant under the $U(1)$ group only instead of $SU(2) \times U(1)$. As a consequence, a scalar particle, the Higgs boson with mass $125.25 \pm 0.17$ [1], emerges.

- The third interaction described by the SM is the strong one. The corresponding theory is QCD and its group is $SU(3)$, meaning that it is invariant under local color transformations [7, 8]. The particles involved in the strong interaction are the 8 massless gauge bosons, the gluons, and a set of fermions, the quarks.

In the following, we will focus on the interaction that is at the core of this work: the strong one.

## 1.2 The strong interaction

In Nature there are in total 6 quark flavours ($N_f = 6$) and 8 gluons. The quantum number of color distinguishes gluons; there are 3 colors ($N_C = 3$): red ($R$), blue ($B$) and green ($G$). Gluons are formed by the combination of these colors with their "complementary" anticolor ($\bar{R}$, $\bar{B}$ and $\bar{G}$). The equivalence:

$$3 \otimes \bar{3} = 8 \oplus 1, \tag{1.1}$$

explains why there are 8 gluons and not 9: since each gluon appears as a combination of color-anticolor states, there are only 8 independent and colored gluons, while the singlet state is white (or colorless) and does not appear in Nature.

It is not rare to find works in which $N_C$ is consider to be different from 3: this does not reflect Nature, but it is a useful extension since the case $N_C \neq 3$ (especially when $N_C$ is large) contains certain simplifications that allow us to understand some phenomenological results of our world with $N_C = 3$. An example of this utility is given by the fact that, at large $N_C$, the decay width of many particles (e.g. $\bar{q}q$ mesons and glueballs) goes to zero. Thus, in this limit, mesons become quasi-stable [9].

The $N_f = 6$ quarks are, instead, divided into two groups: light quarks $u$, $d$, $s$, with masses $m_u = 2.16^{+0.49}_{-0.26}$ MeV, $m_d = 4.67^{+0.48}_{-0.17}$ MeV, $m_s = 93.4^{+8.6}_{-3.4}$ MeV, and heavy quarks $c$, $b$, $t$, with masses $m_c = 1.27 \pm 0.02$ GeV, $m_b = 4.18^{+0.03}_{-0.02}$ GeV, $m_t = 172.69 \pm 0.30$ GeV [1]. In this work, we will concentrate in the light quark sector. Beside flavour, each quark may also appear in 3 different colors $R$, $B$ and $G$.

Two main properties dominate the theory of QCD: *asymptotic freedom* and *confinement*. Experiments of Deep Inelastic Scattering (DIS) have shown that protons are built up of quarks which behave as weakly interacting at really short distances ($\approx 10^{-15}$ m) or at high momentum transferred [10]. The fact that the running coupling goes to zero at short distance is called asymptotic freedom and is reflected in the negative sign of the $\beta$-function discovered in 1973 by Gross, Politzer and Wilczek, Nobel prize winners in 2004. Quarks, despite asymptotic freedom, are trapped because of the increasing of the running coupling (resulting in a confining potential), whose value becomes very large at $\approx 10^{-15}$ m; this phenomenon is called confinement.

The asymptotic states of QCD are hadrons and are divided into two groups, according to a rule which assigns a baryon number $B$ to each quark $q$ and gluon $g$ of the state, namely

$$B_q = \frac{1}{3}, \ B_{\bar{q}} = -\frac{1}{3} \text{ and } B_g = 0\,. \tag{1.2}$$

Thus, hadrons are divided into:

- baryons, fermionic states with $B = 1$,
- mesons, bosonic states with $B = 0$.

We usually refer to conventional baryons if they are formed by three quarks ($|qqq\rangle$) or to conventional mesons if they are formed by a quark and an antiquark ($|\bar{q}q\rangle$).

There are several other possibilities to have the same baryon numbers. For example, $B = 1$ can be also obtained with pentaquarks $|(qq)(qq)\bar{q}\rangle$ and molecular states $|(qqq)(\bar{q}q)\rangle$, while non conventional mesons are states like glueballs $|g...g\rangle$ and tetraquarks $|(\bar{q}\bar{q})(qq)\rangle$. Glueballs are of primary importance during this work. In particular, I focus on:

- Scattering of two scalar glueballs (the lightest according to lattice data), using the formalism of Partial Wave Analysis (PWA) and finding the existence, under certain conditions, of a bound state in the $S$-wave.

- Decays of conventional mesons in a partial wave amplitude covariant helicity formalism [11, 12], focusing on mesons with total angular momentum $J = 1$ and $J = 2$.

- The thermodynamic properties of YM sector of QCD, using the so-called Glueball Resonance Gas (GRG) model.

- Scattering of two tensor glueballs.

## 1.3 QCD symmetries

QCD is a theory based on the local realization of the symmetry group $SU(3)$. The generators $\tau^a$ for this group are given as:

$$\tau^a = \frac{\lambda^a}{2}, \tag{1.3}$$

where $\lambda^a$ are the Gell-Mann matrices, with $a = 1, ..., 8$. The generators $\tau^a$ are $3 \times 3$ matrices, Hermitian and traceless. They are normalized as:

$$\text{Tr}[\tau^a \tau^b] = \frac{1}{2}\delta^{ab}, \tag{1.4}$$

and they obey the following commutation relation:

$$[\tau^a, \tau^b] = if^{abc}\tau^c, \tag{1.5}$$

where $f^{abc}$ are the structure constants of the group. The full QCD Lagrangian is:

$$\mathscr{L}_{QCD} = \frac{1}{2}\text{Tr}[G^{\mu\nu}G_{\mu\nu}] + \sum_{j=1}^{n_f} \bar{q}_j(i\gamma^\mu D_\mu - m_j)q_j, \tag{1.6}$$

where the contribution of the gluons, described by the Yang-Mills (YM) Lagrangian, is:

$$\mathscr{L}_{YM} = \frac{1}{2}\text{Tr}[G^{\mu\nu}G_{\mu\nu}]. \tag{1.7}$$

The gluon field strength tensor is:

$$G_{\mu\nu} = \partial_\mu A_\nu - \partial_\nu A_\mu - ig_0[A_\mu, A_\nu], \tag{1.8}$$

where $A_\mu = A_\mu^a \tau^a$ is the gluon field. The covariant derivative, that links gluons to quarks, is given as:

$$D_\mu = \partial_\mu - ig_0 A_\mu. \tag{1.9}$$

Note, the Lagrangian (1.6) provides the equation of motions, known as Yang-Mills equations [13]

$$[D_\mu, G^{\mu\nu}] = j^\nu, \tag{1.10}$$

where $j^\nu$ is the color current, defined as $j^\nu = j^{a,\nu}$.

The Yang-Mills equations are highly non-trivial even at classical level. In particular various configurations of chromodynamic fields are unstable due to non-Abelian interactions. The problem is still under study, see e.g. [14, 15], where references to older papers can be found.

Several symmetries and, in some cases, their corresponding breakings, are present in QCD.

- ☞ **Local color symmetry.** In the color space, the gluon fields $A_\mu(x)$ and the quark fields $q_j$ are $N_c \times N_c$ and $N_c \times 1$ matrices respectively. In particular, for $N_C = 3$, $q_j$ is the vector

$$q_j = \begin{pmatrix} R_j \\ G_j \\ B_j \end{pmatrix}, \tag{1.11}$$

where $R_j$, $G_j$ and $B_j$ are the color components -red, green and blue- of the quark $q_j$ of flavour $j = u, d, s, ....$. One can prove that Eq. (1.6) is invariant under the local color transformation given by:

(i) $$A_\mu(x) \to A'_\mu(x) = U(x)A_\mu(x)U^\dagger(x) - \frac{i}{g_0}U(x)\partial_\mu U^\dagger(x) \qquad (1.12)$$

and

(ii) $$q_j \to U(x)q_j. \qquad (1.13)$$

The local $SU(3)$ matrix
$$U(x) = e^{i\omega_a(x)\lambda_a} \qquad (1.14)$$

is unitary for any space-time point $x$.

☞ **Dilatation symmetry and trace anomaly.** Dilatation invariance is valid in QCD at the classical level, in the case of vanishing mass of quarks $m_j$, i.e. the so-called chiral limit. Under this condition, the only parameter contained in the Lagrangian of Eq. (1.6) is the dimensionless coupling $g_0$. Therefore, the 4-dimensional space transformation
$$x^\mu \to x'^\mu = \lambda^{-1}x^\mu \qquad (1.15)$$

is an invariance of the theory, provided that the gluon field transforms as:
$$A^a_\mu(x) \to A^{a'}_\mu(x') = \lambda A^a_\mu(x), \qquad (1.16)$$

and the quark field transforms as:
$$q(x) \to q'(x') = \lambda^{\frac{3}{2}} q(x). \qquad (1.17)$$

In this framework, the divergence of the corresponding Noether current vanishes, making it a conserved quantity:
$$\partial_\mu J^\mu_{dil} = T^\mu_\mu = 0, \qquad J^\mu_{dil} = x_\nu T^{\mu\nu}. \qquad (1.18)$$

The energy momentum tensor of the YM Lagrangian is [16]:
$$T^{\mu\nu} = \frac{\partial \mathscr{L}_{YM}}{\partial(\partial_\mu A_\sigma)} \partial^\nu A_\sigma - g^{\mu\nu}\mathscr{L}_{YM} + symmetrization. \qquad (1.19)$$

Dilatation symmetry is broken by gluonic quantum fluctuations [17]. As a consequence, the coupling constant $g_0$ becomes a running coupling $g(\mu)$, which is a function of the energy scale $\mu$. Then, the energy momentum tensor [18], which is the divergence of $J^\mu_{dil}$ in Eq. (1.18), reads:

$$T^\mu_\mu = \frac{\mu}{4g}\frac{\partial g}{\partial \mu}G^{a\mu\nu}G^a_{\mu\nu} \neq 0. \tag{1.20}$$

The $\beta$-function of the YM theory,

$$\beta(g) = \mu\frac{\partial g}{\partial \mu} < 0, \tag{1.21}$$

is responsible of the non-zero value of $T^\mu_\mu$. The fact that the coupling $g$ decreases when the energy scale gets larger is the asymptotic freedom.

☞ **Chiral symmetry and its spontaneous breaking.** The quark fields $q_a$ ($a = R, B, G$) in the flavour space are vectors with $N_f$ components. We consider the lightest flavour states ($N_f = 3$):

$$q_a = \begin{pmatrix} u_a \\ d_a \\ s_a \end{pmatrix}, \tag{1.22}$$

where $u_a$, $d_a$ and $s_a$ are the flavour components -u, d and s- of the quark $q_a$ of color $a$. Since this and the next symmetries act on the flavour space, for simplicity of notation, from now on we denote $q_a$ as $q$, thus omitting the color index:

$$q_a \equiv q = \begin{pmatrix} u \\ d \\ s \end{pmatrix}. \tag{1.23}$$

The quark field can be separated into two parts, left- and right-handed:

$$q = q_R + q_L, \tag{1.24}$$

where

$$q_R = P_R q, \qquad P_R = \frac{1}{2}(1 + \gamma_5),$$
$$q_L = P_L q, \qquad P_L = \frac{1}{2}(1 - \gamma_5). \tag{1.25}$$

The two parts of the quark field transforms independently in the group $U(N_f)_R \times U(N_f)_L$:

$$q = q_R + q_L \to U_R q_R + U_L q_L, \tag{1.26}$$

where $U_R \in U(N_f)_R$ and $U_L \in U(N_f)_L$.

The group $U(N_f)_R \times U(N_f)_L$, under which $\mathscr{L}_{QCD}$ is invariant in the chiral limit, can be rewritten as:

$$U(N_f)_R \times U(N_f)_L \equiv U(1)_V \times U(1)_A \times SU(N_f)_V \times SU(N_f)_A. \quad (1.27)$$

These terms correspond to:

$$\begin{aligned} U(1)_V &\longleftrightarrow U_L = U_R = e^{i\theta t^0}, \\ U(1)_A &\longleftrightarrow U_L^\dagger = U_R = e^{i\nu t^0}, \\ SU(N_f)_V &\longleftrightarrow U_L = U_R = e^{i\theta_a^V t^a}, \\ SU(N_f)_A &\longleftrightarrow U_L = U_R^\dagger = e^{i\theta_a^A t^a}, \end{aligned} \quad (1.28)$$

where $a = 1, ..., N_f^2 - 1$.

The last term of Eq. (1.27), $SU(N_f)_A$, is not a group, but is a well defined transformation and it is spontaneously broken. Namely, the vacuum state of QCD is not invariant under $SU(N_f)_A$ transformations. Consequently, there are, in the chiral limit ($m_u = m_d = m_s = 0$), massless states which are the (pseudo-)Goldstone bosons of the spontaneous chiral symmetry breaking (this result will be seen more clearly when talking about the linear sigma model): for $N_f = 3$ these are the pions, the kaons and the $\eta$-meson. These mesons then acquire a small mass due to the non-zero quark masses, see below.

☞ **Axial symmetry and its anomaly.** The axial transformation $U(N_f = 1)_A$ is a subgroup of $U(N_f)_R \times U(N_f)_L$, that corresponds to

$$U(1)_A \longleftrightarrow U_R = U_L^\dagger = e^{i\nu t^0}. \quad (1.29)$$

The quark field transforms as:

$$q \to Uq = U_R q_R + U_L q_L = e^{i\nu t^0} q_R + e^{-i\nu t^0} q_L = e^{i\nu t^0 \gamma_5} q. \quad (1.30)$$

Quantum fluctuations break this symmetry (axial anomaly), since the divergence of the corresponding Noether current is nonzero (even in the chiral limit):

$$\partial_\mu A^{\mu,0} = \partial_\mu(\bar{q}\gamma^\mu\gamma^5 q) \neq 0. \quad (1.31)$$

☞ **Explicit symmetry breaking because of nonzero quark mass.** The chiral sym-metry is also broken explicitly. This is due to the mass term in Eq. (1.6),

$$\mathscr{L}_{mass} = \sum_{i=1}^{N_f} m_i \bar{q}_i q_i, \tag{1.32}$$

which is non-zero because even if the $u$, $d$ and $s$ quarks are light, they are not massless. This causes pions (as well as kaons and $\eta$) to acquire a mass and to be quasi-Goldstone bosons. The mass coefficient $m_i$, being dimensional, causes $\mathscr{L}_{mass}$ to break some of the symmetries previously mentioned. Indeed, although the symmetry under $U(1)_V \times SU(N_f)_V$ is preserved even for massive quarks, provided that the quarks have the same mass, the symmetry under $U(1)_A \times SU(N_f)_A$ is broken as soon as $m_i \neq 0$.

## 1.4 The QCD running coupling

Because of the behaviour of the strong coupling $g$, that becomes large at low energies, one cannot solve QCD analytically. In general, we can consider the energy spectrum as divided into 2 domains: perturbative and non perturbative [19, 20]. In the high energy domain, perturbative methods (as in QED) can be applied. In the low energy domain, lattice QCD simulations and effective approaches are typically used.
The $\beta$-function controls the scale dependence of $g$; in the UV regime, this can be expressed as a perturbative series:

$$\beta(\alpha_s(\mu^2)) = \frac{\partial \alpha_s(\mu^2)}{\partial \ln\mu^2} < 0, \tag{1.33}$$

where

$$\beta(\alpha_s) = -\sum_{i=0} b_i \alpha_s^{i+2} \tag{1.34}$$

and

$$\alpha_s(\mu^2) = \frac{g^2(\mu)}{4\pi}. \tag{1.35}$$

For a QCD theory with $N_C$ colors and $N_f$ flavours, in the perturbative regime ($\alpha_s << 1$), the leading term (one-loop level) of the series gets the form:

$$b_0 = \frac{11 N_C - 2 N_f}{12\pi}. \tag{1.36}$$

Thus, for YM we have:

$$b_0 = \frac{11 N_C}{12\pi}. \tag{1.37}$$

The renormalization group equation (Eq. (1.33)) can be solved exactly if we keep only the leading term $b_0$:

$$\alpha_s(\mu^2) = \frac{\alpha_s(\mu_0^2)}{1 + b_0 \alpha_s(\mu_0^2) \ln(\mu^2/\mu_0^2)}. \tag{1.38}$$

By considering the YM case ($N_f = 0$), Eq. (1.38) can be rewritten as:

$$\alpha_s(\mu^2) = \frac{1}{b_0 \ln(\mu^2/\Lambda_{YM}^2)}, \tag{1.39}$$

whose behaviour is shown in Fig. 1.1. $\Lambda_{YM}$ is called the Yang-Mills scale (which correspond to the so-called Landau pole in Eq. 1.39). The behaviour of $\alpha_s$ at low energies cannot be directly evaluated within YM in particular and QCD in general. Still, many works provide results on how the running coupling in the two domains, perturbative and non perturbative, is realized. An example is given in the analysis done in [20] *.

For a generic perturbative QCD process, the corresponding amplitude looks like:

$$\text{Amplitude} = \sum_n g^n C_n, \tag{1.40}$$

where $C_n$ are certain coefficients depending on the kinematic parameters.

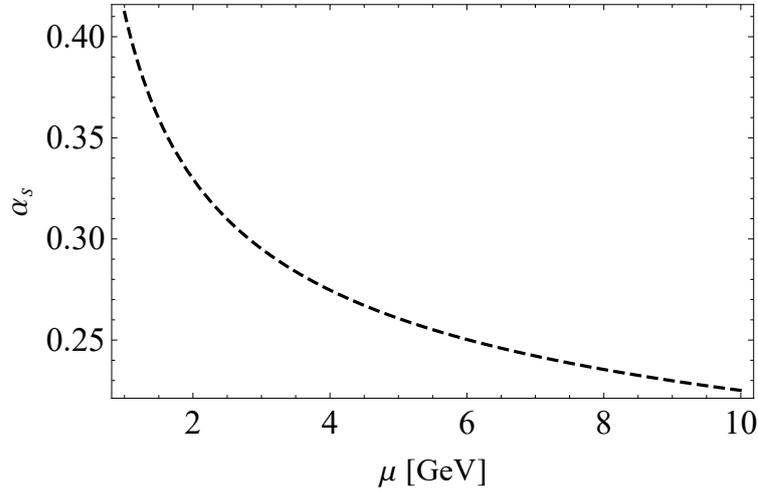

**Fig. 1.1** The plot of the running coupling $\alpha_s$ as a function of $\mu$, according to Eq. (1.39), obtained using $b_0 = 33/(12\pi)$ and $\Lambda_{YM} = 250$ MeV.

The series in Eq. (1.40) does not converge in the strict sense even at high energy, but it converges in an asymptotic sense[†]. The use of perturbation theory at the lowest order leads to an infinite value of the strong coupling when the exchanged momentum approach the value of the Landau Pole

$$\Lambda_{YM} \approx 250 \text{ MeV}, \qquad (1.41)$$

that is the only parameter of YM. This pole is an artifact of the one loop calculation, still it gives an idea of the value of $\mu$ at which non-perturbative phenomena set in.
For low values of $\mu$, the running coupling

$$\alpha_s(\mu^2) \approx 1, \qquad (1.42)$$

therefore the convergence of the series in Eq. (1.40) is not valid (even not asymptotically): in this region perturbation theory fails. Thus, it is convenient to model the low-energy regime of QCD using as degrees of freedom the composite particles -hadrons- instead of quarks and gluons.
The knowledge of the behaviour of the coupling $\alpha_s$ in the low energy region (IR) is important to understand the dynamical chiral symmetry breaking [21]. The non-perturbative QCD (nPQCD) regime ([22–29]) cannot be described by quarks and gluons as fundamental particles, therefore different models have been developed to be applied in the low momentum exchanged region, see below.

---
*The plot shown in Figure 4.1 of [20], was obtained taking into account both experimental and theoretical constraints.

## 1.5 Effective approaches for QCD

Effective models and effective theories are useful to study QCD in the low energy regime. As it is not possible to solve analytically QCD, these descriptions are a powerful tool in our hands, even though we need to keep in mind that the validity of any of these approaches is limited in some sense [30].

### 1.5.1 General discussion on effective approaches

The effective Lagrangians are built in such a way to reproduce some of the symmetries of QCD. In particular, an effective description of the low energy regime is dominated, because of confinement, by the chiral symmetry and its breaking patterns.
Effective models can be also implemented by an additional tool: one can indeed consider the limit at which the number of colors $N_C$ is large (see the review of Witten in Ref. [9]). The large-$N_C$ limit provides us with several simplifications, e.g. the masses of $\bar{q}q$ mesons and gluballs states are constants ($\propto N_C^0$) and become stable. Large-$N_C$ is then used to distinguish dominating and suppressed terms in effective approaches.

---

[†]We remind here that an asymptotic convergence of a series is not a full convergence, but it is a convergence of the first terms of the series.

When talking about the construction of an effective Lagrangian, we may distinguish between theories and models.

In the effective theories, the Lagrangian is obtained via a series expansion around a particular parameter. In the energy region of applicability, one can, in principle, obtain better results by going to the next order of expansion, even though this is linked with the introduction of new parameters. Examples of effective field theories are, for the low energy regime, the chiral perturbation theory ($\chi$PT) and the pionless effective field theory ($\pi\!\!\!/EFT$), and, for the high energy regime, non-relativistic QCD (nRQCD), soft collinear effective theory and heavy quark effective theory (HQET).

The Lagrangian of an effective model is, instead, written starting from some of the symmetries of QCD. Among all the possible effective models, it is worth to cite, for the purpose of this work, the linear sigma model (LSM) and its generalisation, in particular the extended linear sigma model (eLSM), that contains also the dilaton, the vector and the axial-vector mesons, see below.

### 1.5.2 The Extended Linear Sigma Model (eLSM)

The LSM has its basis in 1960 [31], when Gell-Mann and Levy proposed a model in order to study pion-nucleon interactions. That model, which contains (besides the nucleons) only the three pions and the sigma meson, has been gradually extended throughout the years, including more and more particles, thus leading to the eLSM [32–37].

For a didactical introduction of the eLSM see Ref. [7], which we briefly summarize below. The simplest effective Lagrangian, which includes only a complex field $\Phi$, can be written (in the chiral limit and by neglecting the axial anomaly) as:

$$\mathscr{L}_{LSM}^{\Phi} = \frac{1}{2}\left[(\partial^\mu \Phi)^\dagger(\partial_\mu \Phi) - m_0^2 \Phi^\dagger \Phi - \lambda(\Phi^\dagger \Phi)^2\right]. \tag{1.43}$$

The field $\Phi$ contains only the scalar isosinglet $\sigma$ and the pseudoscalar field $\pi$. It corresponds to the case $N_f = 1$. Upon defining

$$\Phi = \sigma + i\pi, \tag{1.44}$$

the chiral symmetry is a simple rotation in the plane $(\sigma, \pi)$, or equivalently a $U(1)$ rotation of $\Phi$: $\Phi \to e^{i\alpha}\Phi$.

The potential, expanded in terms of $\sigma$ and $\pi$, takes the form:

$$V(\sigma, \pi) = \frac{m_0^2}{2}(\pi^2 + \sigma^2) + \frac{\lambda}{4}(\pi^2 + \sigma^2)^2, \tag{1.45}$$

where $\lambda > 0$ to guarantee stability. The form of the potential is directly connected to the sign of $m_0^2$:

- if $m_0 \in \mathbb{R} \longrightarrow \sigma = \pi = 0$ is the unique minimum of the potential and is located in the origin. As a consequence, the two particles $\pi$ and $\sigma$ have the same mass $m_0$.

- if $m_0 = i\mathfrak{m}$, $\mathfrak{m} \in \mathbb{R} \longrightarrow$ the potential has a local maximum in the origin, surrounded by a circle of equipotential global minima located at a distance

$$|\phi| = \phi_N = \sqrt{-\frac{m_0^2}{\lambda}}, \tag{1.46}$$

from the origin. The value of the minima is obtained from $\dfrac{\partial V(r)}{\partial r}$, where $r = |\phi| = \sqrt{\pi^2 + \sigma^2}$.

-Let us take the point $\pi = 0$, thus choosing the minimum at

$$\left(\sigma = \sqrt{-\frac{m_0^2}{\lambda}}, \pi = 0\right). \tag{1.47}$$

Then, the corresponding masses, evaluated as the second derivative at this point, are given by:

$$m_\pi = 0 \qquad \text{and} \qquad m_\sigma = \sqrt{-2m_0^2}. \tag{1.48}$$

We know from PDG that the masses of these two particles have a large difference: the pion ($\approx 135$ MeV) is the lightest hadron and is much lighter than the sigma, which can be identified with $f_0(1370)$ ($\approx 1350$ MeV)[‡]. The only possibility that agrees with Nature is realized for $m_0^2 < 0$, which gives the potential in Fig. 1.2.
This is usually denoted as chiral condensate and is proportional to the vacuum expectation value (VEV) of QCD

$$\langle 0_{QCD}| \bar{q}q |0_{QCD}\rangle \neq 0. \tag{1.49}$$

Since in Nature the pion is light but not massless, the potential of Eq. (1.45) must be modified as:
$$V(\sigma, \pi) = \frac{m_0^2}{2}(\pi^2 + \sigma^2) + \frac{\lambda}{4}(\pi^2 + \sigma^2)^2 - h\sigma. \tag{1.50}$$

---

[‡]The $f_0(500)$ is also a possible candidate for the $\sigma$ state (see Refs. [1, 38]), although often considered to be a four-quark state.

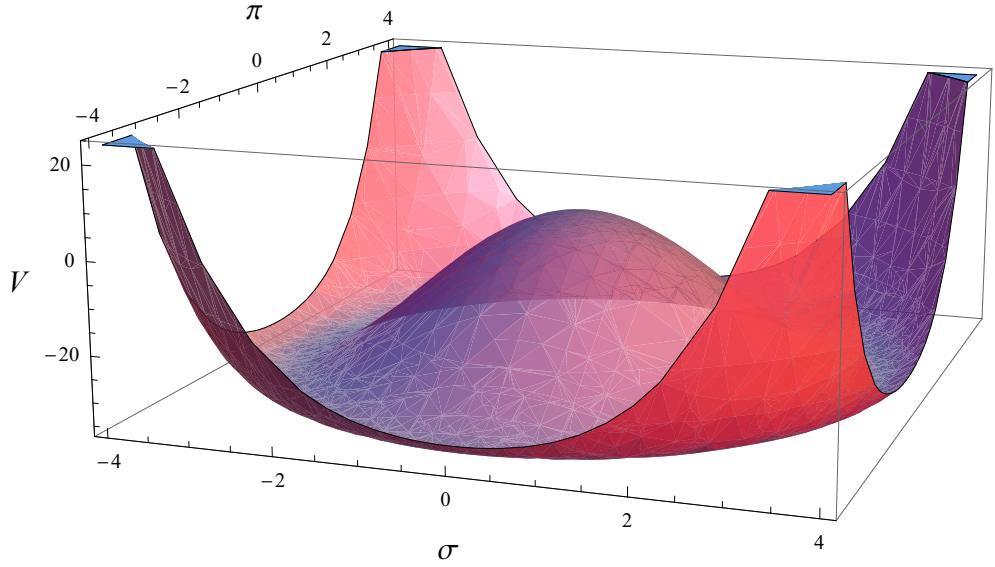

**Fig. 1.2** The shape of the Mexican hat potential, obtained from Eq. (1.45) using $m_0^2 < 0$. The vertical axis is the potential $V(\sigma, \pi)$, while the two horizontal axes are the sigma and the pion fields.

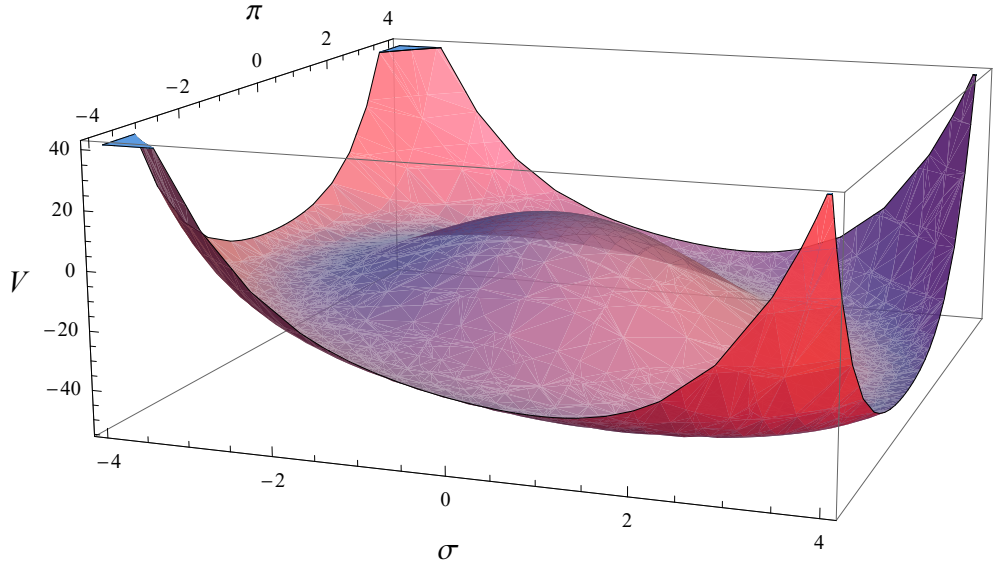

**Fig. 1.3** The shape of the Mexican hat potential, obtained from Eq. (1.50) using $m_0^2 < 0$. The vertical axes is the potential $V(\sigma, \pi)$, while the two horizontal axes are the sigma and the pion fields.

Chiral symmetry is now explicitly broken by the term $-h\sigma$, which modifies the form of the potential as shown in Fig. 1.3. Now the potential has a single absolute minimum at $(\phi_N, 0)$ in the $(\sigma, \pi)$ space, resulting in a nonzero pion mass

$$m_\pi^2 = \left.\frac{\partial^2 V}{\partial \pi^2}\right|_{(\phi_N, 0)} = h\sqrt{-\frac{\lambda}{m_0^2}} > 0, \quad (1.51)$$

as required. Note that
$$m_\pi^2 \xrightarrow{h \to 0} 0 \,. \tag{1.52}$$

Analogously, the value of the mass of the $\sigma$-particle is:

$$m_\sigma^2 = m_\pi^2 + 2\lambda \phi_N^2 \,. \tag{1.53}$$

The Lagrangian of the linear sigma model, $\mathscr{L}_{LSM}^\Phi$, is valid only for pion and sigma fields in the chiral limit and in absence of the axial anomaly. This model corresponds to a linear realization of chiral symmetry, since it contains both the $\sigma$ and the $\pi$ fields; this is different from the $\chi$PT which includes only $\pi$ [39–41].

Namely, one of the possible ways to link the previously discussed $\chi$PT to the eLSM is to study the limit $m_\sigma \to \infty$ by keeping the radius of the minimum of the potential constant (so that $V(0,0) \to \infty$): one obtains a theory based on the only $\pi$-field and where chiral symmetry is non linearly realized.

Moreover, the LSM can be also generalized in order to include additional terms. If we want, for example, to have a model that includes the contribution of the dilaton field $G$ together with its interaction with the field $\Phi$, we can use the following potential:

$$V(G, \sigma, \pi) = V_{dil}(G) + aG^2(\pi^2 + \sigma^2) + \frac{\lambda}{4}(\pi^2 + \sigma^2)^2 \,, \tag{1.54}$$

where
$$V_{dil}(G) = \frac{1}{4}\frac{m_G^2}{\Lambda_G^2}\left(G^4 \ln\left|\frac{G}{\Lambda_G}\right| - \frac{G^4}{4}\right) \tag{1.55}$$

is the so-called dilaton potential [17, 42–47]. The main feature of the potential in Eq. (1.55), which is better explained in Section 2.3.1, is to mimic the trace anomaly in the pure YM sector of QCD.

It is then clear that the Lagrangian of an effective model should be chosen according to the degrees of freedom of the problem that one intend to study.

## 1.6 Aim of this work

Glueballs as bound states of gluons have been widely studied in these last decades, but a question could easily arise: can a bound state of glueballs exist? In order to answer this question, we analyze the scattering of two glueballs, starting from the easiest possible situation: scalar glueballs within a pure YM framework (i.e. considering a world made up of solely gluons and glueballs).

The scalar glueball $J^{PC} = 0^{++}$ is, according to several works [48–50], the lightest bound state component of the YM sector of QCD. Experimentally, scalar glueballs have not been confirmed yet; one of the reasons is their expected mixing with their corresponding conventional mesons. Yet, various works suggest possible candidates for the scalar glueball, e.g. Ref. [51]. This complicated scenario is why it is -at first- easier to study the scattering of two scalar glueballs in pure YM, where the scalar glueball is stable. The results provided by the scattering analysis show that the glueballonium as a bound state of two scalar glueballs may exist.

Next, we study the role of partial wave analysis, together with the so-called covariant helicity formalism, on the decays of conventional mesons, in order to reproduce the value of the ratio between different partial wave amplitudes of the same decay process as reported in PDG [1].

In the final part, we aim to understand the role of various contributions to a Glueball Resonance Gas model, i.e. an analogous version of the so-called Hadron Resonance Gas model, but restricted to the only YM sector of QCD. This model works well below the confinement/deconfinement critical temperature $T_c$. In particular we will show how this model is influenced by the different glueball spectra and by the contribution of additional heavier glueballs and of the interactions. For the interaction part, we consider the scattering of the two lightest glueballs: the scalar, for which we use the results coming from the first part of this work, and the tensor, the we model detailedly.

This thesis is organized in the following chapters:

- Chapter II: We analyze the scattering of two scalar glueballs in the pure YM sector of QCD, first at tree level and then by unitarizing the theory, using an effective Lagrangian (the so-called "*dilaton Lagrangian*" [17]). As a result we get that a bound state of two scalar glueballs ("*glueballonium*") could exist under certain conditions. This result was obtained by the expansion of the scattering amplitude in terms of the angular momentum $\ell$, with a procedure known as partial wave analysis (PWA). This method, introduced in the framework of Quantum Mechanic, is largely used for solving scattering problems in Quantum Field Theory by a decomposition of each wave into its constituent angular-momentum component.

- Chapter III: As mentioned above, PWA is an important method to study scattering. This has been applied to scalar (see above) and tensor (see below) glueballs. Yet, such a method can be also applied to decays. Thus PWA, together with the covariant helicity formalism, will allow us to explain the ratio between different waves in some of the decays present in the PDG. Additionally, it helps us to evaluate the mixing angle for some $\bar{q}q$ nonets, which is an interesting aspect of low energy QCD.

- Chapter IV: After using PWA to solve problems of both scattering and decays, we use the results obtained for the scattering of two glueballs to see how the interaction contributes to the pressure of a Glueball Resonance Gas model. This task uses the results of Chapter II for what concerns scalar glueballs. However, the scattering of two tensor glueballs is modeled in order to evaluate the contribution of this interaction to the GRG model. Additionally, we show also the contribution of a tower of glueballs built with the Regge trajectories. Both contributions, tower of glueballs and interactions, turns out to be negligible when compared with the contribution provided by the lightest non interacting glueballs. We also see that the thermodynamic results provided by lattice could be explained with a better set of lattice masses, without the need of an additional Hagedorn contribution.

This work is based on the following articles:

❶ "Constraints imposed by the partial wave amplitudes on the decays of $J = 1, 2$ mesons."
Vanamali Shastry, Enrico Trotti, Francesco Giacosa
DOI: 10.1103/PhysRevD.105.054022
Phys.Rev.D 105 (2022) 5, 054022

❷ "Glueball–glueball scattering and the glueballonium"
Francesco Giacosa, Alessandro Pilloni, Enrico Trotti
DOI: 10.1140/epjc/s10052-022-10403-z
Eur.Phys.J.C 82 (2022) 5, 487

❸ "On the mass of the glueballonium"
Enrico Trotti, Francesco Giacosa
DOI: 10.31349/SuplRevMexFis.3.0308014
Rev.Mex.Fis.Suppl. 3 (2022) 3, 0308014

❹ "Emergence of ghost in once-subtracted on-shell unitarization in glueball-glueball scattering"
Enrico Trotti
DOI: 10.1051/epjconf/202227403005
EPJ Web Conf. 274 (2022), 03005

❺ "Thermodynamics of the Glueball Resonance Gas"
Enrico Trotti, Shahriyar Jafarzade, Francesco Giacosa
DOI: 10.1140/epjc/s10052-023-11557-0
Eur.Phys.J.C 83 (2023) 5, 390

# CHAPTER 2

# Scalar glueball scattering and partial wave analysis

## 2.1 Theoretical description

Our knowledge of nature is limited and the addition of a new piece to the huge puzzle we are building is anything but easy. The most common way to widen our understanding of the universe is to study how its smallest components interact among each other and what they produce. This process, in particle physics, is called "scattering".

### 2.1.1 Scattering theory

Classically, the problem of scattering is often described considering a particle incident on a target with some probability -$D(\theta)$- that its trajectory is deviated of an angle $\theta$. $D(\theta)$ is the differential scattering cross section and its integral over all the solid angles gives the total cross section $\sigma$, that is the total area under which the incident particle can be scattered by the target.
In the quantum description of the theory, we consider an incident plane wave

$$\psi(x,t) = \psi_0 e^{i(kx-\omega t)}, \qquad (2.1)$$

which produces an outgoing spherical wave when it encounters a scattering potential [52]. In this case, the solution reduces to determine the scattering amplitude $\mathscr{A}(\theta)$.
When discussing about scattering, it is necessary to introduce the $S$-matrix theory. This approach, introduced in 1937 by Wheeler [53], gained popularity in the '60s when it was applied to strong interaction as a possible substitute of the techniques of perturbative QCD (pQCD).
We consider here the Heisenberg picture, for simplicity, but the calculations are often performed in the interaction picture, see Sakurai [54]. We consider a system at a certain initial state ($|i\rangle$) and we ask which is the probability amplitude that it transforms into a certain final state $|f\rangle$. To this end, we introduce an evolution operator $U(t_f, t_i)$ as:

$$U(t_f, t_i) = e^{-iH(t_f - t_i)}, \qquad (2.2)$$

where $H$ is the time-independent Hamiltonian and $t_f > t_i$. We then introduce an operator $S$, defined as:

$$S = \lim_{t \to \infty} U(t, -t), \qquad (2.3)$$

which allows us to get the probability of transition between the two states. In particular, the matrix element $\langle f | S | i \rangle$ of the operator $S$ gives the transition probability from the initial state, measured at the very far past, to the final state, measured at the very far future:

$$P_{i \to f} = |\langle f | S | i \rangle|^2 = \langle i | S^\dagger | f \rangle \langle f | S | i \rangle. \qquad (2.4)$$

Since the sum over all final states of the transition probability equals unity, then the matrix $S$ is unitary:

$$\sum_f |\langle f | S | i \rangle|^2 = \langle i | S^\dagger \left( \sum_f |f\rangle \langle f| \right) S | i \rangle = \langle i | S^\dagger S | i \rangle = 1 \quad \forall |i\rangle,$$
$$\to S^\dagger S = \mathbb{1}. \qquad (2.5)$$

We can finally split the contribution of the interaction by introducing the reaction operator $T$ defined as:

$$S = \mathbb{1} + iT. \qquad (2.6)$$

In case of no interaction, $T = 0$. The operator $T$ is related to the Feynman amplitude $\mathscr{A}$ as:

$$|\langle f | T | i \rangle|^2 = (2\pi)^4 \delta^{(4)}(p_i - p_f)(2\pi)^4 \delta^{(4)}(0)|i\mathscr{A}|^2, \qquad (2.7)$$

where $p_i$ and $p_f$ are the total momenta of the initial and the final configuration. Out of $\mathscr{A}$, one can also calculate the differential and total cross section, see e.g. [55–57].
One of the ways to study scattering is considering the splitting of the amplitude into partial waves, using the technique known as partial wave analysis.

### 2.1.2 Expansion into partial waves

By considering a scattering process of two particles $\Psi_1$ and $\Psi_2$ into two particles $\Psi_3$ and $\Psi_4$, having 4-momenta $p_1$, $p_2$, $p_3$ and $p_4$:

$$\Psi_1(p_1)\Psi_2(p_2) \to \Psi_3(p_3)\Psi_4(p_4), \qquad (2.8)$$

we introduce the Mandelstam variables for this process:

$$\begin{aligned} s &= (p_1 + p_2)^2\,, \\ t &= (p_1 - p_3)^2 = -2k^2(1 - \cos\theta) \leq 0\,, \\ u &= (p_1 - p_4)^2 = -2k^2(1 + \cos\theta) \leq 0\,, \end{aligned} \qquad (2.9)$$

in which $\theta$ is the scattering angle and

$$k = \frac{1}{2\sqrt{s}}\sqrt{[s - (m_3 + m_4)^2][s - (m_3 - m_4)^2]} \qquad (2.10)$$

is the length of the 3-momentum of an outgoing particle in the center of mass frame. Note that, at the threshold for the creation of the two final particles with masses $m_3$ and $m_4$, the 3-momentum is zero.

For the Mandelstam variables, the following general relation holds:

$$s + t + u = \sum_{n=1}^{4} m_n. \qquad (2.11)$$

From Eqs. (2.9), the total amplitude $\mathscr{A}(s,t,u)$ can be written as $\mathscr{A}(s,\cos\theta)$. Moreover, we can also express it as the sum of all the contribution given by the partial amplitudes $\mathscr{A}_\ell$:

$$\mathscr{A}(s,t,u) = \mathscr{A}(s,\cos\theta) = \sum_{\ell=0}^{\infty} (2\ell + 1)\mathscr{A}_\ell(s) P_\ell(\cos\theta)\,, \qquad (2.12)$$

where $\ell$ is the value of the spacial angular momentum and $P_\ell(\cos\theta)$ are the Legendre polynomials. These polynomials are functions commonly used in physics. We list them up to $P_5$ below:

$$\begin{aligned} P_0(x) &= 1 & P_3(x) &= \frac{1}{2}(5x^3 - 3x) \\ P_1(x) &= x & P_4(x) &= \frac{1}{8}(35x^4 - 30x^2 + 3) \\ P_2(x) &= \frac{1}{2}(3x^2 - 1) & P_5(x) &= \frac{1}{8}(63x^5 - 70x^3 + 15x). \end{aligned} \qquad (2.13)$$

The Legendre polynomials, which are defined according to the Rodrigues formula

$$P_\ell(x) := \frac{1}{2^\ell \ell!}\left(\frac{d}{dx}\right)^\ell (x^2 - 1)^\ell, \qquad (2.14)$$

form a complete set of continuous orthogonal functions for $x \in [-1, 1]$:

$$\int_{-1}^{1} P_m(x)P_n(x)dx = \frac{2}{2n+1}\delta_{mn}, \tag{2.15}$$

where $\delta_{mn}$ is the Kronecker delta.

From Eq. (2.12), we find that each partial amplitude $\mathscr{A}_\ell$ is:

$$\mathscr{A}_\ell(s) = \frac{1}{2}\int_{-1}^{1} d\cos\theta \mathscr{A}(s,\cos\theta) P_\ell(\cos\theta). \tag{2.16}$$

The technique of the PWA can also be used to find the differential and the total cross section, given by:

$$\frac{d\sigma}{d\Omega} = \frac{|\mathscr{A}(s,\cos\theta)|^2}{64\pi^2 s} \tag{2.17}$$

and

$$\sigma = \frac{1}{64\pi s}\int_{-1}^{1} d\cos\theta \, |\mathscr{A}(s,\cos\theta)|^2 = \frac{1}{32\pi s}\sum_{\ell=0}^{\infty}(2\ell+1)|\mathscr{A}_\ell(s)|^2. \tag{2.18}$$

For an exhaustive introduction on PWA, see Ref. [58].

The particles I mostly studied during my work, frequently with the help of PWA, are glueballs.

## 2.2 Glueballs

The history of glueballs started about half a century ago, when they were proposed at the early stages of QCD in analogy with $\bar{q}q$ states [59]. From that moment, many works on glueballs appeared [57, 60, 61] and in the last decades the scientific interest has continued to grow [62, 63].

Glueballs are expected to form in the Quantum chromodynamics (QCD) theory of strong intaractions (see Section 1.2). We recall that QCD is a non-Abelian theory where the gauge bosons can self-interact, with the consequence of a possible existence of bound states of only gluons [61, 64–66].

Many approaches have been used to study QCD and, in particular, the existence and the spectrum of glueballs. In all these works, the scalar state $J^{PC} = 0^{++}$ has always been obtained as the lightest glueball (see Ref. [67] and Fig. 16 in Ref. [49]).

The states listed in PDG [1] with a mass below 2 GeV, having $J^{PC} = 0^{++}$, are $f_0(500)$, $f_0(980)$, $f_0(1370)$, $f_0(1500)$ and $f_0(1710)$. Among these states, special attention is re-quired for the three heaviest; they are indeed expected to be an admixture of three pure states $\bar{n}n$, $\bar{s}s$ and glueball[37, 51, 68–72]. For instance, in Ref. [51] is reported that the state $f_0(1710)$ is the one with the largest glueball content.

Among all the approaches used to study glueballs in particular and QCD in general, it is worth to cite:

- ❋ lattice QCD [49, 50, 73–79]
- ❋ Bethe-Salpeter equations [80–83]
- ❋ QCD sum rules [84–86]
- ❋ Hamiltonian QCD [87]
- ❋ flux tube model [88]
- ❋ matrix model approximation [89, 90]
- ❋ anti-de Sitter (AdS) approaches [91–93].

Out of all these approaches, we mainly used the results obtained by Lattice QCD, for both input values and comparison.

## 2.3 Motivation for the investigation

A question can arise: since many models suggest us the existence of glueballs, can these form a bound state? To answer this question we remind that:

(i) One method to study particles is via the analysis of their scattering.

(ii) The scalar glueball, with a predicted mass of $m_G \approx 1.7$ GeV [49, 50, 78, 79], is the lightest particle in pure YM, thus it is also stable within this framework. The $0^{++}$ glueball is therefore the best candidate in order to investigate the possible existence of a bound state of glueballs.

Following these considerations, we study the scattering of two scalar glueballs. We use a Lagrangian that includes the well known dilaton potential (see also Section 1.5.2), already introduced in the 1980s [17, 42–47]. This potential involves a single scalar glueball field $G$, also called dilaton field, and its form naturally arises as an effective description of the trace anomaly of YM part of QCD. According to this phenomenon, because of the gluonic condensate and of the gluonic quantum fluctuations, a scale $\Lambda_{YM} \approx 250$ MeV is developed at low energy. Analogously to YM, also in the framework of the dilaton potential there is a dimentionful constant $\Lambda_G \propto N_c \Lambda_{YM}$, see below.

Here we work in YM (without quarks); yet our results can be relevant for full QCD, since the parameter $\Lambda_G$, which can be constrained via the scattering analysis, is linked to the decay of glueballs into conventional mesons. Thus our results may help to clarify if the glueball width is large or not [47].

### 2.3.1 The dilaton Lagrangian

We have already introduced the QCD Lagrangian in Section 1.5.2. In absence of quarks (or, equivalently, in the quenched limit, i.e. quarks infinitely heavy), QCD reduces to YM Lagrangian, which we rewrite below:

$$\mathscr{L}_{YM} = -\frac{1}{4} G^a_{\mu\nu} G^{a,\mu\nu} \qquad \text{with } G^a_{\mu\nu} = \partial_\mu A^a_\nu - \partial_\nu A^a_\mu + g_0 f^{abc} A^b_\mu A^c_\nu \,, \qquad (2.19)$$

where $G^{a,\mu\nu}$ is the gluon field-strength tensor, $A^a_\mu$ is the gluon field with $a = 1, \ldots, N_c^2 - 1$ ($a = 8$ for $N_c = 3$), $f^{abc}$ are the $SU(N_c)$ structure constants.

Note, the classic $\mathscr{L}_{YM}$ is invariant under dilatation invariance:

$$x^\mu \to \lambda^{-1} x^\mu \longrightarrow \partial_\mu J^\mu_{\text{dil}} = 0. \qquad (2.20)$$

Yet, this symmetry is anomalously broken by quantum fluctuations, resulting in the trace anomaly of YM:

$$\partial_\mu J^\mu_{\text{dil}} = T^\mu_\mu = \frac{\beta(g)}{2g} G^a_{\mu\nu} G^{a,\mu\nu} \neq 0 \,, \qquad (2.21)$$

which expectation value is:

$$\langle T^\mu_\mu \rangle = -2 b_0 \pi^2 C^4, \qquad (2.22)$$

where

$$b_0 = \frac{11 N_C}{12\pi} \qquad (2.23)$$

(see also Eq. (1.37)) and

$$C^4 = \left\langle \frac{\alpha_s}{\pi} G^a_{\mu\nu} G^{a,\mu\nu} \right\rangle. \qquad (2.24)$$

The term obtained is called the gluon condensate. We note that $G^a_{\mu\nu} G^{a,\mu\nu}$ contains a term which includes the gluon field at the fourth power and its dimension is four.

An effective description can be obtained in low-energy YM, where a Lagrangian that mocks the trace anomaly can be written. This is the so-called dilaton Lagrangian [17, 43,

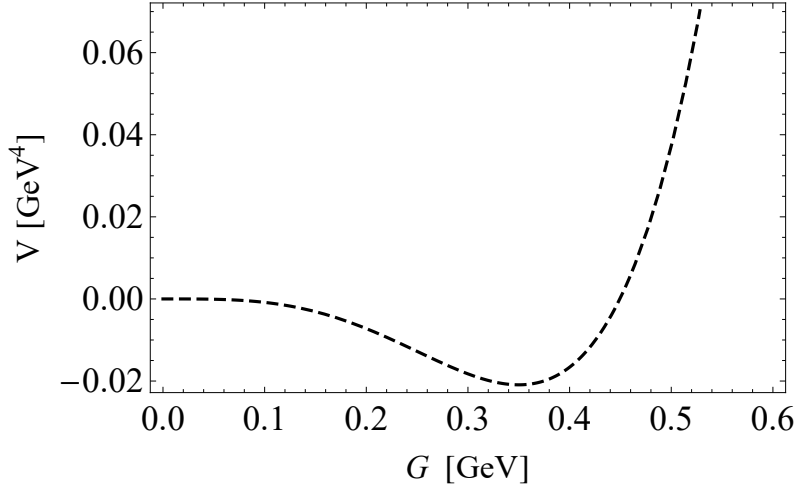

**Fig. 2.1** The form of the dilaton potential obtained using $m_G = 1.653$ GeV and $\Lambda_G = 0.35$ GeV.

45, 46]:

$$\mathscr{L}_{\text{dil}} = \frac{1}{2}(\partial_\mu G)^2 - V(G), \tag{2.25}$$

with

$$V(G) = \frac{1}{4}\frac{m_G^2}{\Lambda_G^2}\left(G^4 \ln\left|\frac{G}{\Lambda_G}\right| - \frac{G^4}{4}\right). \tag{2.26}$$

The minimum of the potential, which is plotted in Fig 2.1, is realized for $G = \Lambda_G$. The parameter $m_G$ can be easily identified as the mass of the glueball as:

$$m_G^2 = \left.\frac{\partial^2 V(G)}{\partial G^2}\right|_{G=\Lambda_G}. \tag{2.27}$$

Next, the shift $G \to \Lambda_G + G$ needs to be performed, see Section 2.4.

Analogously to Eq. (2.21), if we consider the current of the only glueball field, we get:

$$\partial_\mu J_{\text{dil}}^\mu = T_\mu^\mu = 4V - G\partial_G V = -\frac{1}{4}\frac{m_G^2}{\Lambda_G^2}G^4. \tag{2.28}$$

Note, the dilaton potential is the only form which allows us to have $T_\mu^\mu \propto G^4$. This requirement is needed in order to properly reproduce YM, which has a fourth power in the field (see Eq. (2.21)). Namely, this requirement can be written as:

$$\partial_G V - V\frac{4}{G} = \frac{1}{4}\frac{m_G^2}{\Lambda_G^2}G^3, \tag{2.29}$$

which has, as a general solution, the potential given in Eq. (2.26).

The dilaton field is assumed to saturate the trace of the dilaton current (result valid $\forall N_C$):

$$\langle \partial_\mu J^\mu_{\text{dil}} \rangle = \left\langle -\frac{1}{4} \frac{m_G^2}{\Lambda_G^2} G^4 \right\rangle = -\frac{1}{4} m_G^2 \Lambda_G^2 = -2b_0 \pi^2 C^4 \ . \tag{2.30}$$

Thus, $\Lambda_G$ is linked to the gluon condensate $C$ as:

$$\Lambda_G = \sqrt{\frac{11 N_c}{6} \frac{C^2}{m_G}} \ . \tag{2.31}$$

The value of the scalar glueball mass, taken from Ref. [50], is $m_G = 1.653$ GeV, while the value of the gluon condensate $C^4$ is still under debate[*]. As reported in Ref. [94], the lattice fitting for quenched $SU(3)$ gives the value:

$$C^4 = (0.14 \pm 0.02) \text{ GeV}^4 \to C \approx 0.61 \text{ GeV}. \tag{2.32}$$

This value is about an order of magnitude larger than the one extrapolated to the physical quark masses [95],

$$C^4 = (0.022 \pm 0.006) \text{ GeV}^4 \to C \approx 0.39 \text{ GeV}, \tag{2.33}$$

which agrees with the one obtained from phenomenological approaches [96, 97]. If we consider $N_C = 3$, we obtain

$$(\Lambda_G)_{N_c=3} \approx 0.53 \text{ GeV} \tag{2.34}$$

for the quenched value of $C$, and

$$(\Lambda_G)_{N_c=3} \approx 0.22 \text{ GeV} \tag{2.35}$$

for the value from [95]. Because of this uncertainty, it is also important to show how the mass of the bound state of two glueballs, if it exists, is related to the value of $\Lambda_G$ (see later on).

Here we will choose the intermediate value $\Lambda_G \approx 0.35$ GeV as a reference value, which roughly corresponds to the value $C \approx 0.5$ GeV, in the middle between the two values suggested above.

In our work, we show that a bound state of two glueballs (called glueballonium) can exist

---

[*]To be precise, also the mass of the scalar glueball is not exactly fixed. Still, in this case, most of the works are in a quite good agreement, suggesting a mass between 1.65 GeV and 1.8 GeV, both in quenched [49, 50] and in unquenched lattice QCD [79].

below a certain critical value of

$$\Lambda_G \leq \Lambda_{G,crit}, \tag{2.36}$$

which depends on the mass of the glueball (see Ref. [98]). Therefore, the determination of the mass of the scalar glueball together with the one of the glueballonium could in principle provide additional information about the value of the gluon condensate.

The results concerning the glueballonium [99] could be checked in two ways:

- ✽ Lattice YM studies, where both the scalar glueball and the glueballonium are stable. The glueballonium should be then found as an additional scalar state in the lattice spectrum.

- ✽ Experimentally (in full QCD), where, however, two main problems arise: glueballs appear as admixtures of scalar and isoscalar mesons and are not stable, as already discussed. Thus, the glueballonium itself would be unstable, and consequently decay into $4\pi$, $\pi\pi KK$, ... .

Experimentally the glueballonium could be found via its direct production, or indirectly, by producing some heavier particle which could then decay into a glueballonium.

Examples of possible experiments are:

BesIII [100–103],

Belle II [104–106],

LHCb [107–109] and TOTEM [110].

Additionally, in the future, there could be an eventual direct production via proton-antiproton scattering in the planned PANDA experiment [111].

Note, the formalism used in this Chapter for the case of the glueballonium, was analogously used in the case of the Higgs potential (see Ref. [112]), with the difference that in this last case a Higgsonium can form only if the attraction is at least an order of magnitude larger than that provided by the standard model.

## 2.4 Tree-level glueball glueball scattering

Scattering at the tree-level is the first order approximation that we can study and, under certain conditions, it provides the leading contribution to it. In order to study scattering we need first of all a potential, and then to apply QFT and the Feynman rules [55, 56, 113].

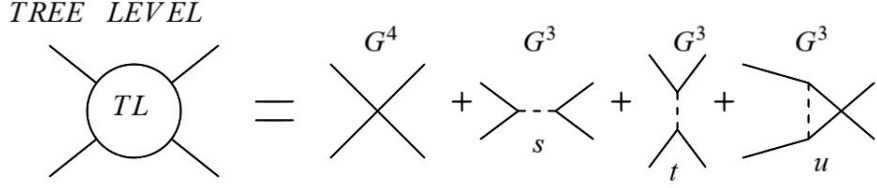

**Fig. 2.2** The terms that are included in the tree-level theory.

2.4.1 The general form of the tree-level amplitude

After the shift $G \to \Lambda_G + G$ is performed, the expansion of the potential of Eq. (2.26) in powers of $G$ provides:

$$V(G) = -\frac{1}{16}\Lambda_G^4 + \frac{1}{2}m_G^2 G^2 + \frac{1}{3!}\left(5\frac{m_G^2}{\Lambda_G}\right)G^3 + \frac{1}{4!}\left(11\frac{m_G^2}{\Lambda_G^2}\right)G^4$$
$$+ \frac{1}{5!}\left(6\frac{m_G^2}{\Lambda_G^3}\right)G^5 + \frac{1}{6!}\left(-6\frac{m_G^2}{\Lambda_G^4}\right)G^6 + ... \quad (2.37)$$

Upon neglecting terms of order $\geq 5$, which are irrelevant for the tree-level $2 \to 2$ scattering, we have:

$$V(G) = -\frac{1}{16}\Lambda_G^4 + \frac{1}{2}m_G^2 G^2 + \frac{1}{3!}\left(5\frac{m_G^2}{\Lambda_G}\right)G^3 + \frac{1}{4!}\left(11\frac{m_G^2}{\Lambda_G^2}\right)G^4 + ... \quad (2.38)$$

This form of the potential allows us to extract the scattering amplitude describing the tree-level contribution, by considering the coefficient in front of the $G^3$ and $G^4$ terms:

$$\mathscr{A}(s,t,u) = -11\frac{m_G^2}{\Lambda_G^2} - \left(5\frac{m_G^2}{\Lambda_G}\right)^2 \frac{1}{s-m_G^2} - \left(5\frac{m_G^2}{\Lambda_G}\right)^2 \frac{1}{t-m_G^2} - \left(5\frac{m_G^2}{\Lambda_G}\right)^2 \frac{1}{u-m_G^2}. \quad (2.39)$$

This form of the amplitude can be understood from the schematization in Fig. 2.2, where we see that the tree-level contribution includes only the 4- and 3-field diagrams, neglecting loop contributions. These contributions will be included later on (see Section 2.5). Note, according to the employed convention, a term with negative sign in the amplitude corresponds to a repulsion, while a positive term an attraction.

Thus, the contact term is clearly repulsive; of the remaining three terms, which originate from the $G^3$-leg, the $s$-channel is also repulsive, while the $t$- and $u$-channels are attractive. We remind that it is possible to write the amplitude as a function of a different set of variables, upon using the relations provided in Eq. (2.9): $\mathscr{A}(s,t,u) = \mathscr{A}(s,\cos\theta)$. Since $\mathscr{A}(s,\cos\theta)$ is symmetric in $\cos\theta$ because of Bose symmetry, only even waves survive.

Next, we will show the form of the lowest non-vanishing waves.

### 2.4.2 $S$-wave

The form of the $\ell = 0$ wave can be obtained as described in Eq. (2.16):

$$\mathcal{A}_0(s) = \frac{1}{2}\int_{-1}^{1} d\cos\theta \mathcal{A}(s,\cos\theta) P_0(\cos\theta) = \frac{1}{2}\int_{-1}^{1} d\cos\theta \mathcal{A}(s,\cos\theta). \quad (2.40)$$

Thus we get:

$$\mathcal{A}_0(s) = -11\frac{m_G^2}{\Lambda_G^2} - 25\frac{m_G^4}{\Lambda_G^2}\frac{1}{s-m_G^2} + 50\frac{m_G^4}{\Lambda_G^2}\frac{\log\left(1+\frac{s-4m_G^2}{m_G^2}\right)}{s-4m_G^2}. \quad (2.41)$$

In Equation (2.41) two singularities are present:

- $s = m_G^2$ corresponds to the single glueball exchange in the $s$-channel. This is a simple pole.

- $s = 3m_G^2$ originates from the projection of the $t$- and $u$-channels into the $s$-channel. This is a branch point.

Both these singularities are needed in certain unitarization procedures, as shown in [114]. The singularity at $s = 3m_G^2$ is the origin of a logarithmic left hand cut. In Figs. 2.3 and 2.4 the two singularities are shown: while the real part of the amplitude $\mathcal{A}_0(s+i\epsilon)$ coincides for the cases $\epsilon > 0$ and $\epsilon < 0$, the splitting of the amplitude into different Riemann sheets is clear from the imaginary part (for a short introduction on Riemann sheets, see App. A).

The knowledge of the tree-level amplitude allows us to derive also the tree-level scattering length $a_0$ and the phase shift $\delta_0$. The scattering length is given by:

$$a_0 = \frac{1}{32\pi m_G}\mathcal{A}_0(s \to s_{th}) = \frac{1}{32\pi m_G}\frac{92 m_G^2}{3\Lambda_G^2} = \frac{23 m_G}{24\pi\Lambda_G^2} \simeq 4.12\,\mathrm{GeV}^{-1}, \quad (2.42)$$

where the threshold is $s_{th} = 4m_G^2$. The phase shift can be obtained by the following general equation:

$$\frac{\eta_\ell(s) e^{2i\delta_\ell(s)} - 1}{2i} = \frac{1}{2}\cdot\frac{k^{2\ell+1}}{8\pi\sqrt{s}}\mathcal{A}_\ell(s), \quad (2.43)$$

where

$$\eta_\ell(s) = \left|1 + i\frac{k^{2\ell+1}}{8\pi\sqrt{s}}\mathcal{A}_\ell(s)\right| \quad (2.44)$$

is the inelasticity and

$$k = \frac{1}{2}\sqrt{s - 4m_G^2} \quad (2.45)$$

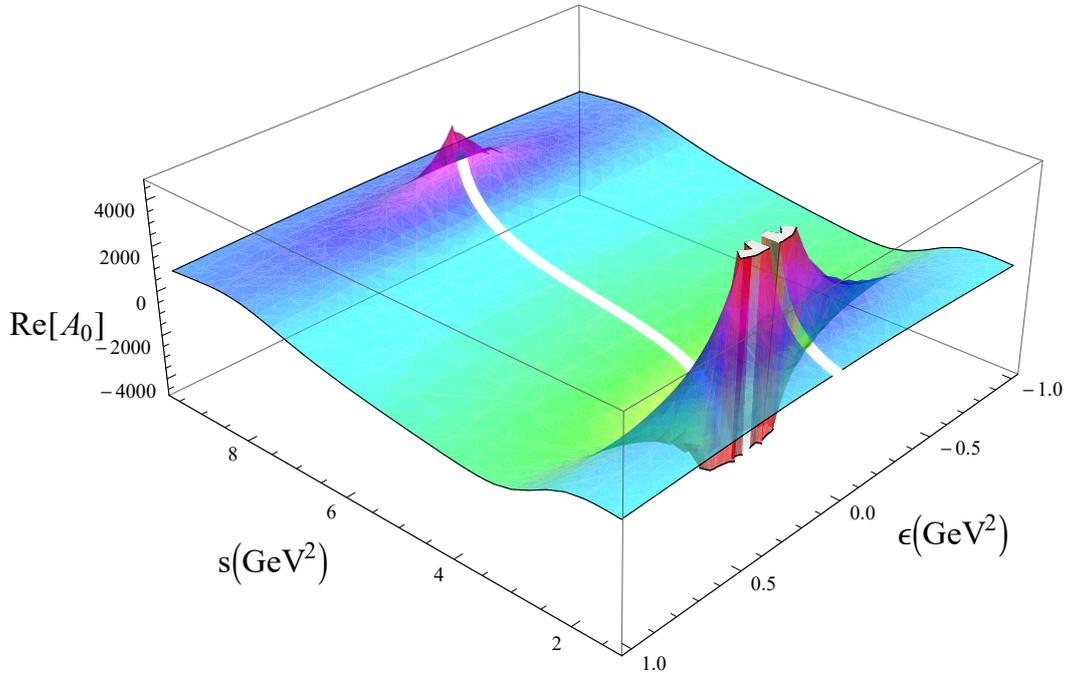

**Fig. 2.3** Real part of the amplitude $\mathscr{A}_0(s+i\epsilon)$.

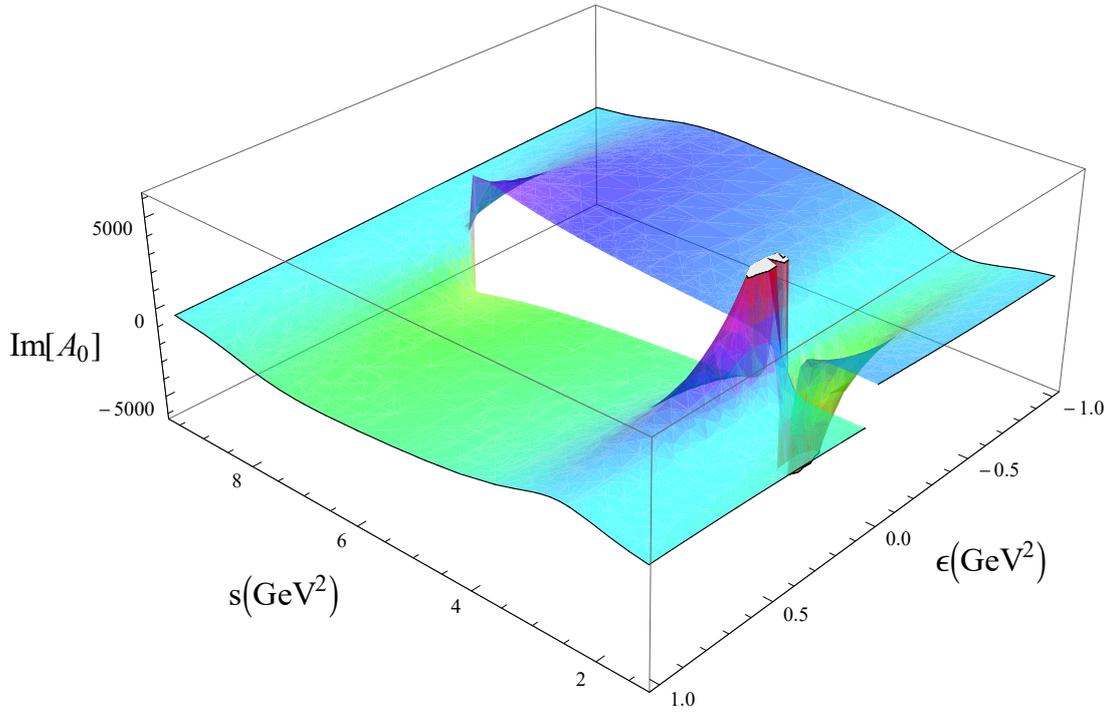

**Fig. 2.4** Imaginary part of the amplitude $\mathscr{A}_0(s+i\epsilon)$.

is the 3-momentum of any particle in the center of mass (for the general version of this equation see Eq. (2.10)). In our case, since we are considering an elastic scattering and then the amplitude is unitary, $\eta_\ell(s) = 1$ and the previous equation gets the form:

$$\frac{e^{2i\delta_\ell(s)} - 1}{2i} = \frac{1}{2} \cdot \frac{k^{2\ell+1}}{8\pi\sqrt{s}} \mathscr{A}_\ell(s). \tag{2.46}$$

In particular, for the $S$-wave, we obtain the following relation:

$$\frac{e^{2i\delta_0(s)} - 1}{2i} = \frac{1}{2} \cdot \frac{\sqrt{s - 4m_G^2}}{16\pi\sqrt{s}} \mathcal{A}_0(s). \tag{2.47}$$

Interestingly, at a certain value $s_c$, the scattering amplitude $\mathcal{A}_0(s = s_c)$ vanishes and the phase shift $\delta_0(s = s_c) \to n\pi$. At this point the attractive and the repulsive contribution compensate making the two glueballs non-interacting. This value, that depends on the mass of the glueball, reads:

$$s_c \simeq 11.86 m_G^2. \tag{2.48}$$

Additionally, it is also a fixed value, in the sense that it does not change with any well behaved unitarization scheme.

### 2.4.3 Higher partial waves

As mentioned above, because of Bose symmetry, odd waves do not survive. Therefore, the second non-zero wave is the $D$-wave, for which $\ell = 2$. Then, one also has $\ell = 4, 6, ...$. We remind that any partial amplitude can be obtained as:

$$\mathcal{A}_\ell(s) = \frac{1}{2} \int_{-1}^{1} d\cos\theta \, \mathcal{A}(s, \cos\theta) P_\ell(\cos\theta). \tag{2.49}$$

In the case of $\ell \geq 2$, we have a general formula for the amplitude, based on the second-kind Legendre function $Q_\ell(x)$:

$$\mathcal{A}_\ell(s) = \frac{50 m_G^4}{\Lambda_G^2} \frac{Q_\ell\left(1 + \frac{m_G^2}{2k^2}\right)}{2k^2}. \tag{2.50}$$

The second-kind Legendre function $Q_\ell(x)$ are closely related to the usual Legendre Polynomials, and the firsts of them are:

$$Q_0(x) = \frac{1}{2}\ln\left(\frac{1+x}{1-x}\right) \qquad Q_2(x) = \frac{3x^2 - 1}{4}\ln\left(\frac{1+x}{1-x}\right) - \frac{3x}{2}$$

$$Q_1(x) = \frac{x}{2}\ln\left(\frac{1+x}{1-x}\right) - 1 \quad Q_3(x) = \frac{5x^3 - 3x}{4}\ln\left(\frac{1+x}{1-x}\right) - \frac{5x^2}{2} + \frac{2}{3}. \tag{2.51}$$

Still, we can see from Eq. (2.50) that the argument of the Legendre function diverges when $k \to 0$. In this case, upon using the definition of $Q_\ell(x)$ and the orthogonality of the

Legendre polynomials (see App. C in [99]), one has:

$$Q_\ell(x) \xrightarrow{k \to 0} \frac{(\ell!)^2}{(2\ell+1)(2\ell)! m_G^2} \left(\frac{4k^2}{m_G^2}\right)^\ell. \tag{2.52}$$

The general scattering "length" (or, more correctly, hyper-volume) reads:

$$a_\ell = \frac{25(\ell!)^2 4^{\ell-2}}{\pi (2\ell+1)(2\ell)!} \frac{1}{\Lambda_G^2 m_G^{2\ell-1}}. \tag{2.53}$$

In particular, the explicit forms of amplitude and scattering length of some of the higher nonzero partial waves are given in App. E.

## 2.5 Unitarization procedures

The results obtained at tree-level cannot produce poles unless these are not explicitly included in the Lagrangian. Thus, the tree-level theory must be implemented by considering loop contribution. This can be done using a process called unitarization. For chiral Lagrangians, many unitarization procedures exist [112, 115–119]. In Ref. [99] two methods were used: on-shell [120, 121] and N/D (in its simplest realization [122, 123]). These two methods provide analogous results near threshold. Here we will focus on the on-shell unitarization. Additionally, as it will be clear later on, a bound state arises in the $S$-wave only. Thus, we will mainly concentrate in the $\ell = 0$ case.

### 2.5.1 On-shell unitarization

This unitarization scheme requires a glueball-glueball self-energy loop function $\Sigma(s)$. Thus, the unitarized $\ell$-th amplitude takes the form:

$$\mathscr{A}_\ell(s) \to \mathscr{A}_\ell^{\text{U}}(s) = \left[\mathscr{A}_\ell^{-1}(s) - \Sigma(s)\right]^{-1} = \frac{\mathscr{A}_\ell(s)}{1 - \mathscr{A}_\ell(s)\Sigma(s)}. \tag{2.54}$$

The choice of $\Sigma(s)$ must take into account some constraints:

❋ The imaginary part is fixed by

$$\text{Im}\Sigma(s) = \theta(s - 4m_G^2)\frac{1}{2}\frac{1}{16\pi}\sqrt{1 - \frac{4m_G^2}{s}}C(U), \tag{2.55}$$

where

$$C(U) = \frac{1}{\left(\frac{s-4m_G^2}{U^2}\right)^2 + 1} \tag{2.56}$$

is a cutoff function. In Ref. [99] the role of the cutoff was not taken into account (this is equivalent to the choice $C(U \gg 2m_G) \approx 1$). The effect of the cutoff in this specific contest is a novel aspect of this thesis.

❋ The pole at $s = m_G^2$ must be preserved. It represents the single glueball exchange in the $s$-channel. Thus, $\Sigma(s = m_G^2) = 0$.

❋ The singularity at $s = 3m_G^2$ comes from the logarithmic term of the tree-level amplitude. It represents the single glueball exchange in the $t$- and $u$-channels. Thus, we also require that $\Sigma(s = 3m_G^2) = 0$.

The role of the cutoff function is to take into account that glueballs are not point-like particles. Roughly speaking, one should integrate over the finite extension, considering the smearing, from the moment at which they start to interact, to the moment in which they end the interaction. It is possible to demonstrate that this effect can be modeled by Eq. (2.55). This form of the cutoff can be also found in the context of nonlocal Lagrangians that consider the finite dimension of glueballs [124–129].

The form of the loop function can be obtained from the dispersion relations [130], as:

$$\Sigma(s) = \frac{(s - m_G^2)(s - 3m_G^2)}{\pi} \int_{4m_G^2}^{\infty} \frac{\text{Im}\Sigma(s')}{(s' - s - i\varepsilon)(s' - m_G^2)(s' - 3m_G^2)} ds'. \qquad (2.57)$$

The loop function is shown in Figs. 2.5 and 2.6, where the influence of the cutoff function on the real and imaginary part of $\Sigma(s)$ is evident. We see that, for a value of the cutoff $U \gtrsim 6 \text{ GeV}$, the form of $\Sigma$ is within the relevant energy range shown in Figs. 2.5 and 2.6, indistinguishable from the case $U \to \infty$.

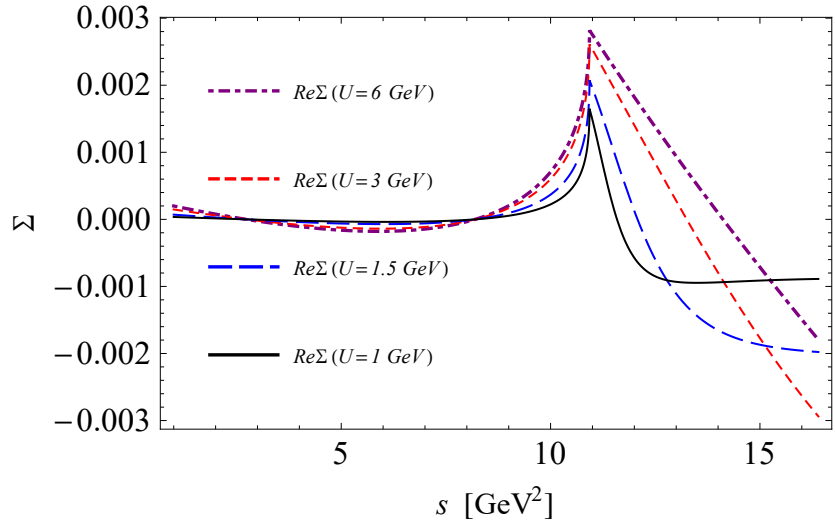

**Fig. 2.5** Real part of the loop function in Eq. (2.57), plotted for different values of the cutoff $U$ (see legend in the plot).

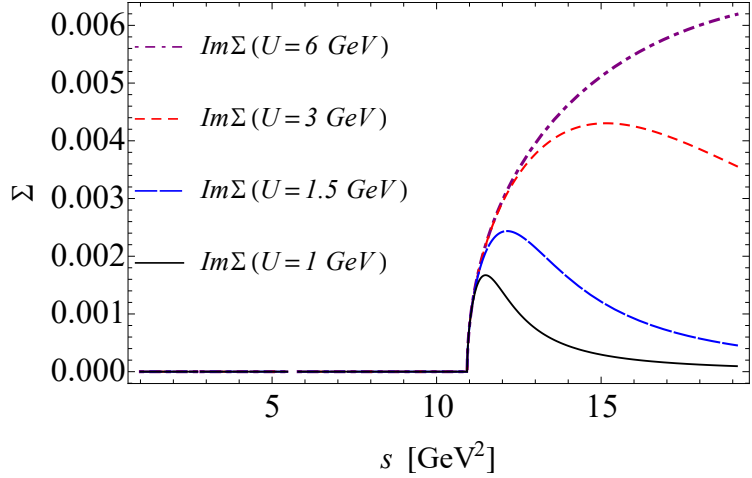

**Fig. 2.6** Imaginary part of the loop function in Eq. (2.55), plotted for different values of the cutoff $U$ (see legend in the plot).

When $U$ is finite, subtractions are not necessary to guarantee convergence. Yet, the subtraction at $m_G$ still guarantees that the glueball mass is fixed. Moreover, as it was exhaustively explained in [114], a single subtracted loop function leads to a state with negative norm, thus not acceptable.

For a certain critical value of $\Lambda_G$, denoted as $\Lambda_{G,crit}$, the unitarized amplitude diverges, together with the scattering length:

$$a_0^{\text{U}} = \frac{1}{32\pi m_G} \mathcal{A}_0^{\text{U}}(s \to s_{th}) = \frac{1}{32\pi m_G} \frac{1}{\frac{3\Lambda_G^2}{92 m_G^2} - \Sigma(4m_G^2)} \,. \tag{2.58}$$

The value of $\Lambda_{crit}$ is obtained from the zero of the denominator of Eq. (2.58):

$$\left. \frac{3\Lambda_G^2}{92 m_G^2} - \Sigma(4m_G^2) \right|_{\Lambda_G = \Lambda_{G,crit}} = 0 \,. \tag{2.59}$$

Rewriting the previous equation, we get (for $U \gg 2m_G$):

$$\Lambda_{G,crit} = m_G \sqrt{\frac{92}{3} \cdot \frac{1}{64\pi\sqrt{3}}} \,. \tag{2.60}$$

The value $\Sigma(4m_G^2) = (64\pi\sqrt{3})^{-1}$, used in Eq. (2.60), is valid only in the limit $U \to \infty$. The form of $\Lambda_{G,crit}$ cannot be expressed analytically for a generic $C(U)$. The critical value of $\Lambda_G$ depends both on the value of the cutoff and on the mass of the scalar glueball, which slightly varies from a lattice work to another [48–50]. Therefore, it can be useful to show the values $\Lambda_{G,crit}$ as function of the cutoff $U$ and of the mass $m_G$. This is done in Table 2.1 and in Figs. 2.7 and 2.8. The value of $\Lambda_{G,crit}$ is relevant since it is the border between the region of existence and non-existence of the bound state for given values of $m_G$ and $U$.

|  | $m_G = 1.475$ [48] | $m_G = 1.653$ [50] | $m_G = 1.710$ [49] |
|---|---|---|---|
| $U = 0.5$ | 0.2674 | 0.2856 | 0.2912 |
| $U = 1$ | 0.3438 | 0.3718 | 0.3804 |
| $U = 1.5$ | 0.3817 | 0.4169 | 0.4277 |
| $U = 3$ | 0.4216 | 0.4676 | 0.4821 |
| $U = 6$ | 0.4348 | 0.4861 | 0.5024 |
| $U \to \infty$ | 0.4375 | 0.4902 | 0.5070 |

**Table 2.1** Values of the critical $\Lambda_G$ for different values of the cutoff $U$ and using the masses from three different lattice works (the errors reported in the papers have not been taken into account). All the values ($m_G$, $U$, $\Lambda_G$) are in GeV.

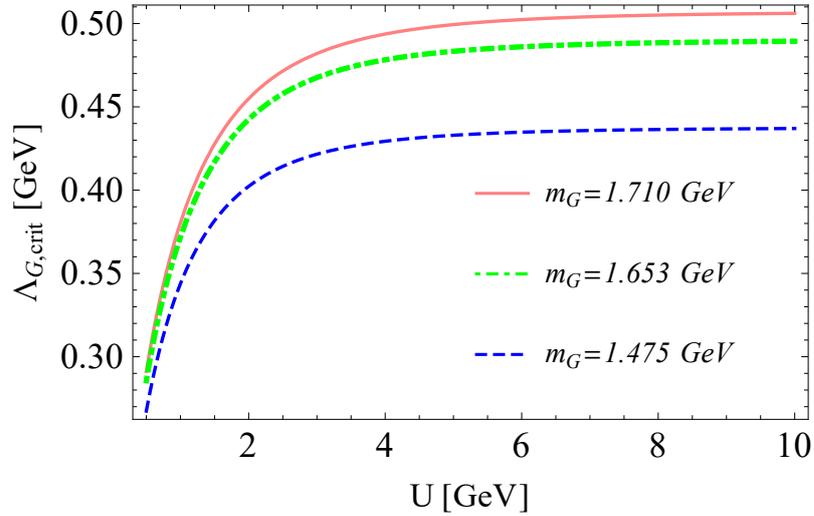

**Fig. 2.7** The value of the $\Lambda_{G,crit}$ as function of the cutoff, for three different values of $m_G$, taken from Refs. [48-50] (see Table 2.1).

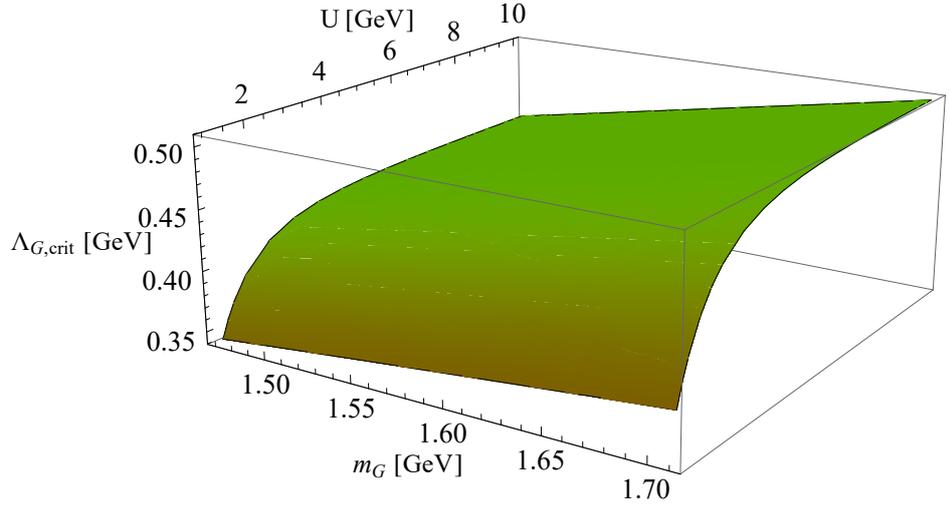

**Fig. 2.8** The value of the $\Lambda_{G,crit}$ as function of the cutoff $U$ and of the scalar glueball mass $m_G$.

In particular, three situations are possible:

- $\Lambda_G > \Lambda_{G,crit} \implies$ no bound state and $a_0^U > 0$
- $\Lambda_G = \lim_{\varepsilon \to 0^+} \Lambda_{G,crit} - \varepsilon \implies$ glueballonium at threshold ($m_B \simeq 2m_G$) and $a_0^U \to \infty$
- $\Lambda_G < \Lambda_{G,crit} \implies$ glueballonium with $m_B \in (\sqrt{3}m_G, 2m_G)$ and $a_0^U < 0$.

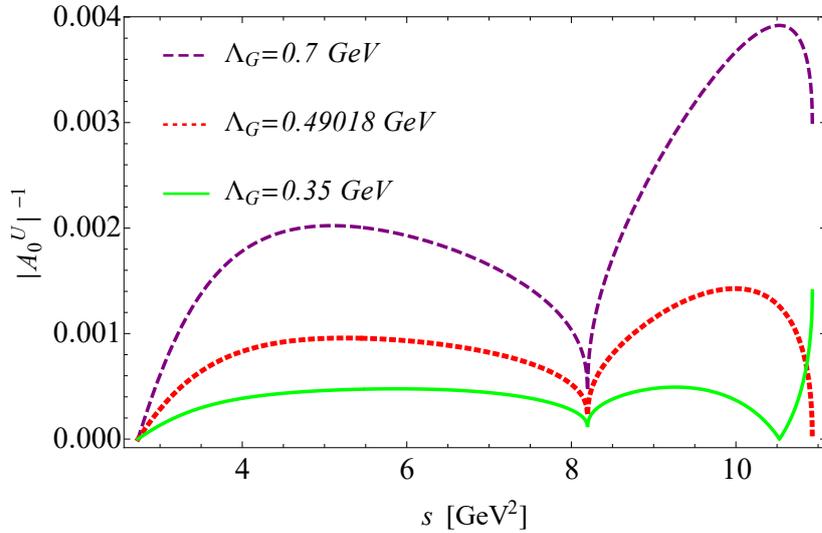

**Fig. 2.9** The inverse of the unitarized amplitude ($\ell = 0$), taken for $U \gg 2m_G$ for three different values of $\Lambda_G$: $\Lambda_G = 0.35$ GeV (green, continuous), $\Lambda_G = \Lambda_{G,crit} = 0.49018$ GeV (red, dotted), $\Lambda_G = 0.7$ GeV (purple, dashed).

We can better visualise these three scenarios in Fig. 2.9, where the inverse of the unitarized amplitude is plotted as a function of the squared energy $s$. In our case, it is better to plot the inverse of the amplitude than the amplitude itself, since this makes it easier to see at which value of $s$ the unitarized amplitude is singular (in this respect, we can compare

Fig. 2.9 with Fig. 2.13). The amplitude is plotted for three different values of $\Lambda_G$: for $\Lambda_G < \Lambda_{G,crit}$ we notice a pole, apart from the one at $s = m_G^2$, below $s = 4m_G^2$. This is the pole representing the mass of the glueballonium $m_B$. If we increase the value of $\Lambda_G$, we reach the point at which $m_B = 2m_G$, corresponding to the critical $\Lambda_{G,crit}$. If we consider a value of $\Lambda_G$ larger than the critical one, we notice that the pole that represents the bound state disappears.

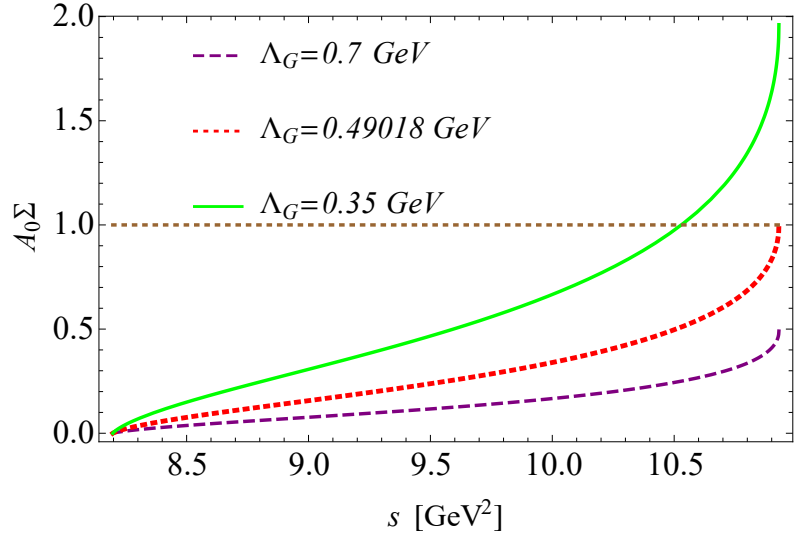

**Fig. 2.10** The value of $\mathscr{A}_0(s)\Sigma(s)$, taken for $U \gg 2m_G$ for three different values of $\Lambda_G$: $\Lambda_G = 0.35$ GeV (green, continuous), $\Lambda_G = \Lambda_{G,crit} = (0.49018 - \epsilon)$ GeV (red, dotted), $\Lambda_G = 0.7$ GeV (purple, dashed). If the curves cross the horizontal line at the value 1, it corresponds to a zero in the denominator of the unitarized amplitude, i.e. $1 - \mathscr{A}_0(s)\Sigma(s)$.

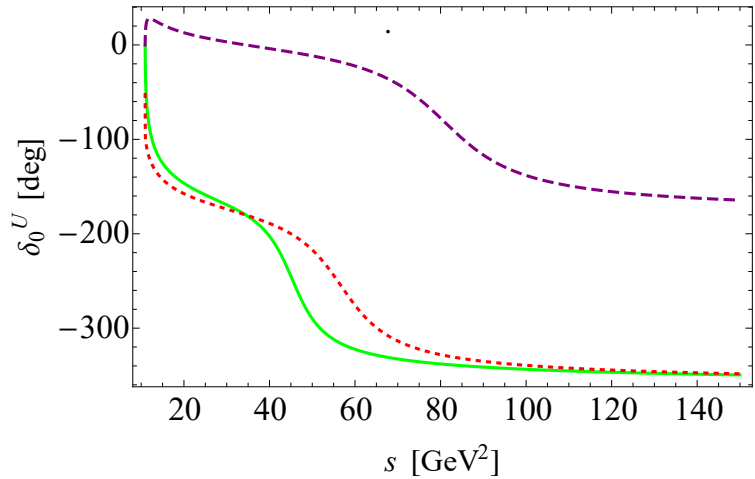

**Fig. 2.11** The unitarized phase shift ($\ell = 0$), taken for $U \gg 2m_G$ for three different values of $\Lambda_G$: $\Lambda_G = 0.35$ GeV (green, continuous), $\Lambda_G \lesssim \Lambda_{G,crit} = 0.49018$ GeV (red, dotted), $\Lambda_G = 0.7$ GeV (purple, dashed).

An analogous description is shown in Fig 2.10, where the denominator of $\mathscr{A}_0^U(s)$ (see

Eq. 2.54) is represented: we see that for $\Lambda_G \leq \Lambda_{G,crit}$ the equation $1 = \mathscr{A}_0(s)\Sigma(s)$ is fulfilled for one value of $s$ within the range $s \in [3m_G^2, 4m_G^2]$. In Fig. 2.9, the value $\Lambda_{G,crit} \approx 0.4902$ GeV was used, which corresponds to the point-like limit situation $C(U \to \infty) \to 1$. The form of the plot would be analogous for any other value of $\Lambda_{G,crit}$ taken from Table 2.1.

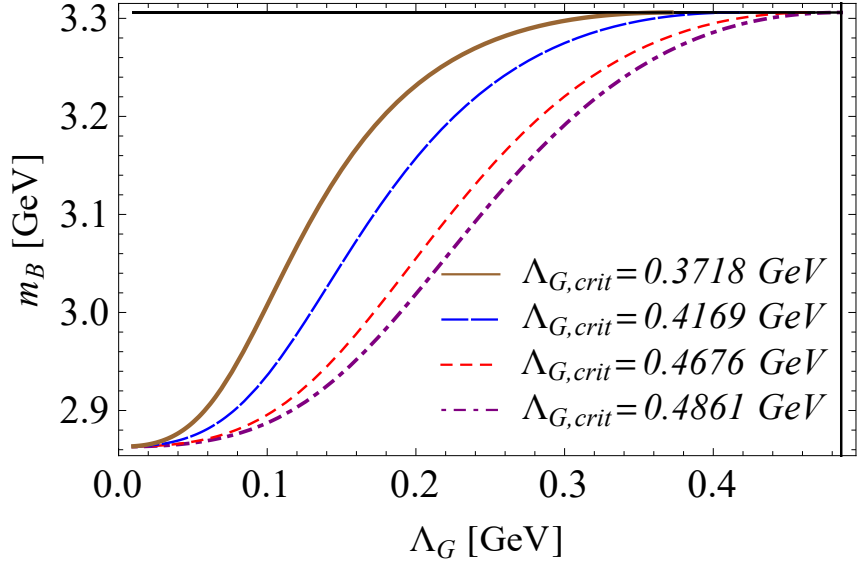

**Fig. 2.12** The mass of the glueballonium as a function of $\Lambda_G$, given for different values of $\Lambda_{G,crit}$ ($m_G = 1.653$ GeV). The values of the cutoff corresponding to the $\Lambda_{G,crit}$ for each curve are reported in Table 2.1. The horizontal line on the top delimits the mass of $2m_G$, while the vertical line on the right corresponds to the value of $\Lambda_{G,crit}$ for $U \to \infty$.

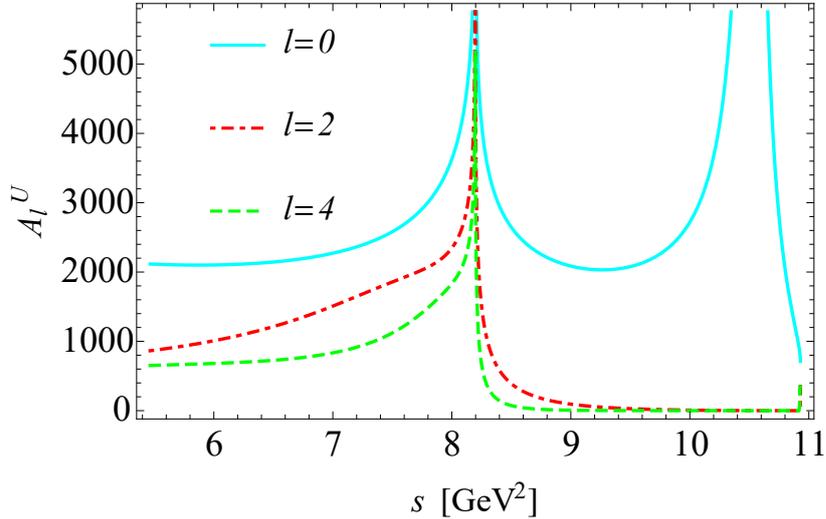

**Fig. 2.13** The unitarized amplitude for three different waves: $\ell = 0$ (cyan, continuous), $\ell = 2$ (red, dotdashed), $\ell = 4$ (green, dashed). Here $m_G = 1.653$ GeV, $\Lambda_G = 0.35$ GeV and $U \to \infty$.

The presence of a second pole below threshold for $\Lambda_G < \Lambda_{G,crit}$ is also supported by the Levinson's theorem, which states that the number of poles $n$ below threshold is related to the difference between the phase shift at threshold and at $s \to \infty$, that must be a multiple of $\pi$:

$$\delta_0^U(s_{th}) - \delta_0^U(\infty) = n\pi, \tag{2.61}$$

with $n \in \mathbb{Z}$. The form of the unitarized phase shift $\delta_0^U(s)$ can be obtained directly from Eq. (2.46), by using $\mathscr{A}_\ell^U(s)$ instead of $\mathscr{A}_\ell(s)$. Indeed, as shown in Fig. 2.11, for values of $\Lambda_G$ below or equal the critical value, $\delta_0^U(s_{th}) - \delta_0^U(\infty) = 2\pi$, since there are two states below the threshold: the single glueball and the glueballonium. For $\Lambda_G > \Lambda_{G,crit}$, we find that the difference equals $n\pi = \pi$, since only the single glueball pole is present.

We now analyse the effects of the different $\Lambda_{G,crit}$ on the mass of the bound state (Fig. 2.12). For the same scalar glueball mass $m_G = 1.653$ GeV, a smaller value of the cutoff $U$ corresponds to a smaller value of $\Lambda_{G,crit}$, resulting in a more squeezed curve compared to a larger value $U$. We notice two limits for the mass of the gluebalonium:

- $m_B(\Lambda_G \to 0) \to \sqrt{3} m_G$: this is a consequence of the applied unitarization scheme and may change for a different scheme, see Refs. [122, 123].

- $m_B(\Lambda_G = \lim_{\epsilon \to 0^+} \Lambda_{G,crit} - \epsilon) \to 2 m_G$, which is expected since at $\Lambda_{G,crit}$ the bound state appears at threshold.

For the values $m_G = 1.653$ GeV and $\Lambda_G = 0.35$ GeV, we find the following values of the gluebalonium mass for different values of the cutoff: $m_B(U \to \infty) = 3.245$ GeV, $m_B(U = 3\,\text{GeV}) = 3.27$ GeV and $m_B(U = 1\,\text{GeV}) = 3.31$ GeV, which can be intuitively understood from Fig. 2.12. Finally, the reason why we consider only the $S$-wave and neglect all the higher $\ell$ waves is that, as it is shown in Fig. 2.13, in the cases of the $D$- and the $G$-waves we do not observe the formation of a bound state[†].

---

[†]We observe the formation of a bound state in the $D$-wave only in the unrealistic limit of very large attraction, i.e. $\Lambda_G \to 0$.

## 2.6 Short summary

In order to study the scattering of two scalar glueballs, we used the dilaton Lagrangian, which reproduces the trace anomaly in YM and depends on a single dimentionful parameter $\Lambda_G$. This Lagrangian allows us to extract the form of the $\ell$-expanded amplitudes in the elastic window at tree-level. We applied a unitarization procedure via a self-energy loop function, introducing also the effect of a cutoff function (that was not done in Refs. [98, 99, 114]).

We saw that this cutoff, apart from the modification of the loop function in its real and imaginary part, influences the critical value $\Lambda_{G,crit}$. This value is important since it tells us whether a bound state of two scalar glueballs exists or not. Interestingly, we found that a bound state exists for the value $\Lambda_G = 0.35$ GeV, that arise as an intermediate value of the gluon condensate (see Section 2.3.1). Additionally, we found the presence of a bound state only in the $S$-wave.

# CHAPTER 3

# Partial wave amplitudes in the decays of mesons

The study of mesons, together with their decays, can provide several information on the interactions and the dynamics between the states involved. A lot of efforts has been put in both the experimental [1, 109, 111, 131–133] and the theoretical [38, 134–138] aspects of the decays of mesons. The coupling constant estimated from most of the models are based on the mass of mesons, the total decay widths and branching fractions. Among all the data available in the PDG, the ratios of the partial wave amplitudes are potentially relevant, even though not always taken into account. This is why we will try to understand how the partial wave amplitudes can be helpful to study the decays of mesons.

## 3.1 Classification of mesons

The classification of mesons starts with the division into nonets, according to their $J^{PC}$ values. Each nonet contain two isosinglet states, two isodoublets and one isotriplet. Up to now, not all the mesonic nonets are completely known as the existence and properties of many particles are still unclear. A well understood nonet is the pseudoscalar ($P$) one -$J^{PC} = 0^{-+}$-; because of that, it will be taken as an example for a description of the nonet structure, which is analogous for each $J^{PC}$.

As it is shown in Fig. 3.1, nine "$P$" mesons are diagrammed according to the values of their charge Q, strangeness S, and third component of the isospin $I_3$. A general representation of the nonet is given in the form of a matrix containing the quark content of the mesons:

$$P = \begin{pmatrix} \frac{\pi^0 + \eta^N}{\sqrt{2}} & \pi^+ & K^+ \\ \pi^- & \frac{-\pi^0 + \eta^N}{\sqrt{2}} & K^0 \\ K^- & \bar{K}^0 & \eta^S \end{pmatrix} \approx \begin{pmatrix} \bar{u}i\gamma_5 u & \bar{d}i\gamma_5 u & \bar{s}i\gamma_5 u \\ \bar{u}i\gamma_5 d & \bar{d}i\gamma_5 d & \bar{s}i\gamma_5 d \\ \bar{u}\gamma_5 s & \bar{d}i\gamma_5 s & \bar{s}i\gamma_5 s \end{pmatrix} \quad (3.1)$$

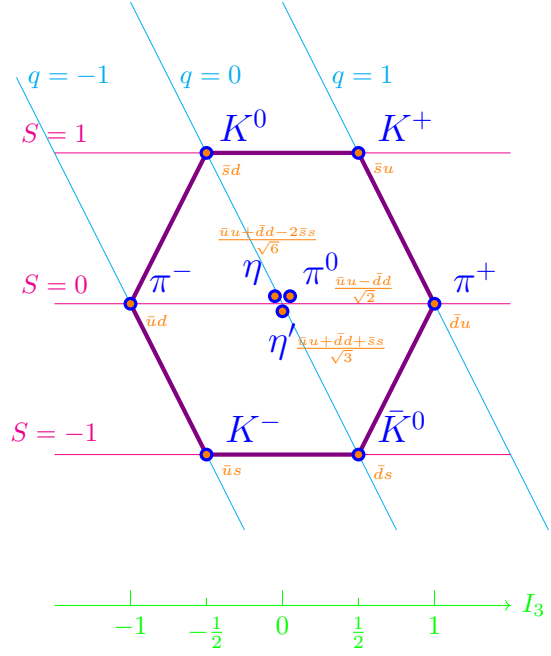

**Fig. 3.1** The pseudoscalar meson nonet, with the various charges $q$, strangeness $S$ and third component of the isospin $I_3$.

Note, we consider only quarks $u$, $d$ and $s$ ($N_f = 3$), as we are interested in the light mesonic sector. In Eq. (3.1) the flavour states, whose combination gives the corresponding quark content, are listed. In the main diagonal there are the non-strange

$$\pi^0 = \frac{1}{\sqrt{2}}(\bar{u}u - \bar{d}d) \qquad (3.2)$$

and

$$\eta^N = \frac{1}{\sqrt{2}}(\bar{u}u + \bar{d}d), \qquad (3.3)$$

and the strange

$$\eta^S = \bar{s}s \qquad (3.4)$$

states. The indices "$S$" and "$N$" in the $\eta$ states stand for strange and non-strange members respectively. The flavour states do not always correspond to the mass (or physical) states. In the pseudoscalar sector, the mixing angle between the flavour states $\eta^N$ and $\eta^S$, caused mostly by the so called axial anomaly, determine the physical states $\eta$ and $\eta'$:

$$\begin{pmatrix} \eta(547) \\ \eta'(958) \end{pmatrix} = \begin{pmatrix} \cos\theta_P & \sin\theta_P \\ -\sin\theta_P & \cos\theta_P \end{pmatrix} \begin{pmatrix} \eta^N \\ \eta^S \end{pmatrix}. \qquad (3.5)$$

The value of the pseudoscalar mixing angle is quite large, $\theta_P = (-40.4 \pm 0.6)°$ [139], which means that each of the physical states is made of a similar admixture of $\eta^N$ and of $\eta^S$. Physically, because of the sign, $\eta(547)$ is close to the octet ($\frac{\bar{u}u+\bar{d}d-2\bar{s}s}{\sqrt{6}}$) and $\eta'(958)$ to the singlet ($\frac{\bar{u}u+\bar{d}d+\bar{s}s}{\sqrt{3}}$), see later on.

Each meson of the nonet has some particular quantum numbers, $I, I_3, S$, that are the isospin, its third component and the strangeness, respectively. The $I$ value splits the components of the nonet into one triplet, two doublets and two singlets of isospin (see Tab. 3.1).

|  | Meson | Quark content | $I$ | $I_3$ | $S$ |
|---|---|---|---|---|---|
|  | $\pi^+$ | $\bar{d}u$ | 1 | 1 | 0 |
| Isotriplet | $\pi^0$ | $\frac{1}{\sqrt{2}}(\bar{u}u - \bar{d}d)$ | 1 | 0 | 0 |
|  | $\pi^-$ | $\bar{u}d$ | 1 | -1 | 0 |
| Isodoublet | $K^+$ | $\bar{s}u$ | 1/2 | 1/2 | 1 |
|  | $K^-$ | $\bar{u}s$ | 1/2 | 1/2 | -1 |
| Isodoublet | $K^0$ | $\bar{s}d$ | 1/2 | -1/2 | 1 |
|  | $\overline{K}^0$ | $\bar{d}s$ | 1/2 | -1/2 | -1 |
| Isosinglet | $\eta^N$ | $\frac{1}{\sqrt{2}}(\bar{u}u + \bar{d}d)$ | 0 | 0 | 0+0=0 |
| Isosinglet | $\eta^S$ | $\bar{s}s$ | 0 | 0 | 1+(-1)=0 |

**Table 3.1** Each component of the pseudoscalar flavour nonet with its values of $I$, $I_3$ and $S$.

Interestingly, the names given to the mesons belonging to different nonets have some regularities. For example, kaonic states are the isodoublets of many nonets, e.g. pseudoscalar ($0^{-+}$), vector ($1^{--}$), axial-vector ($1^{++}$). Another example comes from the isosinglet states, which are called with the letter $\eta$ in the $J^{-+}$ sectors: $\eta^N$ and $\eta^S$ for $0^{-+}$, $\eta_2^N$ and $\eta_2^S$ for $2^{-+}$ (as well as the $f$ states for $J^{++}$, and so on).

The nonets that I used during my work are grouped into matrices and are listed here:

$$P = \begin{pmatrix} \frac{\eta^N+\pi^0}{\sqrt{2}} & \pi^+ & K^+ \\ \pi^- & \frac{\eta^N-\pi^0}{\sqrt{2}} & K^0 \\ K^- & \overline{K}^0 & \eta^S \end{pmatrix}, \quad V^\mu = \begin{pmatrix} \frac{\omega^N+\rho^0}{\sqrt{2}} & \rho^+ & K^{*+} \\ \rho^- & \frac{\omega^N-\rho^0}{\sqrt{2}} & K^{*0} \\ K^{*-} & \overline{K}^{*0} & \omega^S \end{pmatrix}^\mu,$$

$$A^\mu = \begin{pmatrix} \frac{f_{1,A}^N+a_1^0}{\sqrt{2}} & a_1^+ & K_{1,A}^+ \\ a_1^- & \frac{f_{1,A}^N-a_1^0}{\sqrt{2}} & K_{1,A}^0 \\ K_{1,A}^- & \overline{K}_{1,A}^0 & f_{1,A}^S \end{pmatrix}^\mu, \quad B^\mu = \begin{pmatrix} \frac{f_{1,B}^N+b_1^0}{\sqrt{2}} & b_1^+ & K_{1,B}^+ \\ b_1^- & \frac{f_{1,B}^N-b_1^0}{\sqrt{2}} & K_{1,B}^0 \\ K_{1,B}^- & \overline{K}_{1,B}^0 & f_{1,B}^S \end{pmatrix}^\mu,$$

$$T^{\mu\nu} = \begin{pmatrix} \frac{\eta_2^N+\pi_2^0}{\sqrt{2}} & \pi_2^+ & K_{2,T}^+ \\ \pi_2^- & \frac{\eta_2^N-\pi_2^0}{\sqrt{2}} & K_{2,T}^0 \\ K_{2,T}^- & \overline{K}_{2,T}^0 & \eta_2^S \end{pmatrix}^{\mu\nu}, \quad X^{\mu\nu} = \begin{pmatrix} \frac{f_2^N+a_2^0}{\sqrt{2}} & a_2^+ & K_2^{*+} \\ a_2^- & \frac{f_2^N-a_2^0}{\sqrt{2}} & K_2^{*0} \\ K_2^{*-} & \overline{K}_2^{*0} & f_2^S \end{pmatrix}^{\mu\nu}, \quad (3.6)$$

where $P$, $V^\mu$, $A^\mu$, $B^\mu$, $T^{\mu\nu}$, and $X^{\mu\nu}$ are the pseudoscalar ($0^{-+}$), vector ($1^{--}$), axial-vector ($1^{++}$), pseudovector ($1^{+-}$), pseudotensor ($2^{-+}$), and tensor ($2^{++}$) flavour states respectively.

Another, more schematic way to list these nonets is the one presented in the Particle Data Group [1], with the suggested $\bar{q}q$ assignment instead of the pure strange-nonstrange states (here reported in Table 3.2). The symbols $L_k$ and $H_k$ in Table 3.2, refer to:

$$\begin{pmatrix} L_k \\ H_k \end{pmatrix} = \begin{pmatrix} \cos\theta_k & \sin\theta_k \\ -\sin\theta_k & \cos\theta_k \end{pmatrix} \begin{pmatrix} \frac{1}{\sqrt{2}}(\bar{u}u + \bar{d}d) \\ \bar{s}s \end{pmatrix}, \qquad (3.7)$$

$k = P, V, AV, PV, ...$ (see later on).

|  |  | $J^{PC}$ | $I=1$ $\bar{d}u, \bar{u}d,$ $\frac{1}{\sqrt{2}}(\bar{d}d - \bar{u}u)$ | $I=\frac{1}{2}$ $\bar{s}u, \bar{s}d,$ $\bar{d}s, \bar{u}s$ | $I=0$ $L_k$ | $I=0$ $H_k$ |
|---|---|---|---|---|---|---|
| Pseudoscalar | $P$ | $0^{-+}$ | $\pi$ | $K$ | $\eta$ | $\eta'(958)$ |
| Vector | $V^\mu$ | $1^{--}$ | $\rho(770)$ | $K^*(892)$ | $\omega(782)$ | $\phi(1020)$ |
| Axial-vector | $A^\mu$ | $1^{++}$ | $a_1(1260)$ | $K_{1,A}$ ♣ | $f_1(1285)$ | $f_1(1420)$ |
| Pseudovector | $B^\mu$ | $1^{+-}$ | $b_1(1235)$ | $K_{1,B}$ ♣ | $h_1(1170)$ | $h_1(1415)$ |
| Pseudotensor | $T^{\mu\nu}$ | $2^{-+}$ | $\pi_2(1670)$ | $K_2(1770)$ ♣ | $\eta_2(1645)$ | $\eta_2(1870)$ |
| Tensor | $X^{\mu\nu}$ | $2^{++}$ | $a_2(1320)$ | $K_2^*(1430)$ | $f(1270)$ | $f_2'(1525)$ |

**Table 3.2** Suggested assignments for $\bar{q}q$ states, considering mesons made of $u$, $d$, $s$ quarks only.
♣ The marked isospin $\frac{1}{2}$ states mix: the $1^{++}$ with the $1^{+-}$, as well as the state $2^{-+}$ together with the $2^{--}$ (not present in this table).

For an overview on the main properties of the meson fields, see Table 3.3.

The following states are admixtures of the pure strange and non-strange mesons of the corresponding nonet and are reported with the corresponding mixing angle[*]:

- $\eta(547)$ and $\eta'(958)$ for $P$ ($\theta_P \simeq -40.4°$) [131, 139], *status:* well known.

- $\omega$ (782) and $\phi$ (1020) for $V^\mu$ ($\theta_V \simeq -3.9°$) [1, 136], *status:* well known.

- $f_1$ (1285) and $f_1$ (1420) for $A^\mu$ ($\theta_{AV} \simeq 24°$) [108, 141], *status:* angle is small, the precise value is still under debate.

- $h_1$ (1170) and $h_1$ (1415) for $B^\mu$ ($\theta_{PV} \simeq 1°$) [103], *status:* not clarified yet.

- $\eta_2(1645)$ and $\eta_2(1870)$ for $T^{\mu\nu}$ ($\theta_{PT} \simeq -42°$) [142], *status:* angle is still under debate.

---

[*]Here, the mixing angles are given for the nonets expressed in the strange-nonstrange basis. It is also possible to find results for the mixing in the singlet-octet basis. The two mixing angle obtained from these two basis differ for about $35.3°$ (see Appendix A in Ref. [140]).

| Name | $J^{PC}$ | Field | Hermitian quark current operators | Parity ($\hat{P}$) | Charge conjugation ($\hat{C}$) |
|---|---|---|---|---|---|
| Pseudoscalar | $0^{-+}$ | $P$ | $\bar{q}_j i\gamma_5 q_i$ | $-P(t,-\vec{x})$ | $P^t$ |
| Vector | $1^{--}$ | $V^\mu$ | $\bar{q}_j \gamma^\mu q_i$ | $V_\mu(t,-\vec{x})$ | $-(V^\mu)^t$ |
| Axial-vector | $1^{++}$ | $A^\mu$ | $\bar{q}_j \gamma^\mu \gamma_5 q_i$ | $-A_\mu(t,-\vec{x})$ | $(A^\mu)^t$ |
| Pseudovector | $1^{+-}$ | $B^\mu$ | $\bar{q}_j \gamma_5 \overleftrightarrow{\partial}^\mu q_i$ | $-B_\mu(t,-\vec{x})$ | $-(B^\mu)^t$ |
| Pseudotensor | $2^{-+}$ | $T^{\mu\nu}$ | $\bar{q}_j i[\gamma_5 \overleftrightarrow{\partial}^\mu \overleftrightarrow{\partial}^\nu - \frac{2}{3}\tilde{g}^{\mu\nu} \overleftrightarrow{\partial}^\alpha \overleftrightarrow{\partial}_\alpha] q_i$ | $-T_{\mu\nu}(t,-\vec{x})$ | $(T^{\mu\nu})^t$ |
| Tensor | $2^{++}$ | $X^{\mu\nu}$ | $\bar{q}_j i[\gamma^\mu \overleftrightarrow{\partial}^\mu + \gamma^\nu \overleftrightarrow{\partial}^\nu - \frac{2}{3}\tilde{g}^{\mu\nu} \overleftrightarrow{\partial\!\!\!/}] q_i$ | $X_{\mu\nu}(t,-\vec{x})$ | $(X^{\mu\nu})^t$ |

**Table 3.3** Some of the meson fields, their possible quark current assignments, and their behavior under parity and charge conjugation transformations. The quantity $\tilde{g}^{\mu\nu}$ is the projection operator, $\tilde{g}^{\mu\nu} = g^{\mu\nu} + \frac{k^\mu k^\nu}{k^2}$.

- $f_2(1270)$ and $f_2'(1525)$ for $X^{\mu\nu}$ ($\theta_T \simeq 3.16°$) [143], *status:* well known.

The physical state $\eta_2$ can be interpreted as $\eta_2(1645)$, while the nature of $\eta_2'$ is still unclear. Because of the lack of evidence for the $K^*K$ decay mode, it is difficult to identify $\eta_2(1870)$ with the heavier mass state of the pseudotensor nonet [144, 145].

## 3.2 Some issues regarding meson mixing

The world of mesons if quite well known, but some uncertainties still affect our knowledge about this part of QCD.

One open problem regards the value of various mixing (e.g. $\theta_{PV}$, see the discussion below). In general we can distinguish two kinds of mixing angles: one between two mesons of the same nonet (intranonet mixing) and another between mesons belonging to different nonets (internonet mixing). This distinction considers only conventional mesons, since there are also other types of mixing in Nature, e.g. the mixing between $f_0(1370)$, $f_0(1500)$, and $f_0(1710)$, which is expected to include also the scalar glueball [51].

- **Intranonet mixing angle**

  We start considering the isoscalar mixing. In the pseudoscalar sector the mixing angle between strange $\bar{s}s$ and non-strange $\bar{n}n$ states is large, leading to a wide mass difference between $\eta$ and $\eta'$, also because of an additional (instanton driven [146]) mass contribution to the singlet state ($\frac{\bar{u}u+\bar{d}d+\bar{s}s}{\sqrt{3}}$). The mixing angle in the vector sector ($V^\mu$) is instead small, then the isosinglets $\omega(782)$ and $\phi(1020)$ are almost purely non-strange $\bar{n}n$ and strange $\bar{s}s$ states, respectively.

  The mixing between strange and nonstrange isosinglet states presents uncertainties in both the pseudovector and the axial-vector sector. The value of the pseudovector ($B^\mu$) mixing angle between $h_1(1170)$ and $h_1(1415)$ is still unknown. The value found in Ref. [103], using the Gell-Mann-Okubo mass relations, was of a nearly zero mixing in the strange-nonstrange basis. Therefore, according to the results given in Ref. [103], $h_1(1170)$ is predominantly $\bar{u}u + \bar{d}d$, while $h_1(1415)$ has a large $\bar{s}s$ content. The mixing angle of the axial-vector ($A^\mu$) sector, extracted from the LHCb collaboration (Ref. [108]), was measured to be $(24^{+3.4+0.6}_{-2.6-0.8})°$, in the strange-nonstrange basis, for the couple $f_1(1285)$-$f_1'(1420)$. This result, obtained from the the $\bar{B}_s^0 \to J/\psi f_1$ decay, suggests that about $83\%$ of the $\bar{s}s$ component is located in the $f_1'(1420)$ meson.

Various papers state that the result of the pseudovector $\theta_{PV}$ and of the axial-vector $\theta_{AV}$ (intranonets) mixing angle cannot be obtained independently from the value of the (internonet) mixing angle $\theta_{K_1}$ (see below). This angle determines the mixing between the two isodoublets $K_{1,A}$ and $K_{1,B}$ of the $A^\mu$ and $B^\mu$ sectors respectively. In particular, in Ref. [103], the value $\theta_{PV} = (35.9 \pm 2.6)°$ was determined [†].

There are uncertainties also for what concerns the isosinglets of the pseudotensor sector. Koenigstein and Giacosa [142] studied the mixing angle between $\eta_2(1645)$ and $\eta_2(1870)$. In their work they show that, although from the Okubo formula $\theta_{PT} \approx 14.8°$, such a small value is inconsistent in their model. They saw that their result for the branching ratio $\eta_2(1870) \to a_2(1320)\pi$ to $\eta_2(1870) \to f_2(1270)\eta$ agrees with the experimental one obtained by Barberis [147]. The corresponding value of the mixing angle is $\theta_{PT} \in (-40°, -50°)$; in particular, they used the value $\theta_{PT} \approx -42°$ for their predictions.

It is worth noticing that the value of the ratio calculated in Ref. [142] is $23.5$, an order larger than the PDG value of $1.7 \pm 0.4$, close to the value $1.60 \pm 0.40$ extracted in [145]. The identification of $\eta_2(1870)$ as the meson $\eta_2'$ is unsettled due to the missing of the $\eta_2(1870) \to K^*\bar{K}$ decay. This state could also be a $\bar{q}qg$ hybrid [148]. In general, constraints on the mixing angle are difficult, as the lack of data about the branching ratio of $\eta_2'$ is resulting in a large uncertainty in the value of the mixing angle.

- Internonet mixing angle

The two kaons $K_1(1270)$ and $K_1(1400)$ are a mixture of the two flavour states $K_{1,A}$ and $K_{1,B}$. Yet, the value of the mixing angle between them is still under debate, but

---

[†]This value was obtained assuming the value of $\theta_{K_1}$ as $+34°$, for details see Ref. [103].

one of the possible estimated value is $|\theta_{K_1}| = (33.6 \pm 4.3)°$ [149], defined as:

$$\begin{pmatrix} |K_1^+(1270)\rangle \\ |K_1^+(1400)\rangle \end{pmatrix} = \begin{pmatrix} \sin\theta_{K_1} & \cos\theta_{K_1} \\ \cos\theta_{K_1} & \sin\theta_{K_1} \end{pmatrix} \begin{pmatrix} |K_{1,A}^+\rangle \\ |K_{1,B}^+\rangle \end{pmatrix} ; \quad (3.8)$$

for other determinations, see Refs. [134, 150–152]. The results obtained in these works come from many different precesses, e.g. the ratio of the decay widths of $\tau \to K_1(1270)\nu$ and $\tau \to K_1(1400)\nu$ decays. Both the magnitude and the sign of the mixing angle $\theta_{K_1}$ are still an open problem. An accurate determination of this angle is important, also because the isosinglet mixing angles in both the $A^\mu$ and $B^\mu$ sectors depend on $\theta_{K_1}$.

## 3.3 Formalism of partial wave amplitudes

A lot of efforts has been done, both experimentally and theoretically, to improve our knowledge about mesons. Decays as a source of information [131, 132], as well as models, e.g. effective field theories [38, 134, 144], have been largely implemented. Out of the multitude of data available in the PDG, the ratios of the partial wave decay amplitudes is not that often taken into account in models. Yet, they can also provide reliable inputs. The decays that were considered during my work are:

$$\begin{aligned} A_\mu &\to V_\mu P \,, & B_\mu &\to V_\mu P \,, \\ T_{\mu\nu} &\to X_{\mu\nu} P \,, & T_{\mu\nu} &\to V_\mu P \,. \end{aligned} \quad (3.9)$$

The use of partial wave analysis to study decays has been known for more than half a century. An early approach is the tensor formalism introduced by Zemach [153, 154], who used a non-covariant three-dimension spin tensor defined in the rest frame of each particle involved in the decay, in order to write the decay amplitude. The interpretation of the amplitude squared as the decay probability is then difficult, due to the frame dependence of the decay width.

Another non-covariant approach, the helicity formalism, was suggested few years before by Jacob and Wick [155]. This formalism uses the Wigner $D$-matrices $D_{mm'}^J$, which enclose the angular dependence, together with the helicity coupling amplitude, to form the decay amplitude.

As reported in Ref. [156], the analysis via the helicity formalism provides the same result as with the tensor formalism through the introduction of additional terms called *centrifugal barrier factors*. These terms, which are the moduli of the Zemach tensors relative to each state of the decay in its proper rest frame, cause the helicity amplitude to be non-covariant (i.e. $G_{\ell S}^{J} \propto |\vec{k}|^{\ell}$, where $\vec{k}$ is the relative momentum between the two products of the decay).

The helicity amplitude formalism was then proposed in a covariant form by Chung [11, 12]. In the following, we formalize the model that was used in [11, 12], in order to study decays. The starting point is to consider a decay of a meson into two other mesons:

$$A \to BC, \tag{3.10}$$

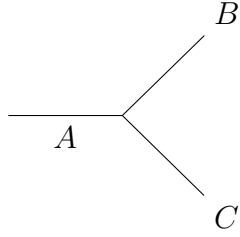

The corresponding total spin of the mesonic fields $A$, $B$, $C$, are $J_A$, $J_B$, $J_C$. If we call $\ell$ the angular momentum between the final states, then the conservation of total angular momentum implies:

$$\vec{J}_A = \vec{J}_B + \vec{J}_C + \vec{\ell}. \tag{3.11}$$

For any decay, we can write the Lagrangian describing it. Because of several constraints, such as charge and parity conjugation invariance and covariance, the interaction term of the Lagrangian for any decay $A \to BC$ can be expressed as:

$$\mathscr{L}_{A \to BC} = \mathscr{L}_{I,A \to BC} + \mathscr{L}_{II,A \to BC} + ... \tag{3.12}$$

The term $\mathscr{L}_{I,A\to BC}$ includes the minimal possible number of derivatives, while $\mathscr{L}_{II,A\to BC}$ has two additional derivatives [‡]. The general form of the interaction Lagrangians that were relevant in my work are:

- $\mathscr{L}_{A_\mu \to V_\mu P} = g_c^A \text{Tr}\{A_\mu [V^\mu, P]\} + g_d^A \text{Tr}\{\mathcal{A}_{\mu\nu}[\mathcal{V}^{\mu\nu}, P]\}$,
- $\mathscr{L}_{B_\mu \to V_\mu P} = g_c^B \text{Tr}\{B_\mu \{V^\mu, P\}\} + g_d^B \text{Tr}\{\mathcal{B}_{\mu\nu}\{\mathcal{V}^{\mu\nu}, P\}\}$,
- $\mathscr{L}_{T_{\mu\nu} \to X_{\mu\nu} P} = g_c^{PT} \text{Tr}\{T_{\mu\nu}\{X^{\mu\nu}, P\}\} + g_d^{PT} \text{Tr}\{\mathcal{T}_{\alpha\mu\nu}\{\mathcal{X}^{\alpha\mu\nu}, P\}\}$,
- $\mathscr{L}_{T_{\mu\nu} \to V_\mu P} = g_v^{PT} \text{Tr}\{T_{\mu\nu}[V^\mu, \partial^\nu P]\} + g_t^{PT} \text{Tr}\{\mathcal{T}_{\alpha\mu\nu}[\mathcal{V}^{\alpha\mu}, \partial^\nu P]\}$,

where

$$\mathcal{A}^{\mu\nu} = \partial^\mu A^\nu - \partial^\nu A^\mu, \qquad \mathcal{B}^{\mu\nu} = \partial^\mu B^\nu - \partial^\nu B^\mu,$$

---
[‡]This fact comes from the necessity to have a covariant Lagrangian: for a fixed number of mesonic fields, the addition of a derivative with index "up" requires an additional derivative with a index "down"

$$\mathcal{V}^{\mu\nu} = \partial^\mu V^\nu - \partial^\nu V^\mu, \qquad\qquad \mathcal{T}_{\alpha\mu\nu} = \partial_\alpha T_{\mu\nu} - \partial_\mu T_{\alpha\nu},$$
$$\mathcal{X}^{\alpha\mu\nu} = \partial^\alpha X^{\mu\nu} - \partial^\mu X^{\alpha\nu}. \qquad(3.13)$$

The coefficients $g_c^A$, $g_d^A$,... are the corresponding coupling constants.

### 3.3.1 Covariant helicity formalism

The covariant helicity formalism was introduced by Chung [11, 12]. Let the total angular momentum quantum numbers of the particles $A$, $B$ and $C$, involved in the decay process $A \to BC$, be $|J, M_J\rangle$, $|s, \lambda\rangle$ and $|\sigma, \nu\rangle$, respectively. The total spin of the final states is given by $S$:

$$|S, m_s\rangle = |s, \lambda\rangle \otimes |\sigma, \nu\rangle. \qquad(3.14)$$

The symbol $\otimes$ means that the state $|S, m_s\rangle$ was built according to the product rules of the angular momenta. The spin of the parent is then formed by the total spin of the daughters and the orbital angular momentum ($\ell$) carried by them:

$$|J, M_J\rangle = |\ell, m_\ell\rangle \otimes |S, m_s\rangle. \qquad(3.15)$$

The value of $\ell$ is limited by the spin of both the parent and the daughter states. All the decays of interest in this work have a pseudoscalar meson in the final state, which we identify with the state $C$. Thus, we can consider $J_C = 0$ in Eq. (3.11), obtaining that $\vec{\ell} = \vec{J}_A - \vec{J}_B$. Although the angular momentum can take the values

$$\ell \in [|J_A - J_B|, J_A + J_B], \qquad(3.16)$$

some of them are forbidden. Because of parity conservation, the freedom of $\ell$ is subject to the relation:

$$P_A = P_B P_C (-1)^\ell = P_B (-1)^{\ell+1}, \qquad(3.17)$$

where $P_A$ and $P_B$ are the parity of the states $A$ and $B$ respectively, and $P_C = -1$ since, in all the decays relevant here, one of the decaying particle is always a pseudoscalar. This determines if $\ell$ is odd or even, resulting in a reduced number of angular momentum channels.

In the rest frame of the parent particle, the general amplitude for the decay can be written as:

$$\mathscr{A}^J(\theta, \phi; M_J) = \sum_{\lambda\nu} \mathscr{A}^J_{\lambda\nu}(\theta, \phi; M_J) \propto \sum_{\lambda\nu} D^{J*}_{M m_s}(\phi, \theta, 0) F^J_{\lambda\nu}, \qquad(3.18)$$

where $\mathscr{A}$, the invariant amplitude for the decay, is related to the product between the complex conjugate of the Wigner $D$-matrix, $D^{J*}_{Mm_s}(\phi, \theta, 0)$, and the helicity amplitude, $F^J_{\lambda\nu}$. The final states with spin $s$ and $\sigma$ have helicity $\lambda$ and $\nu$ respectively, with $m_s = \lambda - \nu$. In the case of identical decay products, parity conservation allow us to write the relationship:

$$F^J_{\lambda\nu} = (-1)^J F^J_{\nu\lambda}. \tag{3.19}$$

The general result of Eq. (3.18) implies that any model dependence will be inside the helicity amplitudes. As the decay is studied in the rest frame of the parent, then the two decaying particles will move along the same line in different direction. If we consider the alignment to be along the $z$ axis, then $D^{J*}_{Mm_s}(0,0,0) = 1$, implying:

$$\mathscr{A}^J_{\lambda\nu}(0,0\,;M_J) \propto F^J_{\lambda\nu}. \tag{3.20}$$

In the case of decaying product with nonzero mass, it is possible to expand the helicity amplitudes in term of the $\ell S$ coupling amplitudes:

$$F^J_{\lambda\nu} = \sum_{\ell S} \sqrt{\frac{2\ell+1}{2J+1}} \langle \ell 0 S m_s | J m_s \rangle \langle s\lambda\sigma - \nu | S m_s \rangle G^J_{\ell S}, \tag{3.21}$$

where $\langle \cdots | \cdots \rangle$ represents the Clebsch-Gordan coefficients, given in App. C, which, as usual, arise when there is an interaction between angular momenta. In particular, the term $\langle s\lambda\sigma - \nu | S m_s \rangle$ in Eq. (3.21) has its origin in Eq. (3.14), while $\langle \ell 0 S m_s | J m_s \rangle$ comes from Eq. (3.15). The sign $-$ in front of $\nu$, as well as the fact that $m_s$ is given by the difference between the two, is understandable since $\lambda$ and $\nu$ are helicities, which contain the information about the direction of the spin along which the particle moves.

The ratio between the partial wave amplitudes is then given by the ratio of $G_{\ell S}$. The decay amplitude can also be expanded as (see Eq. 2.8 in Ref. [157])[§]:

$$i\mathscr{A}(\theta,\phi;M_J) = \sum_{\ell} \sum_{m_\ell=-\ell}^{\ell} G_\ell \langle \ell m_\ell S m_s | J M_J \rangle \langle s\lambda\sigma - \nu | S m_s \rangle Y_{\ell m_\ell}(\theta,\phi). \tag{3.22}$$

Note, in all the decays of our interest, one of the product is a pseudoscalar. Thus the term $\langle s\lambda 00 | S m_s \rangle = 1$.

The partial wave amplitude derived in Eq. (3.22) is related to the one coming from the

---

[§]We should keep in mind that, although the sum of $m_\ell$ is done from $-\ell \to \ell$, not all the possible values of $m_\ell$ are always allowed (for details see App. C).

covariant helicity formalism:

$$G_\ell = \sqrt{\frac{\alpha}{(2J+1)}} G_{\ell S}^J, \qquad (3.23)$$

where the numerical factor $\alpha$ depends on the normalization of the spherical harmonics $Y_{\ell m_\ell}(\theta, \phi)$. The normalization

$$\int d\Omega |Y_{\ell m_\ell}|^2 = 1 \qquad (3.24)$$

gives $\alpha = 4\pi$. With a suitable choice of the helicity amplitudes, we obtain:

$$\sum_{\ell S} |G_{\ell S}^J|^2 = \sum_{\text{spins}} |\mathscr{A}|^2. \qquad (3.25)$$

### 3.3.2 Polarization states

The list given at the very end of Sec. 3.2 includes only decays of mesons $A \to BC$, with total angular momentum states $|J, M_J\rangle$, $|s, \lambda\rangle$ and $|\sigma, \nu\rangle$, where $J \in [1, 2]$ and $s \in [1, 2]$ (the third meson is always a pseudoscalar, then $|\sigma, \nu\rangle = |0, 0\rangle$). The needed polarization states are therefore the polarization vectors (PoV) and polarization tensors (PoT$_J$). In this section we provide a detailed construction of these states.

❊ **PoV**

The PoVs of a spin-1 state $\epsilon^\mu(\vec{k}, M_J)$ depends on the three-momentum $\vec{k}$ and on the third component of the spin of the particle which is described by it. In their rest frame they are given by,

$$\epsilon^\mu(\vec{0}, +1) = -\frac{1}{\sqrt{2}} \begin{pmatrix} 0 \\ 1 \\ i \\ 0 \end{pmatrix}, \quad \epsilon^\mu(\vec{0}, -1) = \frac{1}{\sqrt{2}} \begin{pmatrix} 0 \\ 1 \\ -i \\ 0 \end{pmatrix}, \quad \epsilon^\mu(\vec{0}, 0) = \begin{pmatrix} 0 \\ 0 \\ 0 \\ 1 \end{pmatrix}. \qquad (3.26)$$

The orthonormality conditions satisfied by the PoVs are:

$$k_\mu \epsilon^\mu(\vec{k}, m) = 0, \qquad (3.27)$$
$$\epsilon_\mu^*(\vec{k}, m)\epsilon^\mu(\vec{k}, m') = -\delta_{mm'}. \qquad (3.28)$$

The PoVs given in Eq. (3.26) are useful when applied to the decaying particle, which is taken in its rest frame. The daughters particles, having a non-zero three-momentum, need PoVs evaluated with $\vec{k} \neq 0$. The form of a general PoV, $\epsilon^\mu(\vec{k}, M_J)$, is presented in App.

B. The projection operator, which also appears in Tab. 3.3, is given by:

$$\tilde{g}_{\mu\nu} = \sum_{m=-J}^{J} \epsilon_\mu(\vec{k},m)\epsilon_\nu^*(\vec{k},m) = -g_{\mu\nu} + \frac{k_\mu k_\nu}{M_0^2}, \qquad (3.29)$$

where $k_\mu$ is the 4-momentum and $M_0$ the mass of the corresponding state.

### ❋❋ PoT$_J$

The polarization vectors can be used to construct the PoT$_J$s for higher spins states. The general formula:

$$\epsilon^{\mu_1\mu_2\cdots\mu_J}(\vec{0},m) = \sum_{m_1 m_2 \ldots} \langle 1m_1 1m_2 | 2n_1 \rangle \langle 2n_1 1m_3 | 3n_2 \rangle \ldots \langle J-1 n_{J-2} 1m_J | Jm \rangle$$
$$\otimes \epsilon^{\mu_1}(\vec{0},m_1) \epsilon^{\mu_2}(\vec{0},m_2) \ldots \epsilon^{\mu_J}(\vec{0},m_J). \qquad (3.30)$$

allows us to get the PoT$_J$s for a spin-$J$ state. Analogously to the spin-1 case, the orthonormality relations

$$k_{\mu_i} \epsilon^{\mu_1\mu_2\cdots\mu_J}(m) = 0,$$
$$\epsilon^*_{\mu_1\mu_2\cdots\mu_J}(m) \epsilon^{\mu_1\mu_2\cdots\mu_J}(m') = (-1)^J \delta_{mm'} \qquad (3.31)$$

are satisfied. The generic polarization tensor transforms under rotation, as:

$$\epsilon^{\mu_1\mu_2\cdots\mu_J}(m) \to \sum_{m'} \epsilon^{\mu_1\mu_2\cdots\mu_J}(m') D^J_{m'm}(\phi,\theta,\psi). \qquad (3.32)$$

The form of the polarization tensors PoT$_2$s, boosted along the $z$-axis and written in terms of the PoVs, is (for details see App.B):

$$\epsilon^{\mu\nu}(\vec{k},\pm 2) = \epsilon^\mu(\vec{k},\pm 1)\epsilon^\nu(\vec{k},\pm 1),$$
$$\epsilon^{\mu\nu}(\vec{k},\pm 1) = \frac{1}{\sqrt{2}}\left(\epsilon^\mu(\vec{k},\pm 1)\epsilon^\nu(\vec{k},0) + \epsilon^\mu(\vec{k},0)\epsilon^\nu(\vec{k},\pm 1)\right),$$
$$\epsilon^{\mu\nu}(\vec{k},0) = \frac{1}{\sqrt{6}}\left(\epsilon^\mu(\vec{k},+1)\epsilon^\nu(\vec{k},-1) + \epsilon^\mu(\vec{k},-1)\epsilon^\nu(\vec{k},+1)\right) + \sqrt{\frac{2}{3}}\epsilon^\mu(\vec{k},0)\epsilon^\nu(\vec{k},0).$$
$$(3.33)$$

The boosted form of the PoT$_2$s can be then equally obtained in two ways: (i) by using the PoVs ($\equiv$PoT$_1$s) properly boosted as in Eq. (3.33), or (ii) building the PoT$_2$s from the PoVs in the rest frame of the meson, and then boost the PoT$_J$s obtained. An analogous procedure could be also applied in the case of PoT$_J$s with $J > 2$.

## 3.4 Derivation and analysis of the partial wave amplitudes

The partial wave analysis is applied in this section for two processes:

$a_1(1260) \to \rho\pi$ $\qquad\qquad (A_\mu \to V_\mu P)$,
$b_1(1235) \to \omega\pi$ $\qquad\qquad (B_\mu \to V_\mu P)$.

A similar discussion can be found in Ref. [158] for the decays of:

$\pi_2(1670) \to f_2(1270)\pi$ $\qquad (T_{\mu\nu} \to X_{\mu\nu} P)$,
$\pi_2(1670) \to \rho\pi$ $\qquad\qquad (T_{\mu\nu} \to V_\mu P)$.

The discussion that will be provided is in principle valid for each member of the same nonet of the considered meson.

### 3.4.1 The decay $a_1(1260) \to \rho\pi$ ($1^{++} \to 1^{--}0^{-+}$)

The decay $a_1(1260) \to \rho\pi$ can be described by the following Lagrangian, whose general form was already introduced in Sec. 3.3:

$$\mathcal{L} = ig_c^A \langle a_{1,\mu} \rho^\mu \pi \rangle + ig_d^A \langle a_{1,\mu\nu} \rho^{\mu\nu} \pi \rangle, \qquad (3.34)$$

where $g_c^A$ and $g_d^A$ are the coupling constants, $\langle \, \rangle$ represents the trace over flavour matrices,

$$a_{1,\mu\nu} = \partial_\mu a_{1,\nu} - \partial_\nu a_{1,\mu}, \qquad (3.35)$$

and

$$\rho^{\mu\nu} = \partial^\mu \rho^\nu - \partial^\nu \rho^\mu. \qquad (3.36)$$

Note, the term $a_{1,\mu\nu}$ is a matrix, which can be defined from the Gell-Mann matrices $\lambda^a$, introduced at the beginning of Section 1.3:

$$a_{1,\mu\nu} = \sum_{k=1}^{3} a_{1,\mu\nu}^k \lambda^k. \qquad (3.37)$$

Both terms in Eq. (3.34) are acceptable, as they respect all underlying symmetries. The first interaction term in the Lagrangian is non-derivative, parameterized by the coupling constant $g_c^A$, and the second is derivative, with $g_d^A$ as coupling constant. The full amplitude reads:

$$i\mathcal{A} = g_c^A \, \epsilon_\mu(0, M_J)\epsilon^{\mu*}(\vec{k}, \lambda) + 2g_d^A \big[k_0 \cdot k_1 \, \epsilon^\mu(\vec{0}, M_J)\epsilon_\mu^*(\vec{k_1}, \lambda)$$
$$- k_0^\nu \, k_{1,\mu} \, \epsilon^\mu(\vec{0}, M_J)\epsilon_\nu^*(\vec{k_1}, \lambda)\big]$$

$$= -\begin{cases} g_c^A + 2g_d^A M_{a_1} E_\rho & M_J = \lambda = \pm 1 \\ \gamma(g_c^A + 2g_d^A M_{a_1} E_\rho - 2g_d^A M_{a_1} \beta k) & M_J = \lambda = 0. \end{cases} \quad (3.38)$$

The 4-momentum of the decaying meson and of the vector decay product are $k_0^\mu = (M_{a_1}, \vec{0})$ and $k_1^\mu = (E_\rho, 0, 0, k)$, where $M_{a_1}$ is the mass of the decaying meson, $M_\rho$ is the mass of the decay product, $E_\rho$ is its energy and $k$ is the magnitude of the 3-momentum carried by the vector decay product. For $J = 1$, the case $M_J = 0$ alone corresponds to $M_J < J$. Both the PoVs and the momentum dependence come entirely from the interaction terms, and this is valid for any decay that will be analyzed in this work. From Eq. (3.20) we know that $\mathscr{A}_{\lambda\nu}^J(0,0;M_J) \propto F_{\lambda\nu}^J$, but the proportionality can be written as an equivalence if we multiply the helicity amplitude by a phase.[‡]

Since the ratios of the partial wave amplitudes, that is the quantity we are interested in, do not depend on the phase, as it is the same for each amplitude, we are allowed to arbitrarily choose the phase for each decaying particle. Therefore, by considering Eq. (3.38), we can write:

$$F_{10}^{1,A} = (g_c^A + 2g_d^A M_{a_1} E_\rho) \quad (3.39)$$

$$F_{00}^{1,A} = \gamma(g_c^A + 2g_d^A M_{a_1} E_\rho - 2g_d^A M_{a_1} \beta k), \quad (3.40)$$

arbitrarily setting the phase as a common multiplier. We also know from Eq. (3.21) how the helicity amplitudes and the $\ell S$ coupling amplitudes are related. Hence,

$$F_{10}^{1,A} = \frac{1}{\sqrt{3}} G_0^A + \frac{1}{\sqrt{6}} G_2^A \quad (3.41)$$

$$F_{00}^{1,A} = \frac{1}{\sqrt{3}} G_0^A - \sqrt{\frac{2}{3}} G_2^A, \quad (3.42)$$

where $G_0^A \equiv G_{01}^{1,A}$ and $G_2^A \equiv G_{21}^{1,A}$. The last two equations are derived step by step in App. C. We note that the decay can be entirely due to the $S$- or $D$-wave if $F_{00}^{1,A}$ equals $F_{10}^{1,A}$ or $2F_{10}^{1,A}$ respectively. Now, putting together the equations (3.39)-(3.42), we obtain:

$$G_2^A = \sqrt{\frac{2}{3}} \left[ g_c^A \left( \frac{M_\rho - E_\rho}{M_\rho} \right) + 2g_d^A M_{a_1}(E_\rho - M_\rho) \right] \quad (3.43)$$

$$G_0^A = \frac{1}{\sqrt{3}} \left[ g_c^A \left( \frac{2M_\rho + E_\rho}{M_\rho} \right) + 2g_d^A M_{a_1}(2E_\rho + M_\rho) \right]. \quad (3.44)$$

---

[‡]It is important to remark that the $\nu$ index appearing over the mesons -e.g. $\rho^{\mu\nu}$- is a covariant (or contravariant) index present in the Lagrangian, while the index $\nu$, appearing in the amplitudes together with the index $\lambda$, is the third component of the spin of one of the product of the decay (see Sec. 3.3.1).

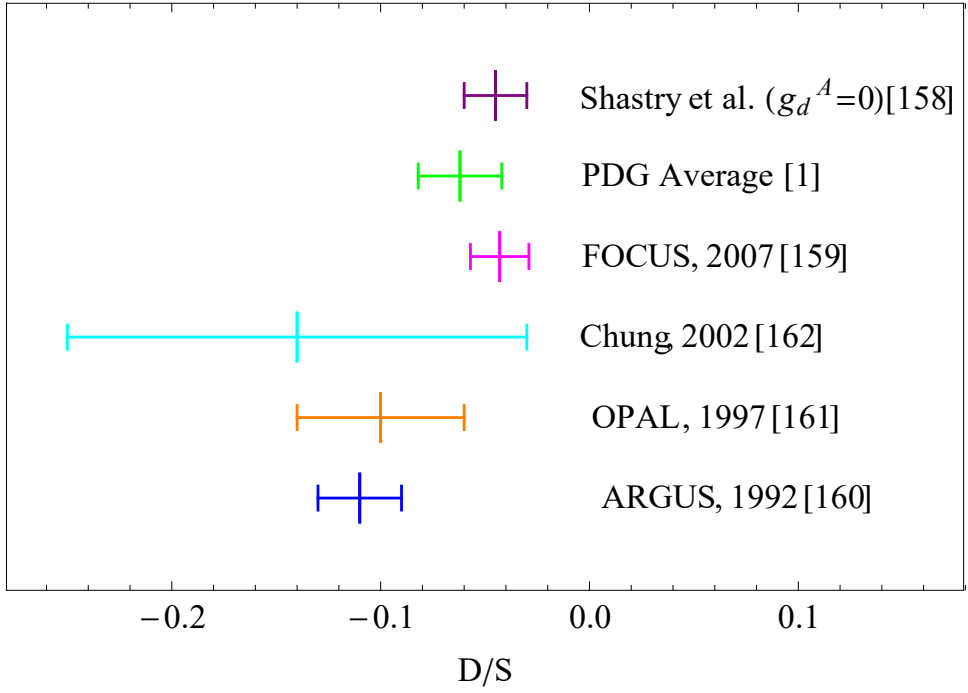

**Fig. 3.2** Comparison among D/S ratios from various experiments and our work [158] for the decay $a_1(1260) \to \rho\pi$. The uncertainties are listed in PDG and, for our work, reported in the article.

The ratio between $G_2^A$ and $G_0^A$ correspond to the ratio between the $D$- and the $S$-wave for the decay of $a_1(1260) \to \rho\pi$. As explained in details later on, the ratio between these two partial wave amplitudes provides good results even by neglecting the derivative interaction. Indeed, by considering $g_d^A = 0$, we obtain:

$$\frac{G_2^A}{G_0^A} = \sqrt{2}\left(\frac{M_\rho - E_\rho}{2M_\rho + E_\rho}\right). \tag{3.45}$$

Using the numerical values, we get

$$\frac{G_2^A}{G_0^A} = -0.045 \tag{3.46}$$

with the proper substitution of the value of the meson masses and the magnitude of the 3-momentum. For a detailed ft that takes into account uncertainties, see later on.

The obtained value of the ratio is in good agreement with the value reported by the FOCUS collaboration [159] and within the error margin [158] of the PDG value of $-0.062 \pm 0.02$ [1]. A comparison among some experimental values obtained and the values of our work are shown in Fig. 3.2, where the references written are: ARGUS [160], OPAL [161], Chung [162], FOCUS [159]. Our result is in good agreement with the most modern evaluations, even with the absence of the derivative interaction.

When dealing with decays, a quantity of primary importance to consider is the decay width (for an overview about it, see App. D). In the case of the decay of the $a_1(1260)$ state, its width range from 250 to 600 MeV [1], which implies that the state $a_1(1260)$ can be considered broad. The final form of the decay width can be obtained from Eqs. (3.43) and (3.44):

$$\Gamma_{a_1 \to \rho\pi} = f_{a_1\rho\pi} \frac{k}{8\pi M_{a_1}^2} (|G_0^A|^2 + |G_2^A|^2), \qquad (3.47)$$

where $f_{a_1\rho\pi}$ is the isospin symmetry factor. Finally,

$$\Gamma_{a_1 \to \rho\pi} = f_{a_1\rho\pi} \frac{k}{24\pi M_{a_1}^2} \left[ (g_c^A)^2 \left( \frac{k^2}{M_\rho^2} + 3 \right) + 12 g_c^A g_d^A E_\rho M_{a_1} \right.$$
$$\left. + 4(g_d^A)^2 M_{a_1}^2 M_\rho^2 \left( \frac{2k^2}{M_\rho^2} + 3 \right) \right]. \qquad (3.48)$$

Equation (3.48) contains three terms with three corresponding coefficients: a term coming from the non-derivative interaction only $((g_c^A)^2)$, another which arises from purely derivative interaction $((g_d^A)^2)$, and the interference term $(g_c^A g_d^A)$. This last term indicates whether the interference between the non-derivative and the derivative interaction is constructive or not.

The same form of the decay widths can be used for any other component of the nonet, by using the appropriate masses, energies and isospin symmetry factors.

- Coupling constant and mixing

We now focus on the value of the coupling constants and of the strange-nonstrange mixing angle. The Lagrangian of the $1^{++}$ sector has the coupling constants $g_c^A$ and $g_d^A$, and the mixing angle $\theta_{AV}$ as parameters. The mixing angle enters the Lagrangian as:

$$\begin{pmatrix} |f_1\rangle \\ |f_1'\rangle \end{pmatrix} = \begin{pmatrix} \cos\theta_{AV} & \sin\theta_{AV} \\ -\sin\theta_{AV} & \cos\theta_{AV} \end{pmatrix} \begin{pmatrix} |\bar{n}n\rangle_{AV} \\ |\bar{s}s\rangle_{AV} \end{pmatrix}, \qquad (3.49)$$

where $|\bar{s}s\rangle_{AV}$ and $|\bar{n}n\rangle_{AV}$ $(= \frac{1}{\sqrt{2}}(|\bar{u}u\rangle_{AV} + |\bar{d}d\rangle_{AV}))$ are the strange and non-strange isosinglet states respectively. The known values are the $D/S-$ratio of the decaying $a_1(1260)$ and the width of the decays $a_1(1260) \to \rho\pi$ and $f_1'(1420) \to K^*K$, given in Table 3.4. The width of the channel $a_1(1260) \to \rho\pi$ was taken as the total width of $a_1(1260)$, since it is the dominant one. The branching fraction from the PDG allows us to estimate the width of the $f_1'(1420) \to K^*K$ decay as $44.5 \pm 4.2$ MeV.

| Decay | Width (MeV) | $D/S$ [1] |
|---|---|---|
| $a_1(1260) \to \rho\pi$ | $420 \pm 35$ | $-0.062 \pm 0.02$ |
| $f_1'(1420) \to K^*K$ | $44.5 \pm 4.5$ | $---$ |

**Table 3.4** Data from PDG used to calculate the values listed in Table 3.5.

We can use these three data to estimate the unknown parameters. Indeed, we have that:

$$\Gamma_{a_1(1260) \to \rho\pi} = \Gamma_{a_1(1260) \to \rho\pi}(g_c^A, g_d^A),$$
$$\frac{G_2^A}{G_0^A} = \frac{G_2^A}{G_0^A}(g_c^A, g_d^A),$$
$$\Gamma_{f_1'(1420) \to K^*K} = \Gamma_{f_1'(1420) \to K^*K}(g_c^A, g_d^A, \theta_{AV}). \quad (3.50)$$

Since there are three unknown parameters in the same numbers of equations, the three parameters can be evaluated without the use of a statistical fit. A $\chi^2$ function was defined in order to find their errors, which are given in Tab. 3.5. The angle $\theta_{AV}$ is in agreement with [108].

| $g_c^A$ (GeV) | $g_d^A$ (GeV$^{-1}$) | $\theta_{AV}$ |
|---|---|---|
| $3.89 \pm 0.75$ | $-0.32 \pm 0.37$ | $(24.9 \pm 3.2)°$ |

**Table 3.5** Values of the parameters obtained in the $1^{++}$ mesonic sector.

We also obtain some predictions for the kaonic decays of $f_1(1285)$ and $f_1'(1420)$, given in Table 3.6.

| Decay | Width (MeV) | | $D/S-$ratio | |
|---|---|---|---|---|
| | Theory | PDG [1] | Theory | PDG [1] |
| $f_1(1285) \to K^*K$ | $4.78 \pm 0.57$ | not seen | $-(0.436 \pm 0.87) \times 10^{-3}$ | $---$ |
| $f_1'(1420) \to K^*K$ | Input | Input | $-0.0116 \pm 0.005$ | $---$ |

**Table 3.6** Predictions based on the parameters listed in Table 3.5.

For what concerns the decay $f_1(1285) \to K^*K$, this is kinematically suppressed since it is below threshold. The values given in Table 3.6 for this decay comes from a spectral integration over the product $K^*$. One should also notice that the $D/S$ ratio for the $1^{++} \to 1^{--}0^{-+}$ decays gets a small value. This indicates that the non-derivative inter-actions are largely predominant compared to the derivative ones. Indeed, the coupling constant $g_d^A$ has a value comparable with zero.

### 3.4.2 The decay of $b_1(1235) \to \omega\pi$ ($1^{+-} \to 1^{--}0^{-+}$))

As for the decay of $a_1(1260)$, we start by introducing the Lagrangian:

$$\mathcal{L} = ig_c^B \langle b_{1,\mu}\omega^\mu \pi \rangle + ig_d^B \langle b_{1,\mu\nu}\omega^{\mu\nu}\pi \rangle, \tag{3.51}$$

where

$$b_{1,\mu\nu} = \partial_\mu b_{1,\nu} - \partial_\nu b_{1,\mu}, \tag{3.52}$$

and

$$\omega^{\mu\nu} = \partial^\mu \omega^\nu - \partial^\nu \omega^\mu. \tag{3.53}$$

The corresponding amplitude is given by:

$$\begin{aligned} i\mathcal{A} &= g_c^B \, \epsilon_\mu(0, M_J)\epsilon^{\mu*}(\vec{k}, \lambda) + 2g_d^B \left[ k_0 \cdot k_1 \, \epsilon^\mu(\vec{0}, M_J)\epsilon_\mu^*(\vec{k_1}, \lambda) \right. \\ &\quad \left. - k_0^\nu \, k_{1,\mu} \, \epsilon^\mu(\vec{0}, M_J)\epsilon_\nu^*(\vec{k_1}, \lambda) \right] \\ &= -\begin{cases} g_c^B + 2g_d^B M_{b_1} E_\omega & M_J = \lambda = \pm 1 \\ \gamma(g_c^B + 2g_d^B M_{b_1} E_\omega - 2g_d^B M_{b_1} \beta k) & M_J = \lambda = 0. \end{cases} \end{aligned} \tag{3.54}$$

Because of Eq. (3.17), $\ell$ can assume only even values, that are, in this case, $\ell = 0,\ 2$. In the same way as done for the decay of $a_1(1260)$, we get:

$$F_{10}^{1,B} = (g_c^B + 2g_d^B M_{b_1} E_\omega), \tag{3.55}$$

$$F_{00}^{1,B} = \gamma(g_c^B + 2g_d^B M_{b_1} E_\omega - 2g_d^B M_{b_1}\beta k), \tag{3.56}$$

and

$$F_{10}^{1,B} = \frac{1}{\sqrt{3}}G_0^B + \frac{1}{\sqrt{6}}G_2^B, \tag{3.57}$$

$$F_{00}^{1,B} = \frac{1}{\sqrt{3}}G_0^B - \sqrt{\frac{2}{3}}G_2^B. \tag{3.58}$$

These relations allow us to obtain the explicit form of the $\ell S$-coupling amplitudes $G_0^B$ and $G_2^B$:

$$G_2^B = \sqrt{\frac{2}{3}}\left[g_c^B\left(\frac{M_\omega - E_\omega}{M_\omega}\right) + 2g_d^B M_{b_1}(E_\omega - M_\omega)\right] \tag{3.59}$$

$$G_0^B = \frac{1}{\sqrt{3}}\left[g_c^B\left(\frac{2M_\omega + E_\omega}{M_\omega}\right) + 2g_d^B M_{b_1}(2E_\omega + M_\omega)\right], \tag{3.60}$$

together with the ratio between them:

$$\frac{G_2^B}{G_0^B} = \sqrt{2}\frac{(M_\omega - E_\omega)(g_c^B - 2g_d^B M_{b_1} M_\omega)}{g_c^B(2M_\omega + E_\omega) + 2g_d^B M_{b_1} M_\omega(M_\omega + 2E_\omega)}. \quad (3.61)$$

We can then extract the form of the decay width:

$$\Gamma_{b_1 \to \omega\pi} = f_{b_1\omega\pi} \frac{k}{24\pi M_{b_1}^2} \left[ (g_c^B)^2 \left(\frac{k^2}{M_\omega^2} + 3\right) + 12 g_c^B g_d^B E_\omega M_{b_1} \right.$$
$$\left. + 4(g_d^B)^2 M_{b_1}^2 M_\omega^2 \left(\frac{2k^2}{M_\omega^2} + 3\right) \right]. \quad (3.62)$$

Because of the similar masses of the mesons involved and the comparable 3-momenta of the products of the decay, one could expect that the ratio of the partial wave amplitudes for the decays $b_1(1235) \to \omega\pi$ and $a_1(1260) \to \rho\pi$ are similar. The value of the partial wave amplitudes ratio, obtained from the Lagrangian describing the $b_1(1235) \to \omega\pi$ decay, is $-0.043$ when considering only the non-derivative term. This value is quite different from the experimental one, $0.277 \pm 0.027$ [1]. Therefore, the derivative interaction is needed in the pseudovector decay, differently from what it was for the decay of $1^{++}$ mesons.

- Coupling constant and mixing

We now focus on the value of the coupling constants and of the strange-nonstrange mixing angle, in analogy to what was done for the decay of an axial-vector particle. Here the three parameters we are interested in are the coupling constants $g_c^B$ and $g_d^B$ and the isoscalar mixing angle $\theta_{PV}$, which is defined as:

$$\begin{pmatrix} |h_1\rangle \\ |h_1'\rangle \end{pmatrix} = \begin{pmatrix} \cos\theta_{PV} & \sin\theta_{PV} \\ -\sin\theta_{PV} & \cos\theta_{PV} \end{pmatrix} \begin{pmatrix} |\bar{n}n\rangle_{PV} \\ |\bar{s}s\rangle_{PV} \end{pmatrix}. \quad (3.63)$$

As for the axial-vector meson, also in this case we extract the values of the parameters in order to get some predictions. In particular, we use the data of Table 3.7 to estimate the parameters of Table 3.8 and to predict the values of Table 3.9.

An interesting comparison can be done by looking at Tabs. 3.5 and 3.8, where it is immediate to notice that $g_c^B \approx 2g_c^A$, showing that here the derivative term is necessary. Note, $\theta_{PV}$ is not yet known from experimental constraints.

| Decay | Width (MeV) | $D/S$ [1] |
|---|---|---|
| $b_1(1235) \to \omega\pi$ | $110 \pm 7$[149] | $0.277 \pm 0.027$ |
| $h_1'(1415) \to K^*K$ | $90 \pm 15$[1] | $---$ |

**Table 3.7** Data from PDG used to calculate the values listed in Table 3.8.

| $g_c^B$ (GeV) | $g_d^B$ (GeV$^{-1}$) | $\theta_{PV}$ |
|---|---|---|
| $6.36 \pm 0.72$ | $-4.37 \pm 0.37$ | $(25.2 \pm 3.1)°$ |

**Table 3.8** Values of the parameters obtained in the $1^{+-}$ mesonic sector.

### 3.4.3 Behaviour of the partial wave amplitudes

In general, we can also study the behaviour of the partial wave amplitudes as function of the velocity of the decaying particles. To do so, we rewrite the amplitudes from Eqs. (3.43) and (3.44) in term of the Lorentz factor $\gamma$. The case dealing with the decay of the pseudotensor has already been discussed in [158]. We call $M_p$ and $M_{d,1}$ the masses of the parent particle and of the heavier daughter, in our case $M_{a_1}$ and $M_\rho$, respectively. The following equations are obtained by considering

$$\gamma = \frac{E_{d,1}}{M_{d,1}} = \frac{1}{\sqrt{1-\beta^2}}, \tag{3.64}$$

where $\beta \in [0,1]$ and $\gamma \in [1,\infty)$:

$$G_2^A = \sqrt{\frac{2}{3}} \left[ g_c^A (1-\gamma) + 2g_d^A M_p M_{d,1}(\gamma-1) \right] \tag{3.65}$$

$$G_0^A = \frac{1}{\sqrt{3}} \left[ g_c^A (2+\gamma) + 2g_d^A M_p M_{d,1}(2\gamma+1) \right]. \tag{3.66}$$

The partial wave amplitudes depend on the mass of the decaying particle only under the condition of nonzero derivative interaction.

Note, the amplitudes of Eqs. (3.65) and (3.66) depends on the factor $\gamma$ (and on $\beta$ analogously), i.e. the relative velocity between the decaying particle and the parent. One could then consider the decay of $a_1(1260) \to \rho\pi$ in the rest frame of the parent. In this specific case, the value of $\gamma$ is fixed within a small range, which depends on the uncertainties of the masses. Changing $\beta$ implies a formal change of $M_p$. While this is not applicable for a specific example, it applies to other decays of the same type, in which different masses are involved.

| Decay | Width (MeV) | | $D/S$−ratio | |
|---|---|---|---|---|
| | Theory | PDG [1] | Theory | PDG [1] |
| $h_1(1170) \to \rho\pi$ | $146 \pm 14$ | seen | $0.281 \pm 0.035$ | − − − |
| $h_1'(1415) \to K^*K$ | Input | Input | $0.021 \pm 0.001$ | − − − |

**Table 3.9** Predictions based on the parameters listed in Table 3.8.

We are then allowed to plot Eqs. (3.65) and (3.66) as function of $\beta$, as it is done in Fig. **??**, where the non-derivative and the derivative interactions are plotted separately and are independent on the masses $M_p$ and $M_{d,1}$. We see that at low energies ($\gamma \to 1$), the higher partial wave vanish (analogously as in [158]).

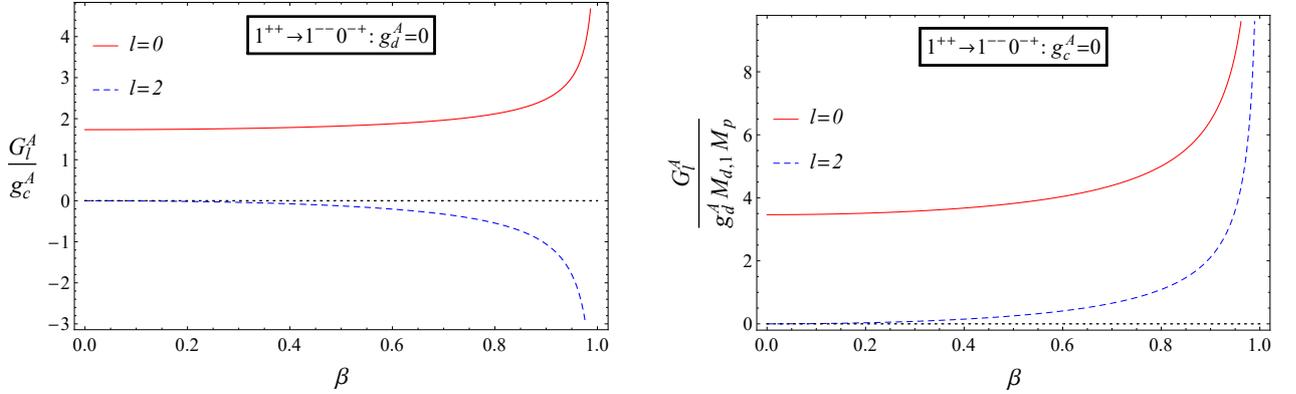

**Fig. 3.3** The plots of the partial wave amplitudes for the $1^{++} \to 1^{--}0^{-+}$ decay, as functions of $\beta$. The left panel shows the behaviour with only low order terms, while the right panel shows the case with only higher order terms. Since the form of the $lS$-coupling amplitude is identical (Eqs. (3.65) and (3.66)), the plots for the decay $1^{+-} \to 1^{--}0^{-+}$ will be the same.

In the ultrarelativistic limit ($\beta \to 1$, $\gamma \to \infty$), the higher partial wave dominate compared to the $\ell = 0$ one. This can be better seen in Fig. 3.4, were the $D/S$-ratio becomes larger (in modulus) as $\beta$ increases.

An analogous approach was used in Ref. [158], where we studied the pseudotensor nonet, as well as nature of $\eta_2(1645)$ and $\eta_2(1870)$ (for more details see that work).

## 3.5 The $b_1(1235) \to \omega\pi$ decay along an arbitrary direction

The decay of the $b_1(1235)$ meson was already discussed in Sec. 3.4.2 in the rest frame of the decaying particle, which is particularly convenient in the case of a 2-body decay. Indeed, in this frame of reference, the two products move along the same axis in opposite directions. In addition, if we consider the decay along the $z$-axis, as it was done before, we largely simplify the form of the polarization matrices (see App. B).

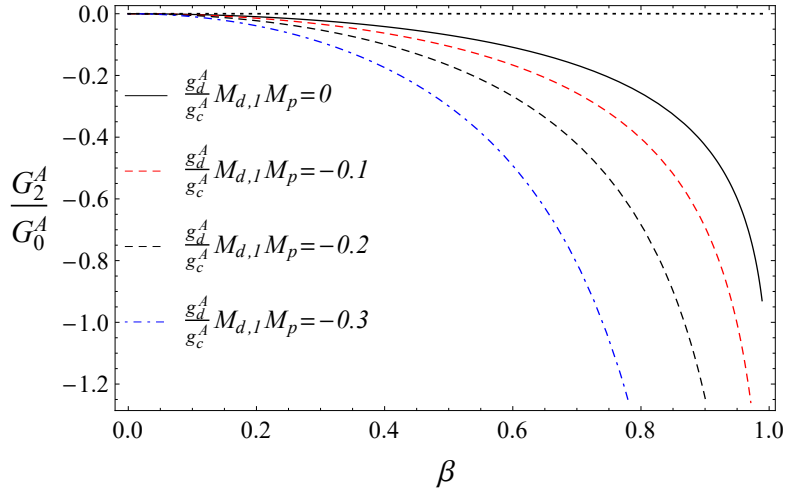

**Fig. 3.4** The $D/S$ ratio as function of $\beta$. Four plots for representative values of $\frac{g_d^A}{g_c^A} M_{d,1} M_p$ are given.

We now provide a study of the $b_1(1235) \to \omega\pi$ decay without any constriction on the direction of the decay (but still in the rest frame of $b_1(1235)$), which was not done in Ref. [158]. In order to do so, it is convenient to write the polarization vectors in spherical coordinates. We will then use spherical coordinates referring to the following diagram:

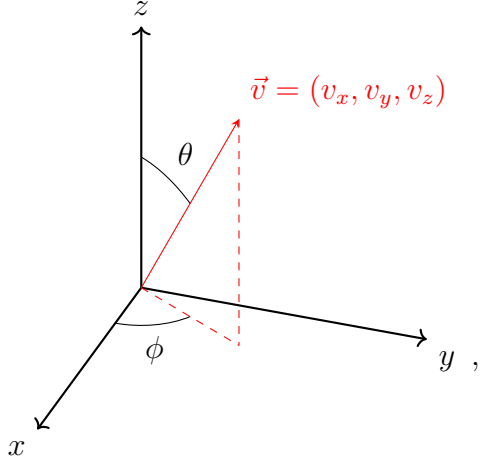

This corresponds to:

$$\begin{cases} v_x = v \cos\phi \sin\theta \\ v_y = v \sin\phi \sin\theta \\ v_z = v \cos\theta \end{cases}, \qquad (3.67)$$

where $v = |\vec{v}|$ is the velocity of a moving outgoing particle. In the spherical coordinates the polarization vectors, taken from App. B, get the form:

$$\epsilon^\nu(\vec{k},+1) = -\frac{1}{\sqrt{2}}\begin{pmatrix} \gamma(\beta_x + i\beta_y) \\ 1+(\gamma-1)(\hat{v}_x^2 + i\hat{v}_x\hat{v}_y) \\ i+(\gamma-1)(\hat{v}_y\hat{v}_x + i\hat{v}_y^2) \\ (\gamma-1)(\hat{v}_z\hat{v}_x + i\hat{v}_z\hat{v}_y) \end{pmatrix} = -\frac{e^{i\phi}}{\sqrt{2}}\begin{pmatrix} \gamma\beta\sin\theta \\ e^{-i\phi}+(\gamma-1)\cos\phi\sin^2\theta \\ ie^{-i\phi}+(\gamma-1)\sin\phi\sin^2\theta \\ (\gamma-1)\cos\theta\sin\theta \end{pmatrix},$$
(3.68)

$$\epsilon^\nu(\vec{k},-1) = \frac{1}{\sqrt{2}}\begin{pmatrix} \gamma(\beta_x - i\beta_y) \\ 1+(\gamma-1)(\hat{v}_x^2 - i\hat{v}_x\hat{v}_y) \\ -i+(\gamma-1)(\hat{v}_y\hat{v}_x - i\hat{v}_y^2) \\ (\gamma-1)(\hat{v}_z\hat{v}_x - i\hat{v}_z\hat{v}_y) \end{pmatrix} = \frac{e^{-i\phi}}{\sqrt{2}}\begin{pmatrix} \gamma\beta\sin\theta \\ e^{i\phi}+(\gamma-1)\cos\phi\sin^2\theta \\ -ie^{i\phi}+(\gamma-1)\sin\phi\sin^2\theta \\ (\gamma-1)\cos\theta\sin\theta \end{pmatrix},$$
(3.69)

and

$$\epsilon^\nu(\vec{k},0) = \begin{pmatrix} \gamma\beta_z \\ (\gamma-1)(\hat{v}_x\hat{v}_y) \\ (\gamma-1)(\hat{v}_y\hat{v}_z) \\ 1+(\gamma-1)\hat{v}_z^2 \end{pmatrix} = \begin{pmatrix} \gamma\beta\cos\theta \\ (\gamma-1)\cos\theta\cos\phi\sin\theta \\ (\gamma-1)\cos\theta\sin\phi\sin\theta \\ 1+(\gamma-1)\cos^2\theta \end{pmatrix},$$
(3.70)

where the Euler's formula, $e^{\pm i\phi} = \cos\phi \pm i\sin\phi$, was used. In both the formalism along the $z$-direction and in an arbitrary direction, the aim of our work is the same: get the ratio between the different $\ell S$-coupling amplitudes. This result can be obtained equivalently with two different choices of the coordinates: along the $z$-axis or along an arbitrary direction.

- **Along the $z$-axis**

In the case of a fixed direction ($\theta = 0$), we used the relation between the two amplitudes $\mathscr{A}^J(\theta,\phi;M_J)$ and $F_{\lambda\nu}^J$ (Eq. (3.18)), which involves the Wigner $D$-matrix $D_{Mm_s}^{J*}(\phi,\theta,0)$ (we remind that $m_s = \lambda - \nu$). In this choice of coordinates, $D_{Mm_s}^{J*}(0,0,0) = 1$. Finally, Eq. (3.21) links $F_{\lambda\nu}^J$ with $G_{\ell S}^J$ [¶].

- **Along an arbitrary direction**

In the case of $\theta \neq 0$ and $\phi \neq 0$, then $D_{Mm_s}^{J*}(\phi,\theta,0) \neq 1$ (as well as $Y_{\ell m_\ell}(\theta,\phi) \neq 1$). Then, the choice between the two equations (Eqs. (3.18) and (3.22)),

$$\mathscr{A}^J(\theta,\phi;M_J) = \sum_{\lambda\nu}\mathscr{A}_{\lambda\nu}^J(\theta,\phi;M_J) \propto \sum_{\lambda\nu} D_{Mm_s}^{J*}(\phi,\theta,0)F_{\lambda\nu}^J \quad (3.71)$$

---
[¶]Also the spherical harmonics $Y_{\ell m_\ell}(0,0) = 1$. Thus, in this case, the use of one or the other method is absolutely equivalent.

and

$$i\mathcal{A}(\theta,\phi;M_J) = \sum_{\ell} \sum_{m_\ell=-\ell}^{\ell} G_\ell \langle \ell m_\ell S m_s | J M_J \rangle Y_{\ell m_\ell}(\theta,\phi), \qquad (3.72)$$

becomes relevant. In particular, the choice of the formula with the spherical harmonics is required because of the information that we are interested in. As it will be clear later on, the method with $Y_{\ell m_\ell}(\theta,\phi)$ requires a spherical integration along $\theta$ and $\phi$, which is not needed in the other case. Thus, the method with $D^{J*}_{M m_s}(\phi,\theta,0)$ provides information about how the amplitude depends on the rotation of the frame of reference, without providing any information on $\ell$ and $m_\ell$. On the other hand, if we use $Y_{\ell m_\ell}(\theta,\phi)$, we know which values (and in which amount) of $\ell$ and $m_\ell$ values contributes.

In other words, we want an expansion in $\ell$, not in $J$, and this can be obtained from $Y_{\ell m_\ell}(\theta,\phi)$ and not from $D^J_{m m_s}$. Summarizing, the two methods give us different information about the decay: from the Wigner $D$-matrix we know how the angular profile change according to the different direction, while with $Y_{\ell m_\ell}(\theta,\phi)$ we know which partial waves and in which amount they contribute to a particular angular profile.

Starting from the Lagrangian of Eq. (3.51)

$$\mathcal{L} = ig_c^B \langle b_{1,\mu} \omega^\mu \pi \rangle + ig_d^B \langle b_{1,\mu\nu} \omega^{\mu\nu} \pi \rangle, \qquad (3.73)$$

we obtain the same amplitude given in the first part of Eq. (3.54), with the difference that now we must use the boosted polarization tensors:

$$i\mathcal{A}_{\lambda\nu}(\theta,\phi;M_J) = g_c^B\, \epsilon_\mu(0,M_J)\epsilon^{\mu*}(\vec{k},\lambda) + 2g_d^B \big[ k_0 \cdot k_1\, \epsilon^\mu(\vec{0},M_J)\epsilon^*_\mu(\vec{k_1},\lambda) \\ - k_0^\nu\, k_{1,\mu}\, \epsilon^\mu(\vec{0},M_J)\epsilon^*_\nu(\vec{k_1},\lambda) \big] \qquad (3.74)$$

Here, differently from the case of $\theta=\phi=0$, also terms with $M_J \neq \lambda$ can be non-zero. The equations resulting from the explicit tensor calculation in Eq. (3.74) must get the same form of Eq. (3.54) if we take $\theta=0$. The following step is to find the relations between $\mathcal{A}$ and

$$G_\ell^{AD} = \sqrt{\frac{\alpha}{(2J+1)}} G_{\ell S}^J, \qquad (3.75)$$

where $G_\ell^{AD}$ is the amplitude calculated along an arbitrary direction. Upon using Eq. (3.22):

$$i\mathcal{A}_{\lambda\nu}(\theta,\phi;M_J) = \sum_{\ell} \sum_{m_\ell=\ell}^{\ell} G_\ell^{AD} \langle \ell m_\ell S m_s | J M_J \rangle Y_{\ell m_\ell}(\theta,\phi). \qquad (3.76)$$

Since we are considering the decay of $b_1(1235)$, which is a $1^{+-} \to 1^{--}0^{-+}$ process, it is immediate to calculate which values of the quantum numbers are allowed. In Eq. (3.74), we can distinguish 3 cases:

- $M_J = 1$
  Since $J = 1$ and $S = 1$ (see App. C for details) $\to \ell = 0$ or $2$.
  The conditions $m_s \in [-S, S]$, $m_\ell \in [-\ell, \ell]$ and $m_\ell + m_s = M_J$ lead to the following cases:
  For $\ell = 0 \to m_\ell = 0$ and $m_s = 1$
  For $\ell = 2 \to \begin{cases} m_\ell = 2 \text{ and } m_s = -1 \\ m_\ell = 1 \text{ and } m_s = 0 \\ m_\ell = 0 \text{ and } m_s = 1 \end{cases}$

- $M_J = -1$
  For $\ell = 0 \to m_\ell = 0$ and $m_s = -1$
  For $\ell = 2 \to \begin{cases} m_\ell = 0 \text{ and } m_s = -1 \\ m_\ell = -1 \text{ and } m_s = 0 \\ m_\ell = -2 \text{ and } m_s = 1 \end{cases}$

- $M_J = 0$
  For $\ell = 0 \to m_\ell = 0$ and $m_s = 0$
  For $\ell = 2 \to \begin{cases} m_\ell = 1 \text{ and } m_s = -1 \\ m_\ell = 0 \text{ and } m_s = 0 \\ m_\ell = -1 \text{ and } m_s = 1 \end{cases}$

We have now enough information to calculate explicitly the values of the $\ell S$-coupling amplitudes $G_0^{AD}$ and $G_2^{AD}$.

We start rewriting Eq. (3.22) as:

$$\iint i\mathcal{A}_{\lambda\nu}(\theta, \phi; M_J) Y_{\ell' m'_\ell}(\theta, \phi) d\cos\theta d\phi = \sum_\ell \sum_{m_\ell = \ell}^{\ell} G_\ell^{AD} \langle \ell m_\ell S m_s | J M_J \rangle \delta_{\ell\ell'} \delta_{m_\ell m'_\ell}$$
$$= G_{\ell'}^{AD} \langle \ell' m'_\ell S m_s | J M_J \rangle. \quad (3.77)$$

Finally,

$$G_\ell^{AD} = \frac{\iint i\mathcal{A}_{\lambda\nu}(\theta, \phi; M_J) Y_{\ell' m'_\ell}(\theta, \phi) d\cos\theta d\phi}{\langle \ell m_\ell S m_s | J M_J \rangle}. \quad (3.78)$$

In order to find the explicit expressions for the $\ell S$-coupling amplitudes, we evaluate the previous equation for all the acceptable combinations of the various quantum numbers, whose allowed values were already discussed in this Section. The evaluation of Eq. (3.78) was done in Table 3.10, where the numerator

$$\mathfrak{N} = \iint i\mathcal{A}_{\lambda\nu}(\theta,\phi;M_J)Y_{\ell'm'_\ell}(\theta,\phi)d\cos\theta d\phi \qquad (3.79)$$

and the denominator

$$C^{Jm_J}_{\ell m_\ell;Sm_s} = \langle \ell m_\ell Sm_s|JM_J\rangle \qquad (3.80)$$

of Eq. (3.78) are considered separately. Additionally, the two symbols ✓and ✗will be used when the values of $\mathfrak{N}$ and $C^{Jm_J}_{\ell m_\ell;Sm_s}$ are zero (✗) or nonzero (✓).

From Table 3.10 it emerges that only in few cases both $\mathfrak{N}$ and $C^{Jm_J}_{\ell m_\ell;Sm_s}$ are different from zero (see the highlighted rows in the table)[‖]. In particular, as it happened for the case along the $z$-direction, we get a finite value for

$$G^{AD}_\ell = \frac{\mathfrak{N}}{C^{Jm_J}_{\ell m_\ell;Sm_s}} \qquad (3.81)$$

only if $M_J = \lambda$ or, analogously, for $m_\ell = 0$, meaning that the whole momentum of the parent was transferred entirely to the spin of the vector product. We can thus focus on six cases: three with $\ell = 0$ and three with $\ell = 2$.

All the cases with $\ell = 0$ have as solution:

$$G^{AD}_0 = \frac{\mathfrak{N}}{C^{Jm_J}_{\ell m_\ell;Sm_s}} = -\frac{2\sqrt{\pi}}{3}\left[g^B_c\left(\frac{2M_\omega + E_\omega}{M_\omega}\right) + 2g^B_d M_{b_1}(2E_\omega + M_\omega)\right], \qquad (3.82)$$

where the $G^{AD}$ means that it was evaluated along an arbitrary direction.
Analogously, all the cases with $\ell = 2$ have as solution:

$$G^{AD}_2 = \frac{\mathfrak{N}}{C^{Jm_J}_{\ell m_\ell;Sm_s}} = -\frac{2\sqrt{2\pi}}{3}\left[g^B_c\left(\frac{M_\omega - E_\omega}{M_\omega}\right) + 2g^B_d M_{b_1}(E_\omega - M_\omega)\right]. \qquad (3.83)$$

Finally, $G^{AD}_0$ and $G^{AD}_2$ must be compared with the respectively formulas evaluated at fixed direction (Eqs. (3.59) and (3.60)):

$$G^B_2 = \sqrt{\frac{2}{3}}\left[g^B_c\left(\frac{M_\omega - E_\omega}{M_\omega}\right) + 2g^B_d M_{b_1}(E_\omega - M_\omega)\right] \qquad (3.84)$$

---

[‖]Eqs. (3.79) and (3.80) do not contain any explicit dependence between $m_s$ and $\lambda$. Thus, the relation $|S,m_s\rangle = |s,\lambda\rangle \otimes |\sigma,\nu\rangle$ from Eq. (3.14) is not considered in the table. Still, since the second decaying particle is a pseudoscalar, $|\sigma,\nu\rangle = |0,0\rangle \to m_s = \lambda$, the contributions of the terms $m_s \neq \lambda$ is zero. Note, if the second decaying particle has non-zero total spin, $m_s$ and $\lambda$ can also be different.

$$G_0^B = \frac{1}{\sqrt{3}} \left[ g_c^B \left( \frac{2M_\omega + E_\omega}{M_\omega} \right) + 2g_d^B M_{b_1}(2E_\omega + M_\omega) \right]. \tag{3.85}$$

Although the different coefficients of the amplitudes:

$$G_0^{AD} = -\frac{2\sqrt{\pi}}{\sqrt{3}} G_0^B \quad \text{and} \quad G_2^{AD} = -\frac{2\sqrt{\pi}}{\sqrt{3}} G_2^B, \tag{3.86}$$

which arise due to the normalization of the spherical harmonics, the ratio of the two amplitudes is unchanged from the fixed direction case. Indeed, as from Eq. (3.61), we get:

$$\frac{G_2^{AD}}{G_0^{AD}} = \frac{G_2^B}{G_0^B} = \sqrt{2} \frac{(M_\omega - E_\omega)(g_c^B - 2g_d^B M_{b_1} M_\omega)}{g_c^B(2M_\omega + E_\omega) + 2g_d^B M_{b_1} M_\omega(M_\omega + 2E_\omega)}, \tag{3.87}$$

as expected.

## 3.6 Short summary

We studied the decay of some conventional mesons using the covariant helicity formalism, which contains the energy dependence in the decay amplitude, in a form that ensures the covariance. Through the relations between the Feynman amplitude $\mathscr{A}_{\lambda\nu}^J$, the helicity amplitude $F_{\lambda\nu}^J$ and the $\ell S$-coupling amplitudes $G_{\ell S}^J$, we were able extract from various interaction Lagrangians the quantities $G_{\ell S}^J$. The ratio between two of these amplitudes, having different $\ell$, was then calculated. This value is composed by the ratio between the partial wave amplitudes that can be found in PDG for conventional mesons. This allowed us also to predict some other possible ratios of partial wave amplitudes that were not present in the PDG.

Finally, as expected, we demonstrated that the ratio between the partial wave amplitudes does not depend on the choice of the set of axes, and we used the $b_1(1235) \to \omega\pi$ decay as an example.

**Table 3.10** The zero (✗) or nonzero (✓) values of $\mathfrak{N}$ and $C^{Jm_J}_{\ell m_\ell; S m_s}$, evaluated for all the possible allowed quantum numbers (see text for details).

| $M_J$ | $\lambda$ | $\ell$ | $m_\ell$ | $S$ | $m_s$ | $\mathfrak{N}$ | $C^{Jm_J}_{\ell m_\ell; S m_s}$ | $M_J$ | $\lambda$ | $\ell$ | $m_\ell$ | $S$ | $m_s$ | $\mathfrak{N}$ | $C^{Jm_J}_{\ell m_\ell; S m_s}$ |
|---|---|---|---|---|---|---|---|---|---|---|---|---|---|---|---|
| 1 | 1 | 0 | 0 | 1 | 1 | ✓ | ✓ | -1 | 1 | 0 | 0 | 1 | -1 | ✗ | ✗ |
| 1 | 1 | 2 | 0 | 1 | 1 | ✓ | ✓ | -1 | 1 | 2 | -2 | 1 | 1 | ✗ | ✓ |
| 1 | 1 | 2 | 1 | 1 | 0 | ✗ | ✗ | -1 | 1 | 2 | -1 | 1 | 0 | ✗ | ✗ |
| 1 | 1 | 2 | 2 | 1 | -1 | ✗ | ✗ | -1 | 1 | 2 | 0 | 1 | -1 | ✗ | ✗ |
| 1 | 0 | 0 | 0 | 1 | 1 | ✗ | ✗ | -1 | 0 | 0 | 0 | 1 | -1 | ✗ | ✗ |
| 1 | 0 | 2 | 0 | 1 | 1 | ✗ | ✗ | -1 | 0 | 2 | -2 | 1 | 1 | ✗ | ✗ |
| 1 | 0 | 2 | 1 | 1 | 0 | ✗ | ✓ | -1 | 0 | 2 | -1 | 1 | 0 | ✗ | ✓ |
| 1 | 0 | 2 | 2 | 1 | -1 | ✗ | ✗ | -1 | 0 | 2 | 0 | 1 | -1 | ✗ | ✗ |
| 1 | -1 | 0 | 0 | 1 | 1 | ✗ | ✗ | -1 | -1 | 0 | 0 | 1 | -1 | ✓ | ✓ |
| 1 | -1 | 2 | 0 | 1 | 1 | ✗ | ✗ | -1 | -1 | 2 | -2 | 1 | 1 | ✗ | ✗ |
| 1 | -1 | 2 | 1 | 1 | 0 | ✗ | ✗ | -1 | -1 | 2 | -1 | 1 | 0 | ✗ | ✗ |
| 1 | -1 | 2 | 2 | 1 | -1 | ✗ | ✓ | -1 | -1 | 2 | 0 | 1 | -1 | ✓ | ✓ |

| $M_J$ | $\lambda$ | $\ell$ | $m_\ell$ | $S$ | $m_s$ | $\mathfrak{N}$ | $C^{Jm_J}_{\ell m_\ell; S m_s}$ |
|---|---|---|---|---|---|---|---|
| 0 | 1 | 0 | 0 | 1 | 0 | ✗ | ✗ |
| 0 | 1 | 2 | -1 | 1 | 1 | ✗ | ✓ |
| 0 | 1 | 2 | 0 | 1 | 0 | ✗ | ✗ |
| 0 | 1 | 2 | 1 | 1 | -1 | ✓ | ✗ |
| 0 | 0 | 0 | 0 | 1 | 0 | ✓ | ✓ |
| 0 | 0 | 2 | -1 | 1 | 1 | ✗ | ✗ |
| 0 | 0 | 2 | 0 | 1 | 0 | ✓ | ✓ |
| 0 | 0 | 2 | 1 | 1 | -1 | ✗ | ✗ |
| 0 | -1 | 0 | 0 | 1 | 0 | ✗ | ✗ |
| 0 | -1 | 2 | -1 | 1 | 1 | ✓ | ✗ |
| 0 | -1 | 2 | 0 | 1 | 0 | ✗ | ✗ |
| 0 | -1 | 2 | 1 | 1 | -1 | ✗ | ✓ |

# CHAPTER 4

# Glueball Resonance Gas model

As already discussed before, YM theory is an important sector of QCD, obtained for the limit $m_q \to \infty$. Its is described by different degrees of freedom (d.o.f.) depending on the energy region under analysis: at low energies, the d.o.f. are glueballs (confined phase of YM), while at high energy the d.o.f. are gluons (deconfined phase of YM). The transition between these two regions occours at the critical temperature $T_c$. In order to study the thermodynamic properties of a YM gas at low energies, we can use a Glueball Resonance Gas model.

The aim of this Section is to compare the thermodynamic results for the pressure and other quantities given in the lattice work by Borsányi [163], obtained in the continuum extrapolation, with the same quantities obtained from a GRG model with the use of masses coming from three lattice spectra [48–50], after a match of the appropriate critical temperatures, $T_c$.

## 4.1 GRG and tensor glueballs

The results obtained in Section 2 can be used in a Glueball Resonance Gas (GRG) model, which, for any glueball spectrum considered, is used to describe a confined gas of glueballs below the corresponding critical confinement/deconfinement temperature $T_c$. In particular, the comparison between the thermodynamic properties of this model and the lattice data from [163] shows that this model well describes the thermal properties of YM theory for a properly matched $T_c$.

The GRG model is conceptually analogous to the well-known Hadron Resonance Gas model (HRG), which is implemented in the study of the low temperature regime in the QCD phase diagram. QCD matter is indeed described, in the low temperature regime, by a weakly interacting hadron gas, and, in the high temperature regime, by the quark-gluon plasma (QGP). The two regimes are connected by a phase transition (crossover or first order, depending on the density) which can be successfully described by lattice QCD

[164, 165].

The lattice glueball masses of [49], when used in a GRG, do not allow for an optimal description of the lattice results from [163] for the pressure and other thermodynamic quantities. When approaching $T_c$, an additional Hagedorn contribution ( $\rho(m) \propto e^{m/T_H}$ [166]) needed to be included [167]. In Ref. [168], we have shown that, when using the more recent lattice results for the glueball masses in [50], there is no need for an additional Hagedorn term. The corresponding value turns $T_c$ to be $323 \pm 18$ MeV.

The GRG model can then be supplemented with two additional contributions:

- Further excited states.

- Interaction between glueballs, especially scalar-scalar and tensor-tensor interactions.

As we shall show, both contributions turn out to be negligible.

## 4.2 Thermodinamic quantities in the GRG

The GRG can be obtained by the sum of the following quantities: pressure $p_i$ and energy density $\epsilon_i$ of the $i$-th glueball. These, in the case of non-interacting particles, read (e.g. Ref. [169]):

$$p_i = -(2J_i + 1)T \int_0^\infty \frac{k^2}{2\pi^2} \ln\left(1 - e^{-\frac{\sqrt{k^2+m_i^2}}{T}}\right) dk \,, \tag{4.1}$$

and

$$\epsilon_i = (2J_i + 1) \int_0^\infty \frac{k^2}{2\pi^2} \frac{\sqrt{k^2+m_i^2}}{\exp\left[\frac{\sqrt{k^2+m_i^2}}{T}\right] - 1} dk \,, \tag{4.2}$$

where $J_i$ is the total spin of the $i$-th state. If we consider $N$ glueballs, the total pressure and energy density of the GRG are given by:

$$p^{GRG} \equiv p = \sum_{i=1}^{N} p_i \,, \tag{4.3}$$

$$\epsilon^{GRG} \equiv \epsilon = \sum_{i=1}^{N} \epsilon_k \,. \tag{4.4}$$

The pressure and energy density allows us to define other two quantities: the trace anomaly $I$ and the entropy density $s$:

$$I = \epsilon - 3p \,, \tag{4.5}$$

$$s = \frac{p + \epsilon}{T} \,. \tag{4.6}$$

The dimensionless versions of these quantities are:

pressure $\quad\quad\quad\quad \hat{p} = p/T^4$

energy density $\quad\quad \hat{\epsilon} = \epsilon/T^4$

trace anomaly $\quad\quad \hat{I} = \hat{\epsilon} - 3\hat{p}$

entropy $\quad\quad\quad\quad \hat{s} = s/T^3.$

## 4.3 Glueball masses from LQCD

Since the thermodynamic quantities in Ref. [163] are given as function of $T/T_c$, the comparison with the GRG is possible only once the $T_c$ for different lattice works [48–50] is determined. The masses of the glueballs were determined in the lattice works [48–50] and are reported in Table 4.1.

| $n\,J^{PC}$ | Mass [MeV] | | | $n\,J^{PC}$ | Mass [MeV] | | |
|---|---|---|---|---|---|---|---|
| | Chen et.al. [49] | Meyer [48] | A and T [50] | | Chen et.al. [49] | Meyer [48] | A and T [50] |
| $1\,0^{++}$ | 1710(50)(80) | 1475(30)(65) | 1653(26) | $1\,1^{--}$ | 3830(40)(190) | 3240(330)(150) | 4030(70) |
| $2\,0^{++}$ | | 2755(30)(120) | 2842(40) | $1\,2^{--}$ | 4010(45)(200) | 3660(130)(170) | 3920(90) |
| $3\,0^{++}$ | | 3370(100)(150) | | $2\,2^{--}$ | | 3740(200)(170) | |
| $4\,0^{++}$ | | 3990(210)(180) | | $1\,3^{--}$ | 4200(45)(200) | 4330(260)(200) | |
| $1\,2^{++}$ | 2390(30)(120) | 2150(30)(100) | 2376(32) | $1\,0^{+-}$ | 4780(60)(230) | | |
| $2\,2^{++}$ | | 2880(100)(130) | 3300(50) | $1\,1^{+-}$ | 2980(30)(140) | 2670(65)(120) | 2944(42) |
| $1\,3^{++}$ | 3670(50)(180) | 3385(90)(150) | 3740(70) | $2\,1^{+-}$ | | | 3800(60) |
| $1\,4^{++}$ | | 3640(90)(160) | 3690(80) | $1\,2^{+-}$ | 4230(50)(200) | | 4240(80) |
| $1\,6^{++}$ | | 4360(260)(200) | | $1\,3^{+-}$ | 3600(40)(170) | 3270(90)(150) | 3530(80) |
| $1\,0^{-+}$ | 2560(35)(120) | 2250(60)(100) | 2561(40) | $2\,3^{+-}$ | | 3630(140)(160) | |
| $2\,0^{-+}$ | | 3370(150)(150) | 3540(80) | $1\,4^{+-}$ | | | 4380(80) |
| $1\,2^{-+}$ | 3040(40)(150) | 2780(50)(130) | 3070(60) | $1\,5^{+-}$ | | 4110(170)(190) | |
| $2\,2^{-+}$ | | 3480(140)(160) | 3970(70) | | | | |
| $1\,5^{-+}$ | | 3942(160)(180) | | | | | |
| $1\,1^{-+}$ | | | 4120(80) | | | | |
| $2\,1^{-+}$ | | | 4160(80) | | | | |
| $3\,1^{-+}$ | | | 4200(90) | | | | |

**Table 4.1** The glueball masses as provided in three lattice works.

These three lattice works do not explicitly provide the corresponding value of $T_c$, hence we need to find them via the following equations [170–172]:

$$T_c = 1.26(7) \cdot \Lambda_{MS} = 1.26(7) \cdot 0.614(2) \cdot r_0^{-1}, \quad (4.7)$$

$$T_c = 0.629(3) \cdot \sqrt{\sigma}. \quad (4.8)$$

Thus, we get the values reported in Table 4.2. The comparison between the GRG with the masses from Table 4.1 and the results from [163] is plotted in Fig. 4.1. Of all the quantities defined in Section 4.2, we consider only the pressure, since it is a key quantity in thermodynamics. Plots like the one in Fig. 4.1 can be done for the trace anomaly and the entropy, see Ref. [168]. It is evident that the pressure obtained using the glueball masses from Athenodorou and Teper [50] is in better agreement with the lattice data from Borsányi [163]. Additionally, we notice that, when approaching $T_c$, a slight disagreement is still present.

| LQCD papers | Number of glueballs | Lattice Parameter | $T_c$ (using Eqs. (4.7)-(4.8)) |
|---|---|---|---|
| Chen et.al [49] | 12 | $r_0^{-1} = 410(20)$ MeV | $317 \pm 23$ MeV |
| Meyer [48] | 22 | $\sqrt{\sigma} = 440(20)$ MeV | $277 \pm 13$ MeV |
| Athenodorou and Teper [50] | 20 | $r_0^{-1} = 418(5)$ MeV | $323 \pm 18$ MeV |

**Table 4.2** Parameters used for the evaluation of the glueball masses in three different works.

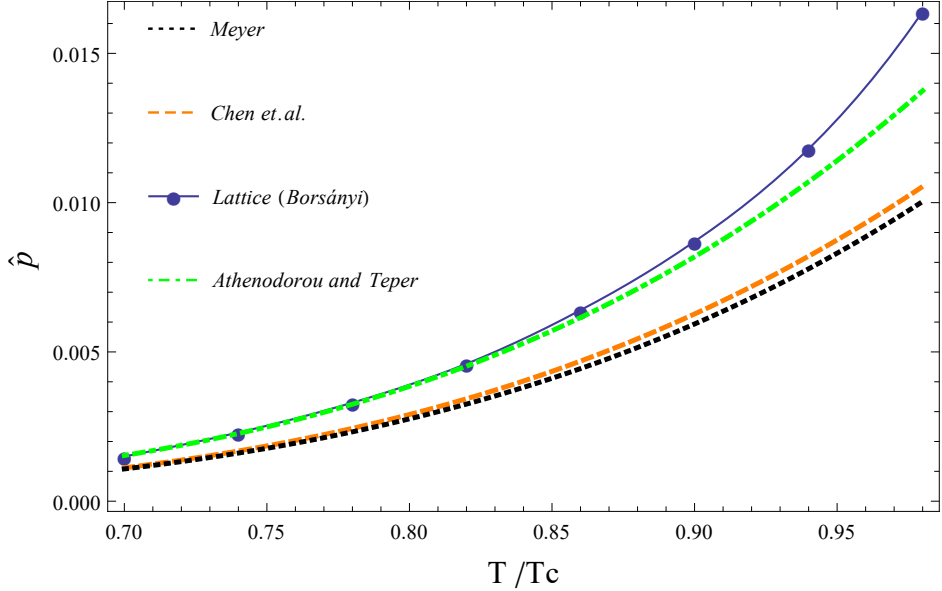

**Fig. 4.1** The GRG pressure, normalized and $T_c$ independent, as function of the ratio $T/T_c$, from three different lattice mass spectra ([48-50]). These are compared with the pressure evaluated in Ref. [163], which lattice data points are taken from the continuum limit (see Table 1 in Ref. [163]).

In the following, we calculate the contributions due to additional states and scattering.

## 4.4 First additional contribution: further excited states

The masses listed in Table 4.1 are the ones seen in lattice, but other glueball states are expected to exist. Indeed, the existence of an infinite tower of glueballs is expected [9], i.e. for any value of $J^{PC}$ allowed, $n$ could get any value from 1 to $M \to \infty$.

The idea of Regge trajectories states that light and conventional mesons can be grouped into three dimensional-$(n, J, M^2)$ planes, as already suggested in several works, e.g. [173, 174]. Therefore, for a given mass $M$, a radial quantum number $n$ and a total angular momentum $J$, we can write:

$$M^2(n, J) = a(n + J) + b. \tag{4.9}$$

This idea can be applied also for glueballs. Since we intend to provide an estimate of the role of higher excited states to the thermodynamic quantities obtained from the lattice spectrum, it will be enough to use a simple approach for the values of the masses of heavier glueballs [48].

Because of the shortage of information about heavy glueballs, together with the fact that the glueballs spectra provided by lattice are not complete, it is impossible at present to rigorously produce a spectrum for glueballs using Regge trajectories. Not all the glueballs belongs to the same Regge plane. Therefore, we should separate the glueballs into groups, according to relevant factors (e.g. $C$-parity) [48]. Still, in order to rigorously describe glueballs, we should have a wider set of ground state glueball masses provided by lattice (the $J^{--}$ sector has only two lattice masses found in [50]). Not only, as we can see from Tab. 4.1, are many ground states missing, but also other are controversial. As an example, we briefly comment on the particular case of the ground state of $1^{-+}$:

($i$) In Ref. [175], the authors report that the relativistic current describing the two-gluon glueball $1^{-+}$ vanishes, suggesting that this state does not exist within the relativistic framework.

($ii$) In Ref. [176], the authors use a relativistic equation for glueballs with two gluons, constructed from the QCD with massive gluons. In their work they also provide a mass for the state $1^{-+}$ (around 2.7 GeV).

From these ambiguities we understand that some of the possible $J^{PC}$ states could not exist. Thus, in our model we only consider the $J^{PC}$ provided by lattice [50], as well as the excitation for these states.

The next problem to solve is how to group the glueballs into Regge planes. We could in principle divide the glueballs for different $PC$ (not enough data to do a fit for the $PC = --$ sector), or divide them into odderon ($C = -$) and pomeron ($C = +$), or consider all the glueballs together in the same Regge plane. In Ref [168], we used this last method, together with a statistical fit. Thus, the following Regge trajectory:

$$M^2(n, J^{PC}) = a(n + J) + b_{J^{PC}}, \tag{4.10}$$

was used for the quantum numbers $J^{PC} = 0^{++}, 2^{++}, 0^{-+}, 2^{-+}, 1^{+-}$. These states are

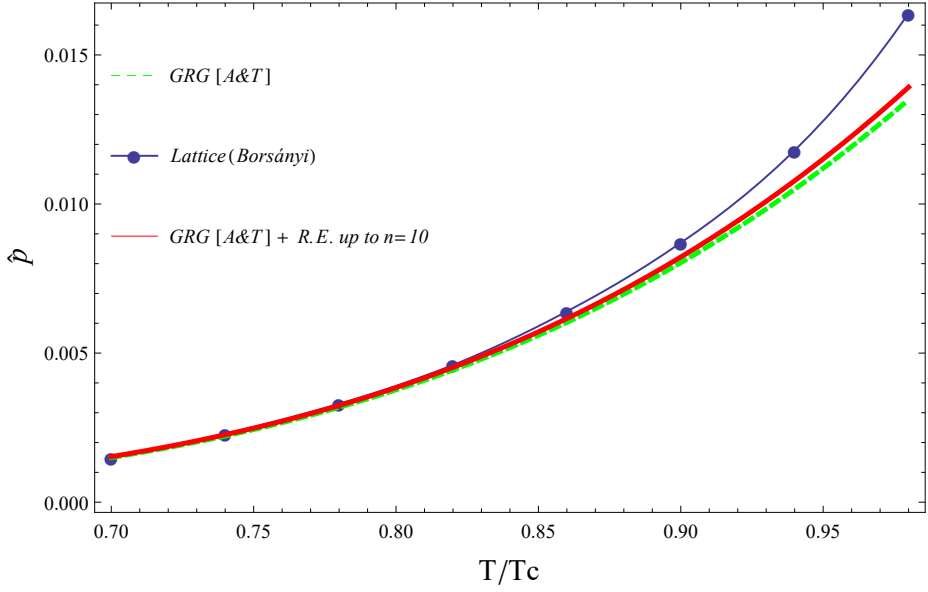

**Fig. 4.2** The pressure from the GRG model using the masses from Athenodorou and Teper [50] (green, dashed), and after including the excited states (red) up to n=10 for any value of $J^{PC}$ in that work, compared with the lattice results from [163].

present in all the three lattice spectra, therefore they are trustable. We then perform the fit:

$$\chi^2(a, b_{0^{++}}, b_{2^{++}}, b_{0^{-+}}, b_{2^{-+}}, b_{1^{+-}}) = \sum_{J^{PC}} \sum_{n=1}^{2} \left( \frac{M(n, J^{PC}) - M^{\text{lat}}(n, J^{PC})}{\delta M^{\text{lat}}(n, J^{PC})} \right)^2. \quad (4.11)$$

By minimizing $\chi^2$, we find $\chi^2_{\text{d.o.f.}} = 0.84$ and $a = 5.49 \pm 0.17\,\text{GeV}^2$.

In Figure 4.2 the effect of further excited glueballs on the GRG pressure is shown. The red curve, which leads to a negligible increment of the pressure, was obtained by using the masses obtained from the Regge trajectories: all the $J^{PC}$ present in [50], excited up to the 10-$th$ level of excitation ($n = 10$). Thus, the free part of the GRG model is largely dominated by the lightest glueballs contributions. For more details, see Ref. [168].

## 4.5 Second additional contribution: interaction between glueballs

In order to estimate the contribution of the interactions to the pressure, we consider the interaction between the two lightest glueballs: scalar ($0^{++}$) and tensor ($2^{++}$).

### 4.5.1 Formalism

In general, the total dimensionless pressure of any $J^{PC}$ state is formed by the sum of three contributions: the non interacting part $\hat{p}_{J^{PC}}^{\text{free}}$, the interacting part $\hat{p}_{J^{PC}J^{PC}}^{\text{int}}$ and, eventually, a bound state contribution $\hat{p}_B$ [177, 178]:

$$\hat{p}^{tot}_{J^{PC}} = \hat{p}^{\text{free}}_{J^{PC}} + \hat{p}^{\text{int}}_{J^{PC}J^{PC}} + \hat{p}_B \,. \tag{4.12}$$

The free part is given as (dimensionless version of Eq. (4.1)):

$$\hat{p}^{\text{free}}_{J^{PC}} = -\frac{(2J+1)}{T^3} \int_0^\infty \frac{k^2}{2\pi^2} \ln\left(1 - e^{-\frac{\sqrt{k^2+m^2}}{T}}\right) dk. \tag{4.13}$$

The interacting part can be better explained by considering two situations separately: one with a bound state and another without.

❏ Absence of bound state ($\hat{p}_B = 0$)

If from the scattering of two $J^{PC}$ particles no bound state forms, then

$$\hat{p}^{\text{int}}_{J^{PC}J^{PC}} = -\frac{1}{T^3} \sum_{\mathfrak{J}=0}^{2J} \sum_{\ell=0}^{\infty} \int_{2m_{J^{PC}}}^{\infty} dx (2\mathfrak{J}+1) \frac{2\ell+1}{\pi} \frac{d\delta_\ell^{J^{PC}J^{PC},\mathfrak{J}}(x^2)}{dx}$$

$$\times \int_k \ln\left(1 - e^{-\frac{\sqrt{k^2+x^2}}{T}}\right), \tag{4.14}$$

where

$$\int_k = \int \frac{d^3k}{(2\pi)^3}, \tag{4.15}$$

$\mathfrak{J} \in [J-J, J+J]$ is the total spin of the system and

$$\frac{d\delta_\ell^{J^{PC}J^{PC},\mathfrak{J}}(x^2)}{dx} \tag{4.16}$$

is the derivative of the phase shift describing the interaction between two $J^{PC}$ states. Note, if, for instance, $\frac{d\delta}{dx}$ changes sign in two regions, we should expect an overall predominance of the sign acquired in the low energy region, since, in the thermal integral, low energies dominate.

❏ Presence of bound state ($\hat{p}_B \neq 0$)

The contribution to the pressure of a bound state corresponds to an additional asymptotic state of the system. In the case of the scattering of two scalar glueballs, we know that a bound state (glueballonium) is expected to exist if the value of the parameter $\Lambda_G$ is below a critical value, denoted as $\Lambda_{G,crit}$ (see Section 2). Therefore, the pressure contribution due to the presence of a glueballonium is:

$$\hat{p}_B = -\frac{\theta(\Lambda_G - \Lambda_{G,crit})}{T^3} \int_k \ln\left(1 - e^{-\frac{\sqrt{k^2+m_B^2}}{T}}\right), \quad (4.17)$$

where the $\theta$ function takes into account that the glueballonium does not form above $\Lambda_{G,crit}$ (we recall that, in the on-shell unitarization, the mass of the bound state $m_B$ vary from $\sqrt{3}m_{0^{++}}$ to $2m_{0^{++}}$, depending on the value of $\Lambda_G$ (see Fig. 2.12)).
Alternatively, the bound state can be also regarded as part of the interaction between the two scalar states. In this optic, the whole interacting part is:

$$\hat{p}^{\text{int}}_{0^{++}0^{++}} + \hat{p}_B = -\frac{1}{T^3} \sum_{\ell=0}^{\infty} \int_{\sqrt{3}m_{0^{++}}}^{\infty} dx \frac{2\ell+1}{\pi} \frac{d\delta_\ell^{0^{++}0^{++}}(x^2)}{dx} \int_k \ln\left(1 - e^{-\frac{\sqrt{k^2+x^2}}{T}}\right), \quad (4.18)$$

where the term $\hat{p}_B$ is responsible for the integration of $x$ in the interval $(\sqrt{3}m_{0^{++}}, 2m_{0^{++}})$, and the phase shift $\delta_\ell^{0^{++}0^{++}}$ (extended below threshold) can be taken from the unitarized version of Eq. (2.46). Note, as already mentioned in Chapter 2, the formation of a bound state is eventually possible only for $\ell = 0$. Consequently, if we compare Eqs. (4.18) and (4.17), we get the following relation:

$$-\frac{\theta(\Lambda_G - \Lambda_{G,crit})}{T^3} \int_k \ln\left(1 - e^{-\frac{\sqrt{k^2+m_B^2}}{T}}\right) =$$
$$-\frac{1}{T^3} \sum_{\ell=0}^{\infty} \int_{\sqrt{3}m_{0^{++}}}^{2m_{0^{++}}} dx \frac{2\ell+1}{\pi} \frac{d\delta_\ell^{0^{++}0^{++}}(x^2)}{dx} \int_k \ln\left(1 - e^{-\frac{\sqrt{k^2+x^2}}{T}}\right). \quad (4.19)$$

### 4.5.2 Scalar interaction

After having introduced the formalism, we focus on analyzing the differences between having or not a bound state, taking the scalar glueball case as an example. As we know from Ref. [179], in the case of absence of a bound state, the interaction generates a positive pressure. If a bound state is present, one has

$$\hat{p}^{\text{int}}_{0^{++}0^{++}} + \hat{p}_B > 0, \quad (4.20)$$

where $\hat{p}^{\text{int}}_{0^{++}0^{++}} < 0$ and $\hat{p}_B > 0$. This fact is shown in Fig. 4.3, where, in the case of the $S$-wave, we see that the combined pressure of interaction above threshold and bound state is positive (due to the bound state contribution). Instead, the pressure generated by the interaction, in the $D$-wave, in which the bound state does not form, is positive.

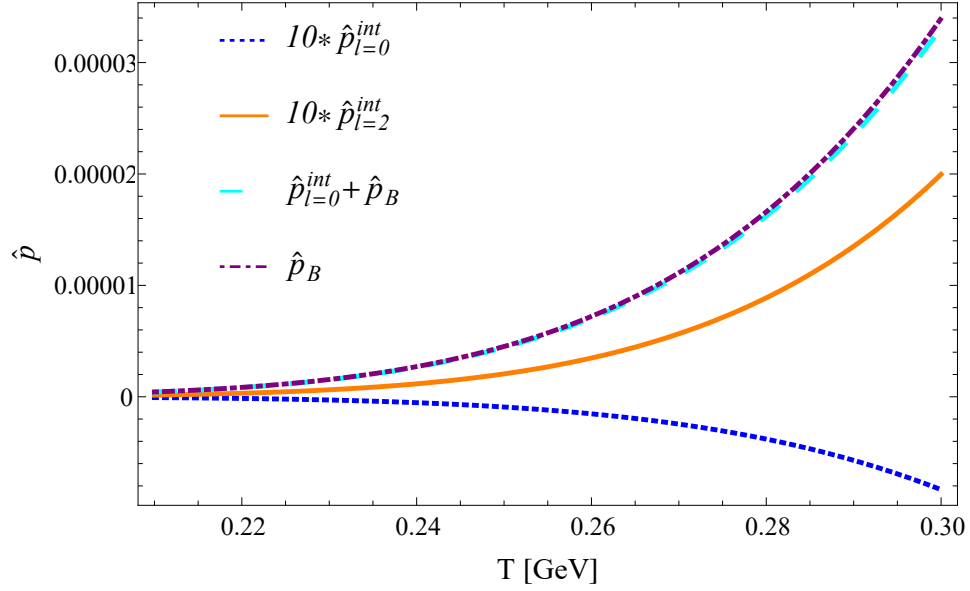

**Fig. 4.3** The pressure of the interaction due to the $\ell = 2$-wave is positive (orange, continuous line). The other curves are generated from the $\ell = 0$-wave: the negative pressure of the interaction (blue, dotted), the pressure of the bound state (violet, dotdashed) and the sum of the two (cyan, dashed). For graphical reasons, the pressures due to the interactions were multiplied by a factor 10.

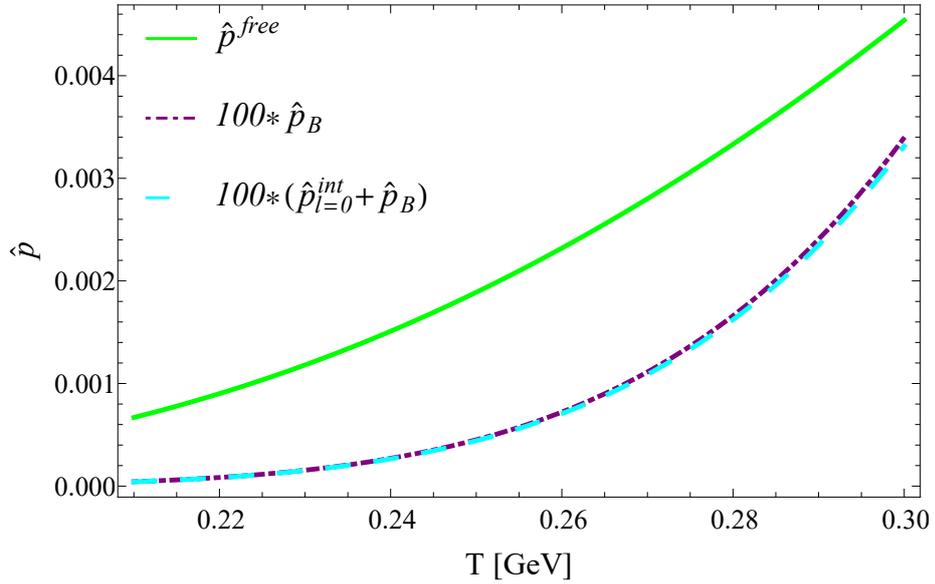

**Fig. 4.4** The pressure of the non-interacting scalar glueball (green, continuous), compared with the pressure of the glueballonium (violet, dotdashed) and the sum of the pressure of the bound state and of the interaction between two scalar glueballs (cyan, dashed). The last two quantities are two orders of magnitude lower than the free curve.

The predominance of the bound state contribution, compared to the interacting part alone, is evident in Fig. 4.3. Despite this fact, $\hat{p}^{\text{int}}_{0^{++}0^{++}} + \hat{p}_B$ is negligible when compared with the pressure obtained from the free scalar glueball (Fig. 4.4). This is an interesting fact, since the free term and the glueballonium term have substantially the same form (see Eqs. (4.13) and (4.17)), but differ of course in the value of the masses. The fact that the mass of the glueballonium is almost twice the mass of the scalar glueball, but its pressure is 100 times lower (Fig. 4.4), helps us also to understand why even the addition of a large number of non-interacting glueballs with increasing masses (Section 4.4) carries an almost negligible change in the GRG pressure.

### 4.5.3 Tensor interaction

The scalar glueball is the lightest state in pure YM. Therefore, if its interaction does not lead to any substantial change to the GRG pressure, why should we consider also the interaction of other, heavier, glueballs? The answer is the spin degeneracy.

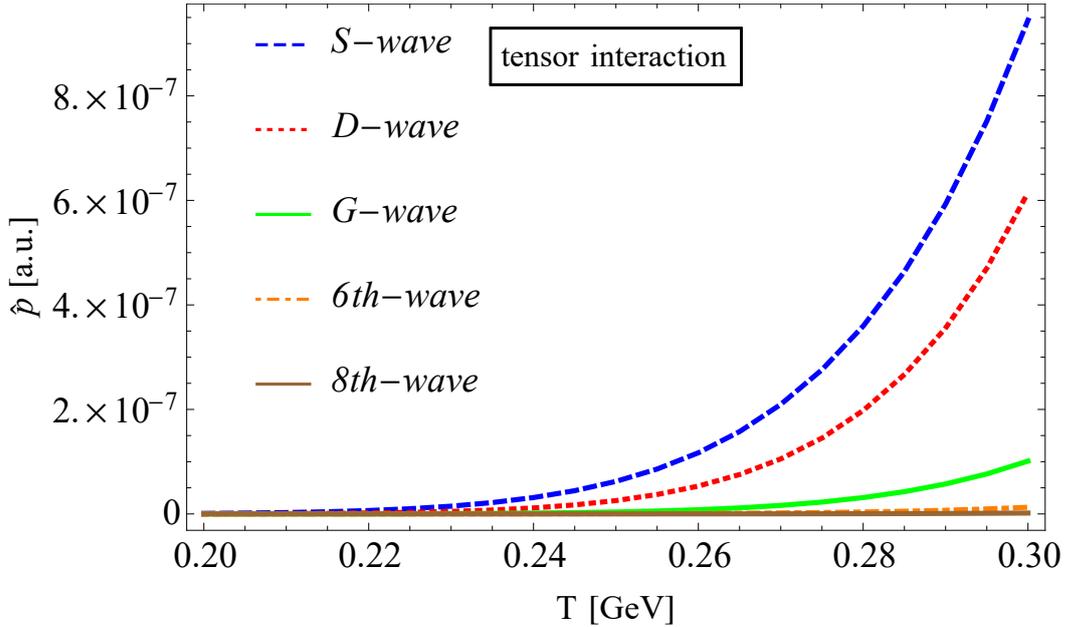

**Fig. 4.5** The contribution to the pressure, from the tensor interaction, for the waves with $\ell = 0, 2, 4, 6, 8$.

The second lightest glueball is, according to lattice, the tensor glueball, and the masses of $0^{++}$ and $2^{++}$ are close enough to compare between their corresponding pressures. The tensor glueball has a total spin $J = 2$, meaning five possible values for its projection along the $z$-axis, i.e. $M_J \in [-2, 2]$. If we consider the process

$$2^{++}2^{++} \to 2^{++}2^{++},$$

we can namely have $5^4 = 625$ possible processes, but only 25 are actually allowed. By applying an effective theory appropriate for the tensor glueball (see Section 4.6.2), we get the pressure shown in Fig. 4.5.

The $S$-wave, that gives the main contribution to the $2^{++}$ pressure, is lower than the pressure from the same wave in the $0^{++}$ scattering (cyan, dashed line in Fig. 4.4).

The contributions to the thermodynamic quantities, given from the interaction terms and the addition of a tower of glueballs, do not affect much the GRG model.

## 4.6 Amplitudes in the presence of isospin or nonzero spin

In this section, we show how to obtain the amplitudes for the $2^{++}$-$2^{++}$ scattering, with the same formalism which was previously used to distinguish the various channels of the $\pi\pi$ scattering. Indeed, the formalism for pion scattering, that involves isospin, is analogous to the one for tensor glueballs, that involves the spin ($J = 2$).

The scattering take the schematic form ($s$-channel):

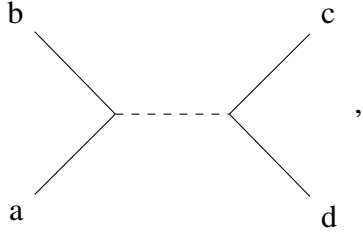

,

where, in Section 4.6.1, *a, b, c, d*=$\pi^i$, and, in Section 4.6.2, *a, b, c, d*=$G_2^i$.

### 4.6.1 The formalism for pions

We analyse the general aspects of pion scattering, following the formalism reported in [180–184]. The Lagrangian used carries the leading order term [183]:

$$\mathscr{L}^{(2)} = \frac{\mathfrak{f}_\pi^2}{4} \langle \partial_\mu U \partial^\mu U^\dagger \rangle, \qquad (4.21)$$

where $\mathfrak{f}_\pi$ is the pion decay constant, $U = exp(i\phi/\mathfrak{f}_\pi')$, and $\phi$ is the matrix of the pseudoscalar nonet. From this Lagrangian, for the case of pion-pion scattering ($a+b \to c+d$), the scattering amplitude can be written as a decomposition of the invariant amplitudes $A$, $B$ and $C$:

$$\langle c,d | \, \mathbb{T} \, | a,b \rangle = A\delta^{ab}\delta^{cd} + B\delta^{ac}\delta^{bd} + C\delta^{ad}\delta^{bc}, \qquad (4.22)$$

where $\mathbb{T}$ is the transition matrix, and

$$A = A(s,t,u) = \frac{s - m_\pi^2}{f_\pi^2}$$
$$B = B(t,u,s) = \frac{t - m_\pi^2}{f_\pi^2}$$
$$C = C(u,s,t) = \frac{u - m_\pi^2}{f_\pi^2}. \tag{4.23}$$

Each scattering amplitude can then be written in term of transition matrices. As a last step, we show how to calculate these factors. When we talk about pions - i.e. particles with isospin $I = 1$ -, we consider a triplet of particles, according to their third component of the isospin $I_z$:

$$|\pi, 1\rangle, \qquad |\pi, 0\rangle, \qquad |\pi, -1\rangle, \tag{4.24}$$

corresponding to:
$\pi^+$ ($|I, I_z\rangle = |1, 1\rangle$),
$\pi^0$ ($|I, I_z\rangle = |1, 0\rangle$) and
$\pi^-$ ($|I, I_z\rangle = |1, -1\rangle$)
respectively.

During their scattering, they produce an intermediate state $|I_{tot}, I_{z,tot}\rangle$, which then decay into the final state. This intermediate state carries the information of the initial particles and determine which will be the products of the decay. Since the intermediate state comes from two pions, we get that $I_{tot} \in [0, 2]$ and $I_{z,tot} \in [-I_{tot}, I_{tot}]$.

The total initial state $|I_{tot}, I_{z,tot}\rangle$ can then be written from the single initial states of the particles $a$ and $b$, $|I^a, I_z^a\rangle$ and $|I^b, I_z^b\rangle$, using the Clebsh-Gordan coefficients:

$$|I_{tot}, I_{z,tot}\rangle = \sum_{I_z^a, I_z^b} \langle I^a\, I_z^a\, I^b\, I_z^b | I_{tot}\, I_{z,tot}\rangle |I^a, I_z^a\rangle \otimes |I^b, I_z^b\rangle. \tag{4.25}$$

As an example, the state $|I_{tot} = 0, I_z = 0\rangle$ is formed by the total contribution of three terms:

$$|I_{tot} = 0, I_z = 0\rangle = \langle 111-1|00\rangle |\pi, 1\rangle \otimes |\pi, -1\rangle + \langle 1010|00\rangle |\pi, 0\rangle \otimes |\pi, 0\rangle +$$
$$\langle 1-111|00\rangle |\pi, -1\rangle \otimes |\pi, 1\rangle. \tag{4.26}$$

It is convenient to rewrite the pion states in terms of the states $|\pi_i\rangle$ ($i = 1, 2, 3$) [184]:

$$|\pi, \pm 1\rangle = \mp \frac{1}{\sqrt{2}}(|\pi_1\rangle \pm i |\pi_2\rangle), \qquad |\pi, 0\rangle = |\pi_3\rangle. \tag{4.27}$$

With Eqs. (4.27) and (4.22), we obtain an explicit form for the transition $I_{tot} = 0 \to I_{tot} = 0$:

$$\langle I_{tot} = 0, I_z = 0 | \mathbb{T} | I_{tot} = 0, I_z = 0 \rangle = 3A + B + C, \tag{4.28}$$

as also reported in [183]. We have now the tools to show one clarifying example. Let rewrite, in our new formalism:

$$\langle \pi_c, \pi_d | \mathbb{T} | \pi_a, \pi_b \rangle = A \delta^{ab} \delta^{cd} + B \delta^{ac} \delta^{bd} + C \delta^{ad} \delta^{bc}. \tag{4.29}$$

✐ $\pi^+ \pi^+ \to \pi^+ \pi^+$

The physical particle $\pi^+$ correspond to the state $|\pi, 1\rangle$. We now write the total final and initial states separately, together with their relation with the intermediate state (via the Clebsh-Gordan coefficients):

$$|\pi, 1\rangle \otimes |\pi, 1\rangle = \langle 1111|22\rangle \frac{1}{2}\bigg( (|\pi_1\rangle + i |\pi_2\rangle)(|\pi_1\rangle + i |\pi_2\rangle) \bigg)$$

$$= \frac{1}{2}\bigg( |\pi_1\rangle |\pi_1\rangle + i |\pi_1\rangle |\pi_2\rangle + i |\pi_2\rangle |\pi_1\rangle - |\pi_2\rangle |\pi_2\rangle \bigg) \tag{4.30}$$

and

$$\langle \pi, 1| \otimes \langle \pi, 1| = \langle 1111|22\rangle \frac{1}{2}\bigg( (\langle \pi_1| - i \langle \pi_2|)(\langle \pi_1| - i \langle \pi_2|) \bigg)$$

$$= \frac{1}{2}\bigg( \langle \pi_1| \langle \pi_1| - i \langle \pi_1| \langle \pi_2| - i \langle \pi_2| \langle \pi_1| - \langle \pi_2| \langle \pi_2| \bigg). \tag{4.31}$$

Then:

$$\bigg( \langle \pi, 1| \otimes \langle \pi, 1| \bigg) \mathbb{T} \bigg( |\pi, 1\rangle \otimes |\pi, 1\rangle \bigg) = \frac{1}{4}\bigg( \langle \pi_1| \langle \pi_1| \mathbb{T} |\pi_1\rangle |\pi_1\rangle -$$

$$\langle \pi_1| \langle \pi_1| \mathbb{T} |\pi_2\rangle |\pi_2\rangle + \langle \pi_2| \langle \pi_1| \mathbb{T} |\pi_1\rangle |\pi_2\rangle + \langle \pi_2| \langle \pi_1| \mathbb{T} |\pi_2\rangle |\pi_1\rangle + \langle \pi_1| \langle \pi_2| \mathbb{T} |\pi_1\rangle |\pi_2\rangle +$$

$$\langle \pi_1| \langle \pi_2| \mathbb{T} |\pi_2\rangle |\pi_1\rangle - \langle \pi_2| \langle \pi_2| \mathbb{T} |\pi_1\rangle |\pi_1\rangle + \langle \pi_2| \langle \pi_2| \mathbb{T} |\pi_2\rangle |\pi_2\rangle \bigg) =$$

$$\frac{1}{4}(A + B + C - A + C + B + B + C - A + A + B + C) = B + C. \tag{4.32}$$

### 4.6.2 The formalism for tensor glueball scattering

The formalism for the calculations of the amplitudes for the $2^{++}$ glueball scattering is analogous to the one provided for the case of pions in Section 4.6.1, with the difference that here we need to use the total spin $J$ instead of the isospin $I$. We again start from a Lagrangian which describes our system. Since we are in pure YM, we use an enlarged dilaton Lagrangian, whose form is the same of Eq. (2.25), with the addition of another part that takes into account the presence of a tensor glueball:

$$\mathscr{L}_{G,G_2} = \frac{1}{2}(\partial_\mu G)^2 + \frac{\alpha}{2} G^2 G_{2,\mu\nu} G_2^{\mu\nu} - \frac{1}{4}\frac{m_G^2}{\Lambda_G^2}\left(G^4 \ln\left|\frac{G}{\Lambda_G}\right| - \frac{G^4}{4}\right) + \mathscr{L}_{G_2}, \quad (4.33)$$

where $\mathscr{L}_{G_2}$ is [185, 186]:

$$\begin{aligned}\mathscr{L}_{G_2} =& \frac{1}{8}\partial_\mu(G_{\alpha\beta} + G_{\beta\alpha})\partial^\mu(G^{\alpha\beta} + G^{\beta\alpha}) - \frac{1}{4}\partial_\alpha(G^{\alpha\beta} + G^{\beta\alpha})\partial^\mu(G_{\mu\beta} + G_{\beta\mu}) \\ & - \frac{1}{2}\partial_\mu G^\rho_\rho \partial^\mu G^\eta_\eta + \frac{1}{2}\partial_\mu(G^{\mu\nu} + G^{\nu\mu})\partial_\nu G^\rho_\rho - \frac{m_{G_2}^2}{8}(G_{\mu\nu} + G_{\nu\mu})(G^{\mu\nu} + G^{\nu\mu}) \\ & + \frac{m_{G_2}^2}{2}(G^\mu_\mu)^2. \end{aligned} \quad (4.34)$$

As before, the scattering amplitude can be written as a decomposition of the invariant amplitudes $A$, $B$ and $C$ [*]:

$$\langle c,d| \, \mathbb{T} \, |a,b\rangle = A\delta^{ab}\delta^{cd} + B\delta^{ac}\delta^{bd} + C\delta^{ad}\delta^{bc}, \quad (4.35)$$

where $\mathbb{T}$ is the transition matrix, and

$$\begin{aligned} A = A(s,t,u) &= \frac{-4(\alpha\Lambda_G)^2}{s - m_G^2} \\ B = B(t,u,s) &= \frac{-4(\alpha\Lambda_G)^2}{t - m_G^2} \\ C = C(u,s,t) &= \frac{-4(\alpha\Lambda_G)^2}{u - m_G^2}. \end{aligned} \quad (4.36)$$

We refer to the following diagram:

---

[*]Here careful is needed: the spin description is slightly different from the isospin one (see Ref.[185]). The formula provided in Eq. 4.35 is valid under the condition that the momenta of the particles are small compared to their masses.

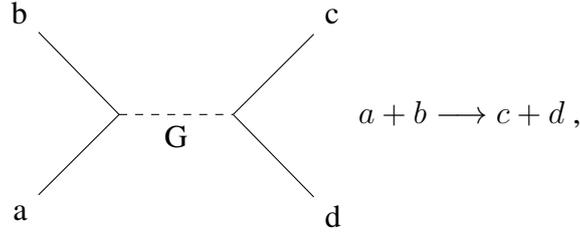

$$a + b \longrightarrow c + d \, ,$$

where the exchanged particle is the scalar glueball and $a, b, c, d$ refer to the third-component of the spin $J$ of the glueballs (incoming and outgoing). Each tensor glueball carries a spin number $J = 2$, namely $J^a = J^b = J^c = J^d = 2$. Thus, the third component can take the values from -2 to +2, i.e. $m_J^a, m_J^b, m_J^c, m_J^d \in [-2, +2]$.

It is convenient to write, for the incoming particles, a global state which contains both the incoming particles in a compact form. The total initial state $\left|J^{ab}, m_J^{ab}\right\rangle$ can then be written from the single initial states $\left|J^a, m_J^a\right\rangle$ and $\left|J^b, m_J^b\right\rangle$, using the Clebsh-Gordan coefficients:

$$\left|J^{ab}, m_J^{ab}\right\rangle = \sum_{m_J^a, m_J^b} \left\langle J^a \, m_J^a \, J^b \, m_J^b \middle| J^{ab} \, m_J^{ab}\right\rangle \left|J^a, m_J^a\right\rangle \otimes \left|J^b, m_J^b\right\rangle \, , \qquad (4.37)$$

where $J^{ab} \in [0, 4]$ and $m_J^{ab} \in [-J^{ab}, +J^{ab}]$.

The same procedure can be done for the outgoing states:

$$\left\langle J^{cd}, m_J^{cd}\right| = \sum_{m_J^c, m_J^d} \left\langle J^{cd} \, m_J^{cd} \middle| J^c \, m_J^c \, J^d \, m_J^d\right\rangle \left\langle J^c, m_J^c\right| \otimes \left\langle J^d, m_J^d\right| \, . \qquad (4.38)$$

The total number of amplitudes that could be in principle calculated is given by considering the combination of all the possible $m_J^\omega$, $\omega = a, b, c, d$, that is $5^4 = 625$. As it happens frequently, many of the possible contribution are zero: in our case, only 25 contributions are nonzero, i.e. those amplitudes $\left\langle J^{cd}, m_J^{cd}\right| \mathbb{T} \left|J^{ab}, m_J^{ab}\right\rangle$, for which $J^{cd} = J^{ab}$ and $m_J^{cd} = m_J^{ab}$. As it was done for the case of the pions, we introduce a new basis $|\hat{\imath}\rangle$, $\hat{\imath} \equiv (\mathbb{1}, \mathbb{2}, \mathbb{3}, \mathbb{4}, \mathbb{5})$ as follows:

$$|2, \pm 1\rangle = \mp\sqrt{\frac{1}{2}}\Big(|\mathbb{1}\rangle \mp i\,|\mathbb{2}\rangle\Big) \qquad |2, \pm 2\rangle = -\sqrt{\frac{1}{2}}\Big(|\mathbb{3}\rangle \mp i\,|\mathbb{4}\rangle\Big) \qquad |2, 0\rangle = |\mathbb{5}\rangle \, . \tag{4.39}$$

Using the equation above, we obtain

$$\left|J^{ab} = 0, m_J^{ab} = 0\right\rangle = \frac{1}{\sqrt{5}}\Big(|\mathbb{1}\rangle \otimes |\mathbb{1}\rangle + |\mathbb{2}\rangle \otimes |\mathbb{2}\rangle + |\mathbb{3}\rangle \otimes |\mathbb{3}\rangle + |\mathbb{4}\rangle \otimes |\mathbb{4}\rangle + |\mathbb{5}\rangle \otimes |\mathbb{5}\rangle\Big), \qquad (4.40)$$

which allows us to have a relation analogous to Eq. (4.28) for the case of pions:

$$\left\langle J^{ab}=0, m_j^{ab}=0 \middle| \mathbb{T} \middle| J^{ab}=0, m_j^{ab}=0 \right\rangle = 5A + B + C. \tag{4.41}$$

We denote $\mathbb{T}^q$ -$q = 0, 1, 2, 3, 4$ as the amplitude where the total incoming (and outgoing) spin is $q = J^{cd} = J^{ab}$. We remind that:

$$|\mathring{\mathbb{i}}_a\rangle \otimes |\mathring{\mathbb{i}}_b\rangle \equiv |\mathring{\mathbb{i}}_a\rangle |\mathring{\mathbb{i}}_b\rangle. \tag{4.42}$$

Thus:

$$\mathbb{T}^q = \sum_{\mathring{\mathbb{i}}_a, \mathring{\mathbb{i}}_b, \mathring{\mathbb{i}}_c, \mathring{\mathbb{i}}_d} \rho_{abcd} \langle \mathring{\mathbb{i}}_c | \langle \mathring{\mathbb{i}}_d | T | \mathring{\mathbb{i}}_a \rangle | \mathring{\mathbb{i}}_b \rangle \qquad \mathring{\mathbb{i}}_a, \mathring{\mathbb{i}}_b, \mathring{\mathbb{i}}_c, \mathring{\mathbb{i}}_d \equiv (\mathbb{1}, \mathbb{2}, \mathbb{3}, \mathbb{4}, \mathbb{5}), \tag{4.43}$$

where $\rho_{abcd} = \left\langle 2\, m_J^a\, 2\, m_J^b \middle| J^{ab}\, m_J^{ab} \right\rangle \cdot \left\langle 2\, m_J^c\, 2\, m_J^d \middle| J^{cd}\, m_J^{cd} \right\rangle$ is the product of the Clebsh-Gordan coefficient of the initial and final state.

Finally, the amplitudes that describe the tensor glueball scattering are (for the explicit calculation, see Section 4.6.3):

$$\boxed{\mathbb{T}^4 = \mathbb{T}^2 = B + C\,, \qquad \mathbb{T}^3 = \mathbb{T}^1 = B - C\,, \qquad \mathbb{T}^0 = 5A + B + C}.$$

These amplitudes, when used in Eq. (4.14), give a pressure contribution which is negligible compared to the free one (see Fig. 4.5).

### 4.6.3 Explicit calculations for tensor glueball scattering

In this section, we provide some examples of the explicit calculation of the amplitudes for the two tensor glueball scattering, as done in the case of pions (see Section 4.6.1). Since the full calculation of all the amplitudes would be long and tedious, we will focus on two illustrative examples of $\left\langle J^{cd}, m_J^{cd} \middle| \mathbb{T} \middle| J^{ab}, m_J^{ab} \right\rangle$:

- $\langle 0, 0 | \mathbb{T} | 0, 0 \rangle$ and $\langle 4, 3 | \mathbb{T} | 4, 3 \rangle$.

In each case, we first build the needed initial and final states and then we calculate the amplitudes, using Eq. (4.22).

✎ $\boxed{\langle 0, 0 | \mathbb{T} | 0, 0 \rangle}$

$|0, 0\rangle =$
$|2, 2\rangle \otimes |2, -2\rangle + |2, 1\rangle \otimes |2, -1\rangle + |2, 0\rangle \otimes |2, 0\rangle + |2, -1\rangle \otimes |2, 1\rangle + |2, -2\rangle \otimes |2, 2\rangle =$
$\langle 22 2 -2 | 00 \rangle \frac{1}{2} \Big( |\mathbb{3}\rangle - i\, |\mathbb{4}\rangle \Big) \Big( |\mathbb{3}\rangle + i\, |\mathbb{4}\rangle \Big) - \langle 21 2 -1 | 00 \rangle \frac{1}{2} \Big( |\mathbb{1}\rangle - i\, |\mathbb{2}\rangle \Big)$

$$\left(|\mathbb{1}\rangle + i\,|\mathbb{2}\rangle\right) + \langle 2020|00\rangle\,|\mathbb{5}\rangle\,|\mathbb{5}\rangle - \langle 2-121|00\rangle\,\frac{1}{2}\left(|\mathbb{1}\rangle + i\,|\mathbb{2}\rangle\right)\left(|\mathbb{1}\rangle - i\,|\mathbb{2}\rangle\right) +$$
$$\langle 2-222|00\rangle\,\frac{1}{2}\left(|\mathbb{3}\rangle + i\,|\mathbb{4}\rangle\right)\left(|\mathbb{3}\rangle - i\,|\mathbb{4}\rangle\right) =$$
$$\frac{1}{\sqrt{5}}\left[\frac{1}{2}\left(|\mathbb{3}\rangle\,|\mathbb{3}\rangle + i\,|\mathbb{3}\rangle\,|\mathbb{4}\rangle - i\,|\mathbb{4}\rangle\,|\mathbb{3}\rangle + |\mathbb{4}\rangle\,|\mathbb{4}\rangle + |\mathbb{1}\rangle\,|\mathbb{1}\rangle + i\,|\mathbb{1}\rangle\,|\mathbb{2}\rangle - i\,|\mathbb{2}\rangle\,|\mathbb{1}\rangle + |\mathbb{2}\rangle\,|\mathbb{2}\rangle\right.\right.$$
$$|\mathbb{3}\rangle\,|\mathbb{3}\rangle - i\,|\mathbb{3}\rangle\,|\mathbb{4}\rangle + i\,|\mathbb{4}\rangle\,|\mathbb{3}\rangle + |\mathbb{4}\rangle\,|\mathbb{4}\rangle + |\mathbb{1}\rangle\,|\mathbb{1}\rangle - i\,|\mathbb{1}\rangle\,|\mathbb{2}\rangle + i\,|\mathbb{2}\rangle\,|\mathbb{1}\rangle + |\mathbb{2}\rangle\,|\mathbb{2}\rangle\Big)$$
$$\left. + |\mathbb{5}\rangle\,|\mathbb{5}\rangle\right] = \frac{1}{\sqrt{5}}\left(|\mathbb{1}\rangle \otimes |\mathbb{1}\rangle + |\mathbb{2}\rangle \otimes |\mathbb{2}\rangle + |\mathbb{3}\rangle \otimes |\mathbb{3}\rangle + |\mathbb{4}\rangle \otimes |\mathbb{4}\rangle + |\mathbb{5}\rangle \otimes |\mathbb{5}\rangle\right),$$
(4.44)

as from Eq. (4.40). Analogously,

$$\langle 0,0| =$$
$$\langle 2,2| \otimes \langle 2,-2| + \langle 2,1| \otimes \langle 2,-1| + \langle 2,0| \otimes \langle 2,0| + \langle 2,-1| \otimes \langle 2,1| + \langle 2,-2| \otimes \langle 2,2| =$$
$$\langle 222-2|00\rangle\,\frac{1}{2}\left(\langle\mathbb{3}| + i\,\langle\mathbb{4}|\right)\left(\langle\mathbb{3}| - i\,\langle\mathbb{4}|\right) - \langle 212-1|00\rangle\,\frac{1}{2}\left(\langle\mathbb{1}| + i\,\langle\mathbb{2}|\right)$$
$$\left(\langle\mathbb{1}| - i\,\langle\mathbb{2}|\right) + \langle 2020|00\rangle\,\langle\mathbb{5}|\,\langle\mathbb{5}| - \langle 2-121|00\rangle\,\frac{1}{2}\left(\langle\mathbb{1}| - i\,\langle\mathbb{2}|\right)\left(\langle\mathbb{1}| + i\,\langle\mathbb{2}|\right) +$$
$$\langle 2-222|00\rangle\,\frac{1}{2}\left(\langle\mathbb{3}| - i\,\langle\mathbb{4}|\right)\left(\langle\mathbb{3}| + i\,\langle\mathbb{4}|\right) =$$
$$\frac{1}{\sqrt{5}}\left(\langle\mathbb{1}| \otimes \langle\mathbb{1}| + \langle\mathbb{2}| \otimes \langle\mathbb{2}| + \langle\mathbb{3}| \otimes \langle\mathbb{3}| + \langle\mathbb{4}| \otimes \langle\mathbb{4}| + \langle\mathbb{5}| \otimes \langle\mathbb{5}|\right). \qquad (4.45)$$

Finally,

$$\langle 0,0|\,\mathbb{T}\,|0,0\rangle =$$
$$\frac{1}{5}\Big(\langle\mathbb{1}|\,\langle\mathbb{1}| + \langle\mathbb{2}|\,\langle\mathbb{2}| + \langle\mathbb{3}|\,\langle\mathbb{3}| + \langle\mathbb{4}|\,\langle\mathbb{4}| + \langle\mathbb{5}|\,\langle\mathbb{5}|\Big)\mathbb{T}\Big(|\mathbb{1}\rangle\,|\mathbb{1}\rangle + |\mathbb{2}\rangle\,|\mathbb{2}\rangle + |\mathbb{3}\rangle\,|\mathbb{3}\rangle + |\mathbb{4}\rangle\,|\mathbb{4}\rangle + |\mathbb{5}\rangle\,|\mathbb{5}\rangle\Big)$$
$$= \frac{1}{5}\Big(\langle\mathbb{1},\mathbb{1}|\mathbb{T}|\mathbb{1},\mathbb{1}\rangle + \langle\mathbb{1},\mathbb{1}|\mathbb{T}|\mathbb{2},\mathbb{2}\rangle + \langle\mathbb{1},\mathbb{1}|\mathbb{T}|\mathbb{3},\mathbb{3}\rangle + \langle\mathbb{1},\mathbb{1}|\mathbb{T}|\mathbb{4},\mathbb{4}\rangle + \langle\mathbb{1},\mathbb{1}|\mathbb{T}|\mathbb{5},\mathbb{5}\rangle +$$
$$\langle\mathbb{2},\mathbb{2}|\mathbb{T}|\mathbb{1},\mathbb{1}\rangle + \langle\mathbb{2},\mathbb{2}|\mathbb{T}|\mathbb{2},\mathbb{2}\rangle + ...\Big) = \text{upon using Eq. (4.35)} = 5A + B + C.$$
(4.46)

✎ $\boxed{\langle 4,3|\,\mathbb{T}\,|4,3\rangle}$

$|4,3\rangle =$

$$|2,2\rangle \otimes |2,1\rangle + |2,1\rangle \otimes |2,2\rangle =$$
$$\langle 2221|43\rangle \frac{1}{2}\Big(|3\rangle - i|4\rangle\Big)\Big(|1\rangle - i|2\rangle\Big) - \langle 2122|43\rangle \frac{1}{2}\Big(|1\rangle - i|2\rangle\Big)\Big(|3\rangle - i|4\rangle\Big) = ... =$$
$$\frac{1}{2\sqrt{2}}\Big(|3\rangle|1\rangle - i|3\rangle|2\rangle - i|4\rangle|1\rangle - |4\rangle|2\rangle + |1\rangle|3\rangle - i|2\rangle|3\rangle - i|1\rangle|4\rangle - |2\rangle|4\rangle\Big).$$
(4.47)

$$\langle 4,3| =$$
$$\langle 2,2| \otimes \langle 2,1| + \langle 2,1| \otimes \langle 2,2| =$$
$$\langle 2221|43\rangle \frac{1}{2}\Big(\langle 3| + i\langle 4|\Big)\Big(\langle 1| + i\langle 2|\Big) - \langle 2122|43\rangle \frac{1}{2}\Big(\langle 1| + i\langle 2|\Big)\Big(\langle 3| + i\langle 4|\Big) = ... =$$
$$\frac{1}{2\sqrt{2}}\Big(\langle 3|\langle 1| + i\langle 3|\langle 2| + i\langle 4|\langle 1| - \langle 4|\langle 2| + \langle 1|\langle 3| + i\langle 2|\langle 3| + i\langle 1|\langle 4| - \langle 2|\langle 4|\Big).$$
(4.48)

Finally,

$$\langle 4,3|\mathbb{T}|4,3\rangle =$$
$$\frac{1}{8}\Big(\langle 3|\langle 1| + i\langle 3|\langle 2| + i\langle 4|\langle 1| - \langle 4|\langle 2| + \langle 1|\langle 3| + i\langle 2|\langle 3| + i\langle 1|\langle 4| - \langle 2|\langle 4|\Big)\mathbb{T}$$
$$\Big(|3\rangle|1\rangle - i|3\rangle|2\rangle - i|4\rangle|1\rangle - |4\rangle|2\rangle + |1\rangle|3\rangle - i|2\rangle|3\rangle - i|1\rangle|4\rangle - |2\rangle|4\rangle\Big)$$
$$\frac{1}{8}\Big(\langle 3,1|\mathbb{T}|3,1\rangle + \langle 3,1|\mathbb{T}|1,3\rangle + \langle 3,2|\mathbb{T}|3,2\rangle + \langle 3,2|\mathbb{T}|2,3\rangle + \langle 4,1|\mathbb{T}|1,4\rangle +$$
$$\langle 4,1|\mathbb{T}|4,1\rangle + \langle 4,2|\mathbb{T}|4,2\rangle + \langle 4,2|\mathbb{T}|2,4\rangle + ...\Big) = \text{upon using Eq. (4.35)} = B + C.$$
(4.49)

The same result can be also obtained with e.g. $\langle 4,-3|\mathbb{T}|4,-3\rangle$.

For the other cases, one gets:

- The 9 amplitudes $\mathbb{T}^4 = \langle 4, m_J^{cd}|\mathbb{T}|4, m_J^{ab}\rangle$ and the 5 amplitudes $\mathbb{T}^2 = \langle 2, m_J^{cd}|\mathbb{T}|2, m_J^{ab}\rangle$ equal $B + C$.

- The 7 amplitudes $\mathbb{T}^3 = \langle 3, m_J^{cd}|\mathbb{T}|3, m_J^{ab}\rangle$ and the 3 amplitudes $\mathbb{T}^1 = \langle 1, m_J^{cd}|\mathbb{T}|1, m_J^{ab}\rangle$ equal $B - C$.

- The single amplitude $\mathbb{T}^0 = \langle 0,0|\mathbb{T}|0,0\rangle$ equal $5A + B + C$.

## 4.7 Short summary

The HRG is a well defined model used to study thermodynamic properties in the low temperature regime of QCD. We used an analogous model for YM, the Glueball Resonance Gas model. In particular, focusing on the pressure, we highlighted that the overall agreement below $T_c$ is good when the glueball masses of Athenodorou and Teper [50] are used, but there is still a discrepancy.

The addition of further excited states not present in the lattice spectra and of the interaction between glueballs provides a small contribution, that does not close the gap between the two methods.

# CHAPTER 5

# Summary and conclusions

In this work, we have explored the following subjects:

- Scattering of two scalar glueballs, within the dilaton potential.
- Decay of conventional meson using the covariant helicity formalism.
- Thermodynamic properties of the YM sector of QCD.
- Scattering of two tensor glueballs.

The partial wave analysis (PWA) was applied in all the studied above. We present below summaries and main results chapter by chapter.

❖ *CHAPTER II*

*Short summary*

After a short introduction, we analyzed the scattering between two glueballs. Since perform this study in pure Yang-Mills (YM), we used the dilaton Lagrangian, which contains a single dimentionful parameter $\Lambda_G$ and a single dilaton scalar field $G$, and has one interesting characteristic that makes it suitable for our case: it is an effective realization of pure YM, i.e. it does not contain quarks, that it reproduces the trace anomaly.

We studied the scattering at tree-level, in which the Levinson's theorem is not fulfilled, showing that a unitarization procedure is required. Upon using a twice-subtracted self-energy loop function, we found that a bound state can form, in the $S$-wave, if the value of $\Lambda_G$ is small enough (thus, if the attraction is large enough). The border between the formation or not of the bound state (glueballonium) is fixed by the parameter $\Lambda_{G,crit}$, which depends on the glueball mass and on the cutoff applied in the loop function.

Some of the results obtained here, e.g. scattering length, can be evaluated in the Lattice QCD, thus allowing for a possible comparison. Moreover, the glueballonium could be eventually searched in experiments, like PANDA [111], BesIII [100–103], Belle II [104–106], LHCb [107–109] and TOTEM [110].

*Main outcomes*

(i) The scattering of two scalar glueballs cannot be properly described at tree-level. Thus, a unitarization is required.

(ii) A bound state can form, in the $S$-wave, below a critical value of the scale (for $m_G = 1.65$ GeV, with the twice-subtracted unitarization, one gets $\Lambda_{G,crit} = 0.49$ GeV). Note, in this work we used the intermediate value $\Lambda_G \approx 0.35$ GeV (see discussion in Section 2.3.1). Thus, our results suggest that a glueballonium might indeed exist.

(iii) A cutoff, encoded in the loop function, numerically affects the value of $\Lambda_{G,crit}$ and the mass of the glueballonium, yet the qualitative picture remains similar.

(iv) Future comparison, with LQCD calculation is possible. An experimental search is also possible, but not easy, because of mixing with $\bar{q}q$ states and decay of the glueball.

❖ **CHAPTER III**

*Short summary*

We studied the decays of axial-vector and pseudovector mesons into a vector and a pseudoscalar particle. Using the covariant helicity formalism for partial wave analysis, a connection between the Feynman, the helicity and the $\ell S$-coupling amplitude was established. The importance of the last of these amplitudes lies in the fact that, for a given decay, the ratio between two $\ell S$ amplitudes corresponds to the ratio between two waves for the same decay. We demonstrated that the derivative interaction becomes relevant only for $J^{PC} = 1^{+-}$ resonances.
We also estimated the iso-singlet mixing angle in the $J^{PC} = 1^{++}$ sector, finding a good agreement with the experimental value [108].
The decays were evaluated in the rest frame of the decaying particle, thus the two products are emitted along opposite directions.

*Main outcomes*

(i) We reproduced the ratio between the $D$- and the $S$-wave for the decays of $a_1(1260) \to \rho\pi$ and $b_1(1235) \to \omega\pi$, in good agreement with PDG.

(ii) We found that for the decay of $a_1(1260)$, the derivative interaction is not relevant, while for $b_1(1235)$, the derivative interaction plays a central role. This is interesting, since all the parameters of the decay are similar and the only main difference is the charge conjugation of the decaying particles.

(iii) Our extracted mixing angle between the isosinglet states:
- for the axial-vector sector, agrees with the experiments [1]
- for the pseudovector sector, disagrees with BesIII (whose results are however very sensitive on the detailed assumptions [187]).

(iv) We have shown that the results for the amplitudes are equivalent, apart from a nu-

merical factor, if we study the case in which the emitted particles move along an arbitrary direction.

### ❖ CHAPTER IV

*Short summary*

We studied the thermodynamic properties of a gas of glueballs, below the critical temperature $T_c$. This was done using a Glueball Resonance Gas model, which is a version of the Hadron Resonance Gas model applied to the YM sector of QCD.

In Ref. [167] the results provided by the GRG using the masses from Chen [49] were not sufficient to reproduce the lattice thermodynamic data (like the pressure), suggesting the need of an additional Hagedorn contribution. Here, we have shown that the most recent lattice spectrum of Ref. [50] can almost fill the gap without the introduction of an Hagedorn term.

In this work, we considered also other contributions: (i) heavier, non interacting, excited glueball states, not found yet in lattice; (ii) the effect of the interaction between two scalar and two tensor glueballs. These contributions turned out to be negligible, thus we conclude that the GRG with free light glueballs well describes data up to almost $T_c$, where however a small gap is still visible.

Finally, the evaluation of the scattering of two tensor glueballs was necessary to study its role in thermodynamics. To this end, we used a formalism based on what was previously done in the 1970's in the case of pion-pion scattering.

*Main outcomes*

(i) A free gas of glueballs dominates the pressure of YM in the confined phase.

(ii) Lattice data are well described by the GRG model with the most recent lattice glueball spectrum, for which $T_c = 323 \pm 18$ MeV.

(iii) The contribution of heavier free glueballs is negligible.

(iv) The contribution of the interaction between scalar and tensor glueballs is also negligible.

(v) A formalism for the description of the scattering between two tensor glueballs was presented.

In conclusion, the results obtained in this work can be tested in future LQCD and experiments. Additionally, all the methods presented can be used in future for studies that require an analogous formalism (as other scattering or decay processes).

# Appendices

9292

**Appendix A**

# Brief summary about complex analysis

Human history benefited by the idea of studying Nature using mathematical tools. Already in the ancient Greece, philosophers like Thales of Miletus understood the importance of mathematics ($\mu\alpha\theta\eta\mu\alpha$). In the following centuries, the methods and the power of calculus developed continuously not only for the pure knowledge itself, but also for the utility of its applications to human life, from medicine to economics and to all natural sciences. In many areas of physics, it is impossible to work without the help of complex analysis, which provides a tool for studying Quantum Mechanics and Quantum Field Theory.

## A.1 Analytic functions

Firstly, a complex number $z \in \mathbb{C}$ is defined as $z = a + ib$, where $a, b \in \mathbb{R}$ and $i^2 = -1$. Thus, $z$ can be represented by a point in a 2-dimensional space spanned by two real axes, called $Argand$ diagram. This allows us to define functions of a complex variable as $w(z) = u(a,b) + iv(a,b)$, which can be separated into a real (Re) and an imaginary (Im) part. Since $u(a,b)$ and $v(a,b)$ are purely real, we may write $\operatorname{Re} w(z) = u(a,b)$ and $\operatorname{Im} w(z) = v(a,b)$.

Once complex functions are introduced, the next step is to differentiate them. An infinitesimal transformation of a vector, that starts in any point of the Argand diagram, can rotate and stretch it. If under the effect of the same transformation, all the infinitesimal vectors with the same initial point are equally rotated and stretched, this transformation is told to be *analytic*. An important theorem in the complex analysis is the Cauchy's integral theorem. This theorem states that, if $w(z)$ is analytic at each point of a simply connected region of the complex plane, then:

$$\oint_{C_0} w(z)dz = 0, \tag{A.1}$$

where $C_0$ is a closed contour within that region. This theorem is valid under the Cauchy-Riemann conditions:

$$\frac{\partial u}{\partial a} = \frac{\partial v}{\partial b}, \qquad \frac{\partial v}{\partial a} = -\frac{\partial u}{\partial b}. \tag{A.2}$$

We now call $\mathfrak{C}$ the area enclosed bounded by the closed contour $C_0$.

A problem for the calculation of integrals is represented by the non-analiticity of points (poles and essential singularities) or curves (branch cuts). Consider now a point $P \in \mathfrak{C}$ that is non-analytic. Remind that the calculation is always possible in a closed and analytic region. Fortunately in this example (and in most of the relevant calculations), this can be done easily by excluding from the region $\mathfrak{C}$ an infinitesimal area nearby the poles or the cuts ($P$ in this case). We call $I_P$ a small neighbourhood of $P$. Thus the region $\mathfrak{C}\backslash\{P\}$ is analytic. and the integral within that region is 0:

$$\oint_{C_0} w(z)dz + \oint_{C_1} w(z)dz = 0, \tag{A.3}$$

where $C_1$ is the contour around $I_P$. Note, the integrals over $C_0$ and $C_1$ must be evaluated over different direction of the path, in our case $C_0$ is traversed in the counterclockwise direction, and $C_1$ in the clockwise direction. Then:

$$\oint_{C_0} w(z)dz = -\oint_{C_1} w(z)dz = 2\pi i \, \text{Res}\{w(z=P)\}. \tag{A.4}$$

As an example, if we have a function singular in the origin, $\omega(z) = \dfrac{a}{z}$, then

$$-\oint_{C_1} \omega(z)dz = -\oint_{C_1} \frac{a}{z}dz = -\int_0^{2\pi} \rho \, i \, d\phi \frac{e^{i\phi}}{\rho e^{i\phi}} = 2\pi \, i \, a, \tag{A.5}$$

where we used the transformation:

$$z = \rho e^{i\phi} \rightarrow dz = \rho \, i \, d\phi \, e^{i\phi}. \tag{A.6}$$

This result can be generalized to any finite number of isolated singularities contained in the region $\mathfrak{C}$. This is the so-called residue theorem:

$$\oint_{C_0} w(z)dz = 2\pi i \sum_{j=1}^{N} \text{Res}\{w(z=P_j)\}. \tag{A.7}$$

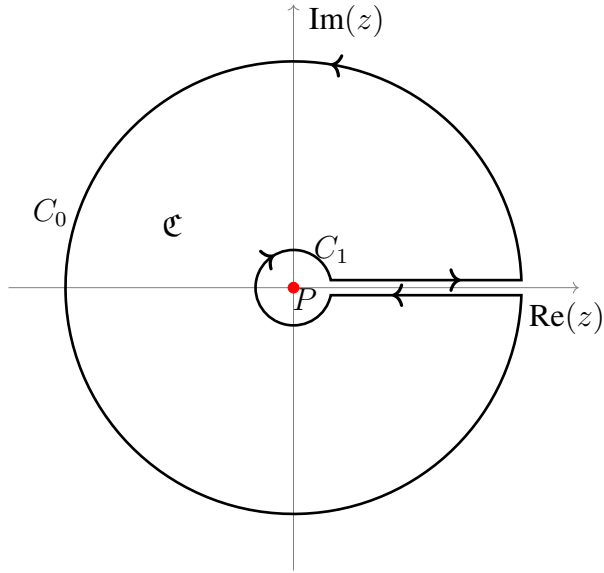

**Fig. A.1** How to exclude an isolated singularity.

## A.2 Multivalued functions

In complex analysis, it is not unusual to find functions which get different values for the same point in its domain. These are called multivalued functions, which are ordinary complex functions locally, but require care when they are described over all their domains. Let us see this with two examples.

- $\log z$

Along the real axis, the logarithmic function is defined as:

$$\log x := \int_1^x \frac{dx'}{x'}, \qquad x \in \mathbb{R}^+. \tag{A.8}$$

In analogy, we write

$$\log z := \int_1^z \frac{dz'}{z'}, \qquad z \in \mathbb{C}. \tag{A.9}$$

If the integral is performed along a path which does not include the origin, then the function $\frac{1}{z'}$ is analytic. We write now $z'$ in terms of polar coordinates $z'(= x' + iy') = \rho' e^{i\phi'}$. We calculate $\log z$ along the path shown in Fig. A.2, which excludes the origin. Thus:

$$\log z = \int_1^\rho \frac{dx'}{x'} + \int_0^\phi \frac{i\rho e^{i\phi'} d\phi'}{\rho e^{i\phi'}} = \log \rho + i\phi. \tag{A.10}$$

We separated the real part from the imaginary part of $\log z$:

$$\mathrm{Re}(\log z) = \log \rho, \qquad \mathrm{Im}(\log z) = i\phi. \tag{A.11}$$

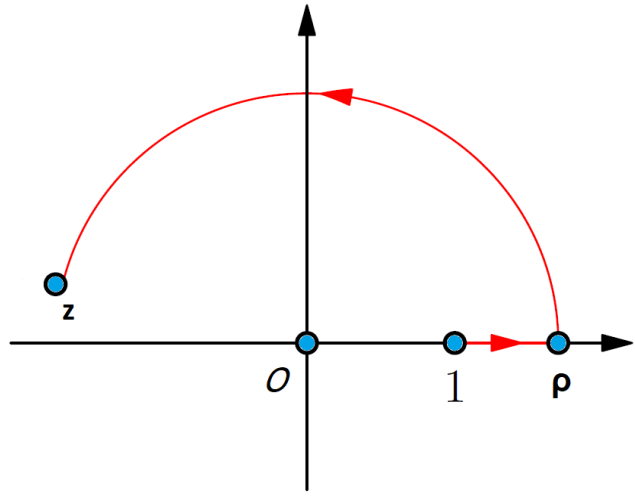

**Fig. A.2** Path used for the calculus of $\log z$.

We immediately notice that:
$$\oint \frac{dz'}{z'} = i2\pi . \qquad (A.12)$$

Although $\log x$ (with $x \in \mathbb{R}^+$) is a real function, $\log z$ is a multivalued function in complex analysis. In fact, if we use Eq. A.10 to evaluate $\log z$ in the same point after a full rotation, we get a different value:

$$\phi = 0 \longrightarrow \log z = \log \rho$$
$$\phi = 2\pi \longrightarrow \log z = \log \rho + i2\pi. \qquad (A.13)$$

Therefore, for a given point $z$, at which $\log z = \log \rho + i\phi$ (with $\phi \in [0, 2\pi)$), the value obtained after $n$ full rotations is:

$$(\log z)_n = \log \rho + i\phi + i2n\pi . \qquad (A.14)$$

Due to our choice of $\phi \in [0, 2\pi)$, there is a branch cut that extends from the origin to infinity along the positive real axis, see below.

- $\sqrt{z}$

Another common example for multivalued functions is $\sqrt{z}$. As before, if we use polar coordinates, we obtain:
$$f(z) = \sqrt{z} = \sqrt{\rho} e^{i\phi/2}. \qquad (A.15)$$

Here:
$$\sqrt{z} = \sqrt{\rho} e^{i\phi/2} \qquad \text{for } \phi \in [0, 2\pi). \qquad (A.16)$$

Then, by applying a full rotation:

$$\phi \to \phi + 2\pi \implies \sqrt{z} = \sqrt{\rho}e^{i\phi/2} \to \sqrt{\rho}e^{i(\phi+2\pi)/2} = \sqrt{\rho}e^{i\phi/2}e^{i\pi} = -\sqrt{\rho}e^{i\phi/2}. \quad (A.17)$$

By a second full rotation we get the original value

$$\phi \to \phi + 4\pi \implies \sqrt{z} = \sqrt{\rho}e^{i\phi/2} \to \sqrt{\rho}e^{i(\phi+4\pi)/2} = \sqrt{\rho}e^{i\phi/2}e^{2i\pi} = \sqrt{\rho}e^{i\phi/2} \quad (A.18)$$

and so on.
In summary,

$$\begin{aligned}(\sqrt{z})_I &= \sqrt{\rho}e^{i\phi/2}, & \phi \in [0, 2\pi) \\ (\sqrt{z})_{II} &= -\sqrt{\rho}e^{i\phi/2}, & \phi \in [0, 2\pi)\,.\end{aligned} \quad (A.19)$$

These examples of multivalued functions are shown in Fig. **??**.

In general, the fact that some functions get a different value after a full rotation can be reformulated by using the idea of the Riemann sheets.

We can take the function $\log z$ as an example to see this in more detail. Consider only one region ($0 \leq \phi \leq 2\pi$). If we project $\log z$ on a horizontal plane, we see that it is a single-valued function for all the values of $\phi \in [0, 2\pi)$, but for $\phi = 0$ it is discontinuous. The line at $\phi = 0$ is a "branch cut". Therefore, we can split the function $\log z$ into different surfaces, where it is always a single-valued function. These surfaces are called Riemann sheets. Thus, if we start from $\phi = 0$, we can easily understand in which single-valued region we are, by enumerating the Riemann sheets.

For instance, in the region $[0, 6\pi)$, we can say that $\log z$ is in the first Riemann sheet for $[0, 2\pi)$, in the second Riemann sheet for $[2\pi, 4\pi)$ and in the third Riemann sheet for $[4\pi, 6\pi)$.

Finally, since there is no discontinuity from a sheet to another (see Fig. **??**), we can interpret the branch cut as the connection between two different Riemann sheets, applying analytic continuation also from one sheet to another.

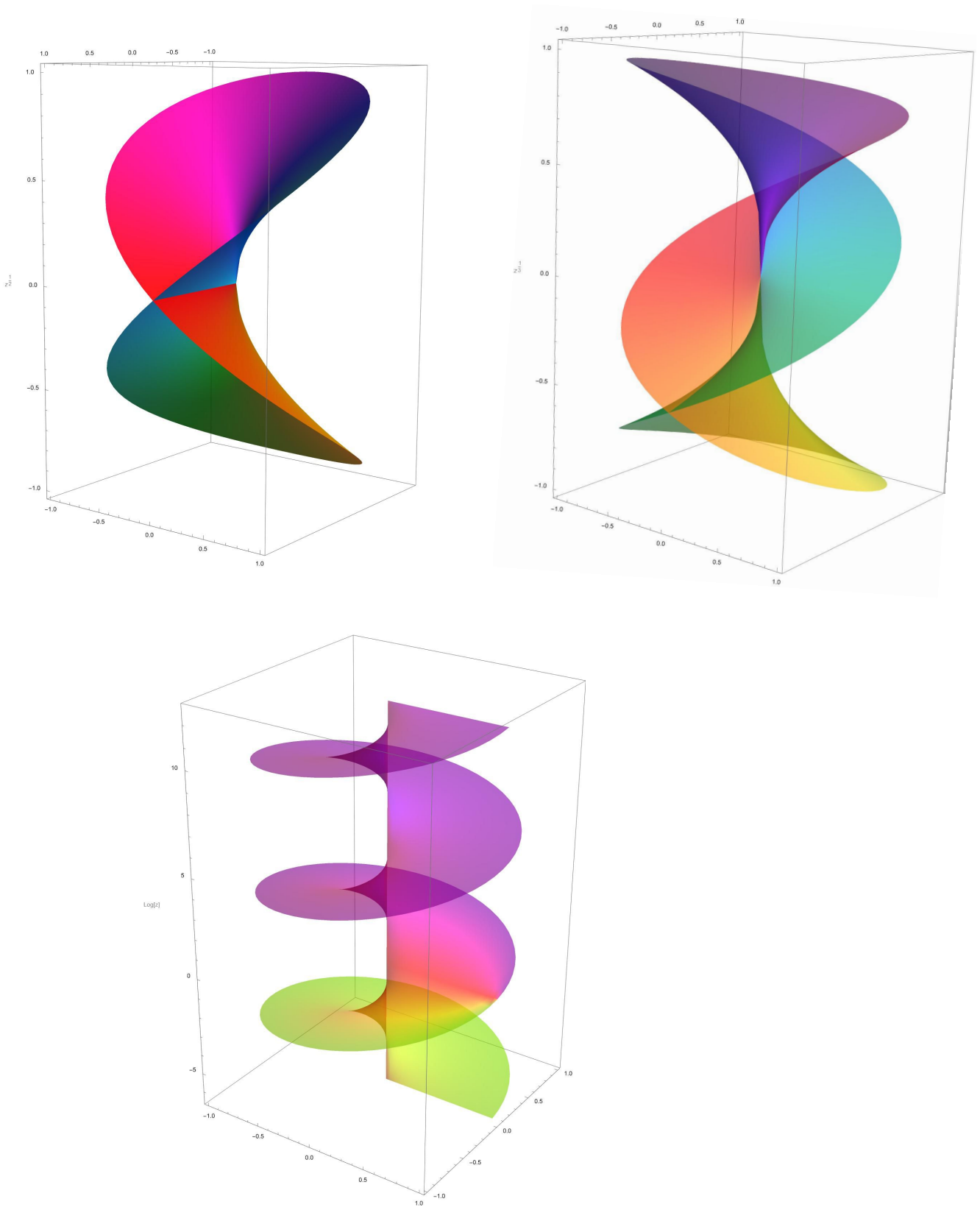

**Fig. A.3** Complex multivalued functions: $f(z) = z^{1/2}$ (up, left), $f(z) = z^{1/3}$ (up, right) and $f(z) = \log z$ (down).

# Appendix B

# Decays along an arbitrary direction

## B.1 Polarization states

The form of the PoVs states given in Eq. (3.26) does not describe a moving particle with arbitrary velocity $\vec{v} = (v_x, v_y, v_z)$, but can be boosted using the symmetric Lorentz matrix $\Lambda^\nu{}_\mu$, given by:

$$\Lambda^\nu{}_\mu(\vec{v}) = \begin{pmatrix} \gamma & -\gamma\beta_x & -\gamma\beta_y & -\gamma\beta_z \\ -\gamma\beta_x & 1+(\gamma-1)\hat{v}_x^2 & (\gamma-1)\hat{v}_x\hat{v}_y & (\gamma-1)\hat{v}_x\hat{v}_z \\ -\gamma\beta_y & (\gamma-1)\hat{v}_y\hat{v}_x & 1+(\gamma-1)\hat{v}_y^2 & (\gamma-1)\hat{v}_y\hat{v}_z \\ -\gamma\beta_z & (\gamma-1)\hat{v}_z\hat{v}_x & (\gamma-1)\hat{v}_z\hat{v}_y & 1+(\gamma-1)\hat{v}_z^2 \end{pmatrix}, \quad \text{(B.1)}$$

where

$$\hat{v}_i = \frac{v_i}{v} = \frac{v_i}{\sqrt{v_x^2+v_y^2+v_z^2}},$$

$$\beta_i = \frac{v_i}{c} = \hat{v}_i \beta$$

and

$$\gamma = \frac{1}{\sqrt{1-\beta^2}} = \frac{1}{\sqrt{1-\frac{v^2}{c^2}}}$$

is the Lorentz factor. The metric is chosen to be $g^{\mu\nu} = diag(+,-,-,-)$.

Note, the polarization states can be written in the canonical or in the helicity basis (see Ref. [188]). This distinction gains relevance when the decay is evaluated along a non-fixed direction. Yet, the results obtained in Section 3.5 were calculated using the states written in a canonical basis, which brings to the correct results since one of the final states has $J = 0$. Thus, the sign in front of $\nu$ (Eq. (3.21), for instance) becomes irrelevant since $\nu = 0$.

The boosted PoVs gets then the form[‡]:

$$\epsilon^\nu(\vec{k},+1) = \Lambda^\nu{}_\mu(-\vec{v})\epsilon^\mu(\vec{0},+1) = -\frac{\Lambda^\nu{}_\mu}{\sqrt{2}}\begin{pmatrix}0\\1\\i\\0\end{pmatrix} = -\frac{1}{\sqrt{2}}\begin{pmatrix}\gamma(\beta_x + i\beta_y)\\1+(\gamma-1)(\hat{v}_x^2 + i\hat{v}_x\hat{v}_y)\\i+(\gamma-1)(\hat{v}_y\hat{v}_x + i\hat{v}_y^2)\\(\gamma-1)(\hat{v}_z\hat{v}_x + i\hat{v}_z\hat{v}_y)\end{pmatrix}, \tag{B.2}$$

$$\epsilon^\nu(\vec{k},-1) = \Lambda^\nu{}_\mu(-\vec{v})\epsilon^\mu(\vec{0},-1) = \frac{\Lambda^\nu{}_\mu}{\sqrt{2}}\begin{pmatrix}0\\1\\-i\\0\end{pmatrix} = \frac{1}{\sqrt{2}}\begin{pmatrix}\gamma(\beta_x - i\beta_y)\\1+(\gamma-1)(\hat{v}_x^2 - i\hat{v}_x\hat{v}_y)\\-i+(\gamma-1)(\hat{v}_y\hat{v}_x - i\hat{v}_y^2)\\(\gamma-1)(\hat{v}_z\hat{v}_x - i\hat{v}_z\hat{v}_y)\end{pmatrix} \tag{B.3}$$

and

$$\epsilon^\nu(\vec{k},0) = \Lambda^\nu{}_\mu(-\vec{v})\epsilon^\mu(\vec{0},0) = \Lambda^\nu{}_\mu\begin{pmatrix}0\\0\\0\\1\end{pmatrix} = \begin{pmatrix}\gamma\beta_z\\(\gamma-1)(\hat{v}_x\hat{v}_y)\\(\gamma-1)(\hat{v}_y\hat{v}_z)\\1+(\gamma-1)\hat{v}_z^2\end{pmatrix}. \tag{B.4}$$

The case of one particle decaying into 2 states greatly simplifies the situation, as the two products move in opposite direction in the parent rest frame. If we identify the moving direction with the $z$-axis, then

$$\hat{v}_x = \hat{v}_y = 0,$$
$$\hat{v}_z = 1 \quad \text{and}$$
$$\beta = \sum_i \sqrt{\beta_i^2} = \beta_z. \tag{B.5}$$

In this case:

$$\Lambda^\nu{}_\mu = \begin{pmatrix}\gamma & 0 & 0 & \gamma\beta\\0 & 1 & 0 & 0\\0 & 0 & 1 & 0\\\gamma\beta & 0 & 0 & \gamma\end{pmatrix}, \tag{B.6}$$

---

[‡]In general, one should define two PoVs for each value of M, in two oppostite directions. This is not necessary in our case (see later on in Appendix B.2).

and

$$\epsilon^\mu(\vec{k},+1) = -\frac{1}{\sqrt{2}}\begin{pmatrix}0\\1\\i\\0\end{pmatrix}, \quad \epsilon^\mu(\vec{k},-1) = \frac{1}{\sqrt{2}}\begin{pmatrix}0\\1\\-i\\0\end{pmatrix}, \quad \epsilon^\mu(\vec{k},0) = \begin{pmatrix}\gamma\beta\\0\\0\\\gamma\end{pmatrix}. \quad (B.7)$$

One can also notice that, if $\vec{v} = (0,0,v_z)$, one gets $\epsilon^\mu(\vec{k},\pm 1) = \epsilon^\mu(0,\pm 1)$. For convenience, we will use from now on the case of $v_z$ as the only non-zero component of $\vec{v}$. Analogously, for the case $J = 2$, the form of the five polarization tensors boosted along the $z$-axis can be obtained from the PoT$_2$s in the rest frame together with the $\Lambda^\nu{}_\mu$ given in Eq. (B.6). Using Eq. (3.33), with $\vec{k} = \vec{0}$, one gets:

$$\epsilon^{\mu\nu}(\vec{0},\pm 2) = \frac{1}{2}\begin{pmatrix}0 & 0 & 0 & 0\\0 & 1 & \pm i & 0\\0 & \pm i & -1 & 0\\0 & 0 & 0 & 0\end{pmatrix}, \quad \epsilon^{\mu\nu}(\vec{0},\pm 1) = -\frac{1}{2}\begin{pmatrix}0 & 0 & 0 & 0\\0 & 0 & 0 & 1\\0 & 0 & 0 & \pm i\\0 & 1 & \pm i & 0\end{pmatrix}, \quad (B.8)$$

$$\epsilon^{\mu\nu}(\vec{0},0) = -\frac{1}{\sqrt{6}}\begin{pmatrix}0 & 0 & 0 & 0\\0 & 1 & 0 & 0\\0 & 0 & 1 & 0\\0 & 0 & 0 & -2\end{pmatrix}.$$

These matrices can be compared with the ones present in Ref. [185]. The boosted PoT$_2$s can be equivalently obtained by the use of Eq. (3.33) or by boosting the tensors in Eq. (B.8). Thus, we obtain:

$$\epsilon^{\mu\nu}(\vec{k},\pm 2) = \frac{1}{2}\begin{pmatrix}0 & 0 & 0 & 0\\0 & 1 & \pm i & 0\\0 & \pm i & -1 & 0\\0 & 0 & 0 & 0\end{pmatrix}, \quad \epsilon^{\mu\nu}(\vec{k},\pm 1) = \mp\frac{1}{2}\begin{pmatrix}0 & \gamma\beta & \pm i\gamma\beta & 0\\\gamma\beta & 0 & 0 & \gamma\\\pm i\gamma\beta & 0 & 0 & \pm i\gamma\\0 & \gamma & \pm i\gamma & 0\end{pmatrix},$$

$$\epsilon^{\mu\nu}(\vec{k},0) = -\frac{1}{\sqrt{6}}\begin{pmatrix}-2\gamma^2\beta^2 & 0 & 0 & -2\gamma^2\beta\\0 & 1 & 0 & 0\\0 & 0 & 1 & 0\\-2\gamma^2\beta & 0 & 0 & -2\gamma^2\end{pmatrix}. \quad (B.9)$$

For completeness, it is important to recall that in the rest frame of the decaying particle:

$$k_A^\mu = \begin{pmatrix} m_A \\ 0 \\ 0 \\ 0 \end{pmatrix}, \quad k_B^\mu = \begin{pmatrix} E_B \\ 0 \\ 0 \\ |\vec{k}| \end{pmatrix} \quad \text{and} \quad k_C^\mu = \begin{pmatrix} E_C \\ 0 \\ 0 \\ -|\vec{k}| \end{pmatrix}, \tag{B.10}$$

where the indices $A$ $B$ and $C$ refers to the particles in the decay $A \to BC$. Additionally, we have that $\beta = \frac{k_B}{E_B}$ and $\gamma = \frac{E_B}{M_B}$.

## B.2 The choice of the boost

One should be careful when defining the boosted PoVs and PoT$_J$s: the two decaying particles are moving in opposite directions, then the matrix $\Lambda^\nu{}_\mu$ should be different for the two products. As an example, the form of $\epsilon^\mu(\vec{k}, 0)$ in Eq. (B.7) should be $(-\gamma\beta, 0, 0, \gamma)$ for one particle and $(+\gamma\beta, 0, 0, \gamma)$ for the other. The fact that we consider only one boost direction is a simplification allowed in our particular case; thus, since one of the decaying products is always a spin-0 state, no PoV or PoT$_J$ will be related to it. Therefore, the boost is relevant only for the particles with non-zero spin, whose velocity can be taken along the same arbitrary direction for each decay analyzed in this work. This point justifies the use of only one direction for $\Lambda^\nu{}_\mu$. Our choice is therefore based on the following diagram (given in the case of the decay $a_1(1260) \to \rho\pi$):

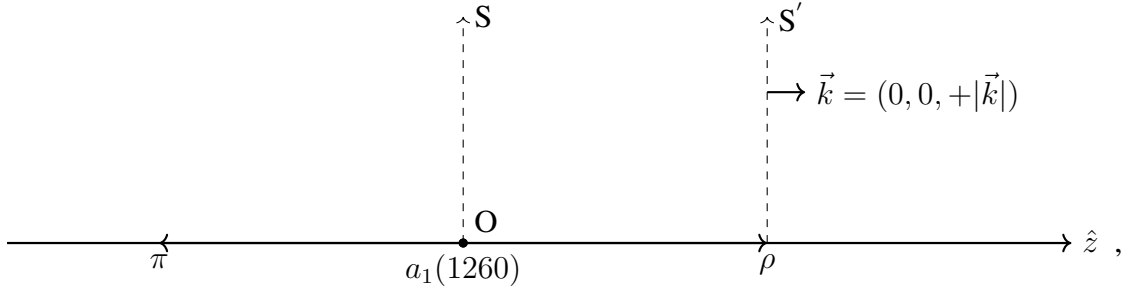

where O is the origin in the rest frame of reference S of $a_1(1260)$, whose decay into the particle $\rho$, moving along the positive direction of $\hat{z}$ with momentum $\vec{k} = (0, 0, +|\vec{k}|)$. The magnitude of the momentum is given by $k = |\vec{k}|$, as the measurements are done in the frame S. The PoVs are known in the frame of reference S$'$, but we have to boost them into the frame of reference S, where the calculations are performed. Therefore, $\epsilon^\mu|_{S'}$ must be boosted into the reference frame S. The polarization states must then be boosted from S$'$ to S, in the opposite direction of $\hat{z}$.

# Appendix C

# Clebsch-Gordan coefficients

The Clebsch-Gordan coefficients appear when dealing with angular momenta. In Sec. 3.3.1 we introduced the two relations:

$$|S, m_s\rangle = |s, \lambda\rangle \otimes |\sigma, \nu\rangle \tag{C.1}$$

and

$$|J, M_J\rangle = |\ell, m_\ell\rangle \otimes |S, m_s\rangle . \tag{C.2}$$

The value in the second place in each ket represents the projection on the third axis ($\hat{z}$) of the quantity given in the first place of the same ket. The last two relations give, as consequence, the two Clebsch-Gordan coefficients,

$$\langle \ell 0 S m_s | J M_J \rangle = \langle \ell, 0, S, m_s | J, M_J \rangle$$

and

$$\langle s\lambda\sigma - \nu | S m_s \rangle = \langle s, \lambda, \sigma, -\nu | S, m_s \rangle ,$$

in Eq. (3.21):

$$F^J_{\lambda\nu} = \sum_{\ell S} \sqrt{\frac{2\ell + 1}{2J + 1}} \langle \ell 0 S m_s | J M_J \rangle \langle s\lambda\sigma - \nu | S m_s \rangle G^J_{\ell S}.$$

As explained in Section 3.3.1, the opposite sign in front of $\nu$ is due to the fact that they are helicity and include the information that the two particles created by the decay move in opposite direction. The general Clebsch-Gordan coefficient for a decay

$$|JM_J\rangle \to |j_1 m_1\rangle |j_2 m_2\rangle ,$$

is
$$\langle j_1 m_1 j_2 m_2 | J M_J \rangle.$$

We consider now the decay $a_1(1260) \to \rho\pi$. The quantum numbers of this process are $J^{++} \to s^{--}\sigma^{-+}$, with $J = 1$, $s = 1$ and $\sigma = 0$. The Eqs. (C.1) and (C.2) require the conditions:

$S \in [|s - \sigma|, s + \sigma] \to S = 1$,
$\nu \in [-\sigma, +\sigma] \to \nu = 0$,
$\lambda \in [-s, +s] \to \lambda = -1, 0, +1$,
$m_s = \lambda - \nu \to m_s = -1, 0, +1$.

Finally, since $\vec{\ell} + \vec{S} = \vec{J}$, the permitted values for $\ell$ are 0 and 2. We can therefore have 6 combinations considering the freedom of $\lambda$ and $\ell$ ($m_s$ is not free to vary, since it is equal to the value of $\lambda$). The cases with $\lambda = -1$ and $+1$ differ only for the sign of the amplitude, while the magnitude is the same. Since our goal is to find a ratio between two amplitudes, the identical change of sign in both the numerator and the denominator together is irrelevant on the result. We thus neglect the values with negative $\lambda$ (this statement is valid for the other decay analyzed too). We can then rewrite the Clebsch-Gordan coefficient for the decay of $a_1(1260)$ as

$$\langle \ell 0 S m_s | J M_J \rangle = \langle \ell 0 1 \lambda | 1 \lambda \rangle$$

and

$$\langle s \lambda \sigma - \nu | S m_s \rangle = \langle 1 \lambda 0 0 | 1 \lambda \rangle.$$

Thus, the possible helicity amplitudes are $F^1_{10}$ and $F^1_{00}$. Using Eq. (3.21) we get:

$$F^1_{10} = \sum_{\ell 1} \sqrt{\frac{2\ell + 1}{3}} \langle \ell 0 1 1 | 1 1 \rangle \langle 1 1 0 0 | 1 1 \rangle G^1_{\ell 1}$$

$$= \sqrt{\frac{1}{3}} \langle 0 0 1 1 | 1 1 \rangle \langle 1 1 0 0 | 1 1 \rangle G^1_{01} + \sqrt{\frac{5}{3}} \langle 2 0 1 1 | 1 1 \rangle \langle 1 1 0 0 | 1 1 \rangle G^1_{21}$$

$$= \sqrt{\frac{1}{3}} \cdot 1 \cdot 1 \cdot G^1_{01} + \sqrt{\frac{5}{3}} \cdot \frac{1}{\sqrt{10}} \cdot 1 \cdot G^1_{21} = \sqrt{\frac{1}{3}} G^1_{01} + \sqrt{\frac{1}{6}} G^1_{21}, \quad \text{(C.3)}$$

$$F^1_{00} = \sum_{\ell 1} \sqrt{\frac{2\ell + 1}{3}} \langle \ell 0 1 0 | 1 0 \rangle \langle 1 0 0 0 | 1 0 \rangle G^1_{\ell 1}$$

$$= \sqrt{\frac{1}{3}} \langle 0 0 1 0 | 1 0 \rangle \langle 1 0 0 0 | 1 0 \rangle G^1_{01} + \sqrt{\frac{5}{3}} \langle 2 0 1 0 | 1 0 \rangle \langle 1 0 0 0 | 1 0 \rangle G^1_{21}$$

$$= \sqrt{\frac{1}{3}} \cdot 1 \cdot 1 \cdot G^1_{01} + \sqrt{\frac{5}{3}} \cdot -\sqrt{\frac{2}{5}} \cdot 1 \cdot G^1_{21} = \sqrt{\frac{1}{3}} G^1_{01} - \sqrt{\frac{2}{3}} G^1_{21}. \quad \text{(C.4)}$$

# Appendix D

# Decay formalism

## D.1 General features

When dealing with decays, we should keep in mind the following set-up: if $N_0$ is the number of unstable particles at the time $t = 0$, then the number of that particles after a time $t \geq 0$ is given by:

$$N(t) = N_0 e^{-\Gamma t}. \tag{D.1}$$

One can also define (in natural units) the mean lifetime $\tau$ as the inverse of the decay width $\Gamma$, namely $\tau = \frac{1}{\Gamma}$. For a single particle, the term $e^{-\Gamma t}$ is the survival probability $p(t)$. From a QM point of view, $p(t) = |A(t)|^2$, with $A(t)$ being the survival probability amplitude. If we consider the state $|\psi\rangle$ as created at time $t = 0$, then

$$A(t) = \langle \psi | e^{-iHt} | \psi \rangle, \tag{D.2}$$

where $H$ is the Hamiltonian that determines the dynamical evolution of the system under study. We now focus on the most relevant features of the decay theory for this work, and for more details we refer to Ref. [189].

Let $d_\psi(E)$ be the spectral function i.e. the energy distribution of the state $\psi$. Then (up to a phase):

$$|\psi\rangle = \int_{-\infty}^{+\infty} \sqrt{d_\psi(E)} |E\rangle \, dE. \tag{D.3}$$

It follows that

$$1 = \langle \psi | \psi \rangle = \int_{-\infty}^{+\infty} d_\psi(E) dE, \tag{D.4}$$

and

$$A(t) = \int_{-\infty}^{+\infty} dE \, d_\psi(E) e^{-iEt}. \tag{D.5}$$

The form of the spectral function is often chosen to be a Breit-Wigner distribution (BWd), given as:

$$d_\psi(E) = \frac{\Gamma}{2\pi} \frac{1}{(E - m_\psi)^2 + \Gamma^2/4}. \tag{D.6}$$

The Fourier transform of this distribution leads to

$$A(t) = e^{-im_\psi t - \Gamma t/2} \to p(t) = e^{-\Gamma t}. \tag{D.7}$$

Yet, BWd cannot be physical, since it does not fulfill these two conditions:

1) $d_\psi(E) = 0$ below a certain minimal energy $E_{min}$.

2) The mean energy must be finite, i.e.

$$\int_{-\infty}^{+\infty} d_\psi(E) E \, dE < I, \quad I \to \infty. \tag{D.8}$$

They are instead fulfilled by e.g. the so-called Sill distribution, which takes the form (see also later on):

$$d_\rho^{Sill}(E) = \frac{2E}{\pi} \frac{\tilde{\Gamma}\sqrt{E^2 - E_{th}^2}}{(E^2 - m^2)^2 + (\tilde{\Gamma}\sqrt{E^2 - E_{th}^2})^2} \theta(E - E_{th}), \tag{D.9}$$

where

$$\tilde{\Gamma} = \frac{\Gamma m}{\sqrt{m^2 - E_{th}^2}}. \tag{D.10}$$

## D.2 Two-body decays

Using Eq. (2.7), it is possible to derive the general form of the decay width of a decaying particle $A$ into $n$-particles (written in the rest frame of the decaying particle):

$$\Gamma = \int \frac{\mathfrak{s}_f}{2m_A} \left( \prod_{j=1}^{n} \frac{d^3\vec{k}_j}{(2\pi)^{3n} 2E_j} \right) (2\pi)^4 \delta^4(k_A - \sum_{j=1}^{n} k_j) |i\mathcal{A}|^2, \tag{D.11}$$

where $m_A$ is the mass of the parent particle, $\vec{k}_j$ and $k_j$ are the 3- and 4- momenta of the $j$-particle, and $\mathfrak{s}_f$ is a symmetry factor. In the case of a two body decay, $A \to BC$, the tree-level decay width gets the final general expansion:

$$\Gamma_{A \to BC} = \mathfrak{s}_f \frac{|\vec{k}|}{8\pi m_A^2} |i\mathcal{A}|^2, \tag{D.12}$$

where, in the parent particle's rest frame, $\vec{k} = \vec{k}_B = -\vec{k}_C$. This decay is schematized by the following Feynman diagram:

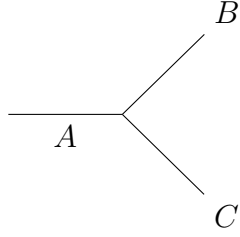

For pedagogical purposes, it is convenient to analyze the decay of a scalar into two scalar particles, and then see how the form of the decay width changes in some particular cases. In the following examples, the fields will be labeled as $\varphi_A$, $\varphi_B$ and $\varphi_C$. For convenience, the Lagrangian will be explicitly divided into two parts: the free term,

$$\mathscr{L}_{free} = \frac{1}{2}[(\partial_\mu \varphi_A)^2 - m_A^2 \varphi_A^2] + \frac{1}{2}[(\partial_\mu \varphi_B)^2 - m_B^2 \varphi_B^2] + \frac{1}{2}[(\partial_\mu \varphi_C)^2 - m_C^2 \varphi_C^2], \quad \text{(D.13)}$$

and the interaction term $\mathscr{L}_I$, which will vary according to the example given.

✽ Case $\mathscr{L}_I = g\varphi_A \varphi_B \varphi_C$

In the rest frame of the decaying particle, we have:

$$(m_A, \vec{0}) = (\sqrt{\vec{k}_B^2 + m_B^2}, \vec{k}_B) + (\sqrt{\vec{k}_C^2 + m_C^2}, \vec{k}_C). \quad \text{(D.14)}$$

Since $\vec{k} = \vec{k}_B = -\vec{k}_C$, we get:

$$|\vec{k}_B| = \frac{1}{2m_A}\sqrt{[m_A^2 - (m_B + m_C)^2][m_A^2 - (m_B - m_C)^2]}. \quad \text{(D.15)}$$

In this case, $|i\mathscr{A}| = |ig|$. We have then the form of the decay width for this process:

$$\Gamma_{\varphi_A \to \varphi_B \varphi_C} = \frac{g^2}{16\pi m_A^3}\sqrt{[m_A^2 - (m_B + m_C)^2][m_A^2 - (m_B - m_C)^2]}. \quad \text{(D.16)}$$

Note, this decay width is tacitly a function of the parameter $m_A$. If the decaying state is broad, one should take this dependence into account.

✽ Case $\mathscr{L}_I = g(\partial_\mu \varphi_A)\varphi_B(\partial^\mu \varphi_C)$.

This form of the Lagrangian leads to a similar form of the decay width. The only difference is the Feynman amplitude, that, because of the presence of the derivatives in the Lagrangian, leads to:

$$|i\mathscr{A}| = |ig(ik_{A\mu})(ik_C^\mu)|. \quad \text{(D.17)}$$

Finally:

$$\Gamma_{\varphi_A \to \varphi_B \varphi_C} = \frac{[g(k_A \cdot k_C)]^2}{16\pi m_A^3}\sqrt{[m_A^2 - (m_B + m_C)^2][m_A^2 - (m_B - m_C)^2]}, \quad \text{(D.18)}$$

where
$$k_A \cdot k_C = \frac{m_A^2 - m_B^2 + m_C^2}{2}. \tag{D.19}$$

✤ Case $\mathscr{L}_I = g_1 \varphi_A \varphi_B^2 + g_2 \varphi_A \varphi_C^2 + g_3 \varphi_A \varphi_B \varphi_C$.

This is the common case in which the particle has various decay channels. Here we should consider each part of the Lagrangian separately. Using Eq. (D.15), with $B \equiv C$, the first term reads:

$$\Gamma_{\varphi_A \to \varphi_B^2} = \mathfrak{s}_f \frac{|\vec{k}|_{B \equiv C}}{8\pi m_A^2} |-i\mathscr{A}|^2 = 2 \frac{g_1^2}{16\pi m_A^2} \sqrt{m_A^2 - 4m_B^2}, \tag{D.20}$$

where $\mathfrak{s}_f = 2$ because of identical particles[¶]. Analogously, we can calculate the other decay widths $\Gamma_{\varphi_A \to \varphi_C^2}$ and $\Gamma_{\varphi_A \to \varphi_B \varphi_C}$; the total decay width associated with the Lagrangian is then given by:

$$\Gamma^{TOT} = \Gamma_{\varphi_A \to \varphi_B^2} + \Gamma_{\varphi_A \to \varphi_C^2} + \Gamma_{\varphi_A \to \varphi_B \varphi_C}. \tag{D.21}$$

This case becomes very important when dealing with particles with isospin. If we consider, for example, the process $\sigma \to \pi\pi$, we find that:

$$\Gamma_{\sigma \to \pi\pi} = \Gamma_{\sigma \to \pi^+ \pi^-} + \Gamma_{\sigma \to \pi^0 \pi^0}. \tag{D.22}$$

✤ Case $\varphi_A$ decaying into non-scalar particles

As a final example, we show the case of the decay of a scalar particle $\varphi_A$ into two non scalar particles, $\varphi_B$ and $\varphi_C$. Let us consider $B = C$ and vector particles. The free Lagrangian is given by

$$\mathscr{L}_{free} = \frac{1}{2}[(\partial_\mu \varphi_A)^2 - m_A^2 \varphi_A^2] - \frac{1}{2}\left[\frac{1}{2}(\partial_\mu \varphi_{B\nu} - \partial_\nu \varphi_{B\mu})^2 - m_B^2 \varphi_{B\mu}^2\right]. \tag{D.23}$$

For the calculation of the decay width we are interested in the form of the interacting term:

$$\mathscr{L}_I = g \varphi_A \varphi_{B\mu} \varphi_B^\mu. \tag{D.24}$$

Again, the form of the decay width is given by:

$$\Gamma_{\varphi_A \to \varphi_{B\mu} \varphi_B^\mu} = 2 \frac{|\vec{k}|}{8\pi m_A^2} |i\mathscr{A}|^2. \tag{D.25}$$

---

[¶]Actually, one has $\frac{1}{2} \cdot |2|^2 = 2$.

In this case we have to sum over the polarization states, as:

$$i\mathcal{A}^{a,b} = ig\epsilon_\mu^{a*}(\vec{k}_B)\epsilon^{b,\mu*}(\vec{k}_B) ,$$

$$\sum_{a,b}|i\mathcal{A}^{a,b}|^2 = g^2 \sum_{a,b}[\epsilon_\mu^{a*}(\vec{k}_B)\epsilon_\nu^{a*}(\vec{k}_B)][\epsilon^{b,\mu*}(\vec{k}_B)\epsilon^{b,\nu*}(\vec{k}_B)]$$

$$= g^2\left(2 + \frac{(m_A^2 - 2m_B^2)^2}{4m_A^4}\right). \quad (D.26)$$

## D.3 Decays of the scalar glueball

We now calculate the tree-level decay of a scalar glueball into two particles, referring to:

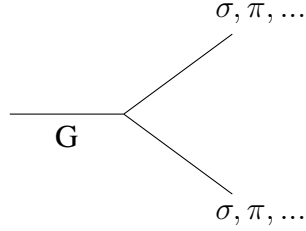

We start from a chiral and dilaton invariant potential build up using a toy model with the dilaton, $\sigma$ and $\pi$ fields:

$$V(G,\sigma,\pi) = V_{dil}(G) + aG^2(\pi^2 + \sigma^2) + \frac{\lambda}{4}(\pi^2 + \sigma^2)^2. \quad (D.27)$$

In the $(\sigma,\pi)$ space, chiral transformations are $O(2)$ rotations. Since the parameters $a$ and $\lambda$ are dimensionless, the only parameter which is dimentionful is encoded in the dilaton potential $V(G)$, i.e. $\Lambda_G$. Both the fields $G$ and $\sigma$ condense if $a < 0$, thus realizing spontaneous symmetry breaking. If we neglect the $\sigma$-$G$ mixing for illustrative purposes, we get the vacuum expectation value (v.e.v.) for both $G$ and $\sigma$. The v.e.v. for G is set at $G_0 = \Lambda_G$, while the v.e.v. for $\sigma$ is:

$$\sigma_0^2 = -2\frac{a}{\lambda}\Lambda_G^2 \simeq f_\pi^2 , \quad (D.28)$$

where $f_\pi$ is the pion decay constant. We then rescale the field $G$ around its minimum, namely $G = G + \Lambda_G$, and then rewrite $aG^2(\pi^2 + \sigma^2)$ as

$$a(G^2 + \Lambda_G^2 + 2G\Lambda_G)(\pi^2 + \sigma^2). \quad (D.29)$$

The term we are interested in is then clearly $2a\Lambda_G[G(\pi^2+\sigma^2)]$. Since the mass of the sigma state is $m_\sigma^2 = 2\lambda\sigma_0^2$, we have all the information to find the expression of the decay amplitude of a scalar glueball. Finally, using the formalism of this appendix, we obtain that with a nonzero pion mass, the decay of $G$ into two pion states is:

$$\Gamma_{G\to\pi\pi} = 2\frac{\sqrt{\frac{m_G^2}{4}-m_\pi^2}}{8\pi m_G^2}\left(\frac{m_\sigma^2}{2\Lambda_G}\right)^2. \tag{D.30}$$

The previous equation is valid in the illustrative case of a single pseudoscalar field $\pi$. In the reality, where the isospin of the pion equals 1, Eq. (D.30) becomes:

$$\Gamma_{G\to\pi\pi} = 6\frac{\sqrt{\frac{m_G^2}{4}-m_\pi^2}}{8\pi m_G^2}\left(\frac{m_\sigma^2}{2\Lambda_G}\right)^2. \tag{D.31}$$

The values given in [99] are $\Lambda_G \simeq 0.4$ GeV, $m_G \approx 1.7$ GeV and $m_\sigma \simeq 1.3$ respectively (the latter roughly corresponds to the $f_0(1370)$). Using these values, one obtain $\Gamma_{G\to\pi\pi} \simeq 0.310$ GeV. The extension to $SU(3)$ carries two other possible decays: $G \to KK$ and $G \to \eta\eta$. They are:

$$\Gamma_{G\to KK} = 8\frac{\sqrt{\frac{m_G^2}{4}-m_K^2}}{8\pi m_G^2}\left(\frac{m_\sigma^2}{2\Lambda_G}\right)^2, \quad \Gamma_{G\to\eta\eta} = 2\frac{\sqrt{\frac{m_G^2}{4}-m_\eta^2}}{8\pi m_G^2}\left(\frac{m_\sigma^2}{2\Lambda_G}\right)^2. \tag{D.32}$$

The term $\frac{m_\sigma^2}{2\Lambda_G}$ in the parenthesis is identical in each case, since it comes from the flavour blindness of $G$.

The symmetry factor, which is 6, 8 and 2 for the decay of $G$ into two $\pi$, $K$ and $\eta$ respectively, reads:

$$\mathfrak{s}_f = 2(2I+1). \tag{D.33}$$

We notice that $\mathfrak{s}_f$ includes two contributions: the term $(2I+1)$ is related to flavour symmetry, while the factor $2$ in front comes from the counting of identical particles. In particular, this last factor is the product of $\frac{1}{2}$, due to the integration over half of the solid angle because of identical particles, and a factor 2 in the amplitude. Since the numerical values of the last decays are $\Gamma_{G\to KK} \simeq 0.340$ GeV and $\Gamma_{G\to\eta\eta} \simeq 0.080$ GeV, we get that the sum of all the three decays is about 0.729 GeV. This approach does not allow the calculation of the decay $G \to \rho\rho \to 4\pi$, expected to be also sizable. Although the lack of this last information, we can estimate that, within this simple model, the total decay of our glueball should be not smaller than 1 GeV, meaning that the glueball would be too wide to be observed [47]. Yet, the model described above serves only as an illustration, since it is too simple to be considered as realistic. For a more advanced approach see [51], where the glueball turns out to be narrow.

For what concerns the behaviour at large $N_C$ of the decay width, this scales as $N_C^{-1}$ for the process $\sigma \to \pi\pi$, and as $N_C^{-2}$ for any decay of the glueball into a couple of pseudoscalar mesons. These scaling can be easily calculating by considering how the parameters scales: $\Lambda_G \propto N_c$, $m_G \propto N_c^0$ and $\lambda \propto N_c^{-1}$. The parameter $a$, describing the $GG \to \pi\pi$ scattering, scales as $a \propto N_c^{-2}$. Additionally, $\sigma_0 \propto N_c^{1/2}$ and $m_\sigma \propto N_c^0$. All these large-$N_c$ dependence are in agreement with the literature [9].

## D.4 Chain decay

In the case of decays involving unstable final states, we cannot describe their full dynamics using only tree-level diagrams. This kind of decays are also relevant for this work; examples are $a_1(1260) \to \rho\pi$, $h_1(1170) \to \rho\pi$, and $\pi_2(1670) \to \rho\pi$.

We start from a general decay process $A \to BC$. We call $\mathfrak{m}_a$, $\mathfrak{m}_b$ and $\mathfrak{m}_c$ the masses of the three particles. If the masses are fixed, then the decay width:

$$\Gamma_{A \to BC}(\mathfrak{m}_a, \mathfrak{m}_b, \mathfrak{m}_c) = \Gamma^{tl}_{A \to BC}(\mathfrak{m}_a, \mathfrak{m}_b, \mathfrak{m}_c) \tag{D.34}$$

is a well defined quantity, whose value is the tree-level decay width $\Gamma^{tl}_{A \to BC}$. Let us consider now the case in which $B$ is unstable, i.e. $B \to \phi_1\phi_2$. Then $\mathfrak{m}_b$ is not anymore a fixed quantity and we can replace it with a variable $y$: $\mathfrak{m}_b \to y$. Thus, the total decay width, which considers both the decays of $A$ and $B$, is

$$\begin{aligned}\Gamma_{A \to \phi_1\phi_2 C} &= \int_0^\infty dy\, \Gamma^{tl}_{A \to BC}(\mathfrak{m}_a, y, \mathfrak{m}_c)\, d_B(y)\, \theta(\mathfrak{m}_a - y - \mathfrak{m}_c) \\ &= \int_{\mathfrak{m}_{\phi_1}+\mathfrak{m}_{\phi_2}}^{\mathfrak{m}_a - \mathfrak{m}_c} dy\, \Gamma^{tl}_{A \to BC}(\mathfrak{m}_a, y, \mathfrak{m}_c)\, d_B(y)\,,\end{aligned} \tag{D.35}$$

which is still a fixed value, but it depends on the spectral function $d_B(y)$ of the particle $B$. Table D.1 shows that the result from the correction to the tree-level width is not much different from the tree level result itself. For this calculation we use the Sill distribution, which is not only normalized to unity, but has also a built-in threshold [190].

All the decays in Table D.1 contains the $\rho$-meson, which decays further into two pions. Therefore, it is enough to use the single channel Sill distribution, whose explicit form,

| Decay | Tree-level width (MeV) | Integrated width (MeV) |
|---|---|---|
| $a_1(1260) \to \rho\pi$ | $420 \pm 35$ | $354 \pm 30$ |
| $h_1(1170) \to \rho\pi$ | $146 \pm 14$ | $142 \pm 14$ |
| $\pi_2(1670) \to \rho\pi$ | $80.6 \pm 10.8$ | $93.5 \pm 13$ |

**Table D.1** Comparison of the tree-level widths of the three decays with the integrated widths.

written in term of the squared energy in the center of mass, is given by:

$$d^{Sill}_{B=\rho}(s) = \frac{1}{\pi} \frac{\tilde{\Gamma}\sqrt{s - s_{th}}}{(s - m_\rho^2)^2 + (\tilde{\Gamma}\sqrt{s - s_{th}})^2}, \tag{D.36}$$

where

$$\tilde{\Gamma} = \frac{\Gamma_\rho m_\rho}{\sqrt{m_\rho^2 - s_{th}}}, \tag{D.37}$$

$m_\rho$ is the mass of the $\rho$-meson and $\Gamma_\rho$ is its total width. Using this spectral function in Eq. (3.48), we obtain the integrated widths listed in Table D.1.

# Appendix E

# Amplitudes and scattering lengths of glueball-glueball interaction

In this Appendix the explicit form of the tree level amplitudes and scattering length for the $D$, $G$, and $I$-waves are reported.

- $D$-wave

$$\mathscr{A}_2(s) = \frac{1}{2} \int_{-1}^{1} d\cos\theta \, \mathscr{A}(s, \cos\theta) P_2(\cos\theta)\,, \tag{E.1}$$

$$\mathscr{A}_2(s) = \frac{50 m_G^4}{(s - 4m_G^2)^3 \Lambda_G^2} \bigg[ -3(8m_G^4 - 6m_G^2 s + s^2) \\ + (s^2 - 2m_G^4 - 2m_G^2 s) \log\left(1 + \frac{s - 4m_G^2}{m_G^2}\right) \bigg] \tag{E.2}$$

and

$$a_2 = \frac{5}{6\pi m_G^3 \Lambda_G^2}\,. \tag{E.3}$$

- $G$-wave

$$\mathscr{A}_4(s) = \frac{1}{2} \int_{-1}^{1} d\cos\theta \, \mathscr{A}(s, \cos\theta) P_4(\cos\theta)\,, \tag{E.4}$$

$$\mathscr{A}_4(s) = \frac{-25 m_G^4}{3(s - 4m_G^2)^5 \Lambda_G^2} \bigg[ -5\left(46 m_G^4 - 2m_G^2 s - 5s^2\right)\left(s - 2m_G^2\right)\left(s - 4m_G^2\right) \\ + 6\left(74 m_G^8 - 124 m_G^6 s + 54 m_G^4 s^2 - 4 m_G^2 s^3 - s^4\right) \log\left(1 + \frac{s - 4m_G^2}{m_G^2}\right) \bigg] \tag{E.5}$$

and
$$a_4 = \frac{40}{63\pi m_G^7 \Lambda_G^2}. \tag{E.6}$$

- *I*-wave

$$\mathcal{A}_6(s) = \frac{1}{2}\int_{-1}^{1} d\cos\theta \mathcal{A}(s,\cos\theta) P_6(\cos\theta), \tag{E.7}$$

$$\mathcal{A}_6(s) = \frac{5m_G^4}{(s-4m_G^2)^7 \Lambda_G^2}\bigg[7(52m_G^8 + 648m_G^6 s - 428m_G^4 s^2 + 48m_G^2 s^3 + 7s^4)$$
$$(s-2m_G^2)(s-4m_G^2) + 10(1324m_G^{12} - 1692m_G^{10} s + 270m_G^8 s^2 +$$
$$400m_G^6 s^3 + 180m_G^4 s^4 + 18m_G^2 s^5 + s^6)\log\bigg(1 + \frac{s-4m_G^2}{m_G^2}\bigg)\bigg] \tag{E.8}$$

and
$$a_6 = \frac{2400}{3003\pi m_G^{11} \Lambda_G^2}. \tag{E.9}$$

# ACKNOWLEDGEMENT

I would like to take the opportunity to thank all the people who helped to bring forward my work on this thesis. First, I thank my supervisor, Francesco Giacosa, for being my guide during these years, for all the support and the useful discussion, and for all the nice and funny moments spent together ("vai, aiuta il bambino finito nella neve").

I thank my good friend Vanamali Shastry, who was carefully explaining me things any time I needed. I hope once he will become brave enough to learn how to ski.

Obviously I thank my colleague, girlfriend and future wife Sylwia, who was supporting me so much in all these years.

A good luck for his career in physics and in cooking to my roommate and colleague Shahriyar Jafarzade, together with his beloved tuna fish.

I thank also my other collaborators, Alessandro Pilloni, for the useful discussion, and Nodoka Yamanaka, who was so kind during my internship in Nagoya: it was a pleasure to meet you and discuss with you.

Ringrazio mamma e papà, che mi hanno sempre aiutato e mi sono stati sempre vicini. Vi voglio bene.

I also thank the professors from my university, as well as my colleagues: Leonardo Tinti, Stanisław Mrówczyński, Wojciech Broniowski, Giorgio (iron-stomach) Torrieri, Regina Stachura, Haradhan Adhikary and all the others.

I have also to thank for the useful discussion Markus Huber, Robert Kamiński and Stanisław Mrówczyński.

A particular thanks for all the time spent together is due to the nice people I met during the various schools: Cyrille and Lorenzo ("wtf!"), Guilherme, Fabrizio, Giuseppe.

Ringrazio i miei amici, per i bei momenti spesi insieme, nonché per il supporto ricevuto in questi anni: Silvio, Giova, Ghio, Danilo, Derrik, Sara, Bea, Gigi, Ila, Cri, Riccardo, Boss, Eli, Marcello, Alessandro e tutti gli altri qui non menzionati, ma che ringrazio comunque.

Infine ringrazio le mie care Mora e Pimpolina (o Chewbecca o Baby Yoda), per la loro gioia che mi ha sempre reso felice. Ringrazio anche tutti i gatti: Czarnuszka, Miri, Puffetta e Mruczek.

Questa tesi è dedicata a chi ho recentemente visto, ma che ora vedrò solo nei miei ricordi e nel mio cuore: il mio caro maestro, Mirco e i giovanissimi Pecorina e Grande Puffo, oltre al mio amato Beethoven.